\documentclass[reprint,aps,preprintnumbers,amsmath,amssymb,nofootinbib,epsf,epsfig]
{revtex4-2}
\usepackage[utf8]{inputenc}
\usepackage{graphicx}
\usepackage{amsmath}
\usepackage{amssymb}
\usepackage{array}
\usepackage{float}
\usepackage{booktabs}
\usepackage{multirow}
\usepackage{color}
\usepackage{caption}
\usepackage{subcaption}
\usepackage{natbib}
\usepackage{pifont}
\usepackage{braket}
\usepackage[colorlinks, citecolor=cyan]{hyperref}
\usepackage{epsfig}
\usepackage{slashed}
\usepackage{appendix}
\usepackage{comment}
\begin{document}
\newcommand{\bea}{\begin{eqnarray}}
\newcommand{\eea}{\end{eqnarray}}
\newcommand{\ldm}{\Delta m_{31}^2}
\newcommand{\sdm}{\Delta m_{21}^2}
\newcommand{\dcp}{\delta_{CP}}
\newcommand{\nn}{\nonumber}

\global\long\def\d{\partial}
\def\ds{\displaystyle}
\def\s1{\hat s}
\def\para{\parallel}
\def\CP{{\it CP}~}
\def\cp{{\it CP}}
\def\ml{m_\mu}

\title{\large Exploring supernova neutrino mass ordering at DUNE via quantum entanglement}
\author{Adikiran Johny$^a$}
\email{adikirann@gmail.com}
\author{Athulkrishna R$^a$}
\email{athulkrishnar35@gmail.com}
\author{Rudra Majhi$^b$}
\email{rudra.majhi95@gmail.com}
\author{Suchismita Sahoo$^a$}
\email{suchismita@cuk.ac.in}
\affiliation{$^a$Department of Physics, Central University of Karnataka, Kalaburagi-585367, India \\ $^b$Department of Physics,  Nabarangpur College, Nabarangpur, Odisha- 764059, India}
\begin{abstract}
The Deep Underground Neutrino Experiment (DUNE) offers strong sensitivity to neutrinos from a Galactic core collapse supernova, providing a powerful probe of neutrino flavor conversion and the neutrino mass ordering. In this work, we study supernova neutrino oscillations at DUNE using quantum entanglement as an organizing framework. Treating the three flavor neutrino system as an effective multipartite quantum state, we quantify flavor correlations through the entanglement of formation, concurrence, and negativity, expressed directly in terms of flavor survival and transition probabilities. Benchmark scenarios defined by representative variations of the electron neutrino survival probability are constructed for each entanglement measure. Event rates and fluences are computed for a supernova at 10 kpc, and the mass ordering sensitivity is evaluated using detector-level simulations performed with the \texttt{SNOwGLoBES} framework, employing the Garching supernova flux model and including the dominant detection channels in liquid argon: $\nu_e$ and $\bar{\nu}_e$ charged-current interactions on argon and elastic scattering on electrons. We analyze both individual and combined detection channels and incorporate $5\%$ normalization and energy calibration systematic uncertainties. Our results show that DUNE achieves a $5\sigma$ determination of the neutrino mass ordering for a supernova at distances of $\sim 20$~kpc for the $\nu_e$ charged current channel and $\sim 2$~kpc for the $\bar{\nu}_e$ channel, with the reach depending on the entanglement scenario considered. These results demonstrate that entanglement based observables provide a complementary and robust framework for probing supernova neutrino oscillations and the neutrino mass ordering.
\end{abstract}

\maketitle
\flushbottom

\section{Introduction}
The nucleosynthesis process during the final stage of a massive star results in the collapse of the stellar core, accompanied by an explosion that releases a large amount of energy. This process is known as the core collapse of a supernova \cite{Burrows:2000mk, Horiuchi:2017qja} and generally occurs for massive stars with masses above $8M_\odot$, where $M_\odot$ denotes the mass of the Sun. During the core collapse, most of the released energy is emitted in the form of neutrinos of all flavours over a timescale of a few seconds. These neutrinos carry crucial information about the supernova core collapse and open up the possibility of probing the inner nucleosynthesis processes of stellar bodies. As neutrinos interact only feebly with matter, they can easily escape the dense stellar environment, unlike photons. Consequently, neutrinos can provide early information about a supernova well before the arrival of electromagnetic radiation. The gravitational binding energy liberated by compressing the stellar core to nuclear densities amounts to about $10\%$ of its rest mass, corresponding to roughly $100~\mathrm{MeV}$ per nucleon. Nearly $99\%$ of this energy is carried away by neutrinos, while only $\sim 1\%$ remains available to power the explosion and the photon luminosity \cite{Raffelt:2025wty}. Following core bounce, the newly formed proto-neutron star cools via neutrino emission over a timescale of a few tens of seconds, during which neutrinos continue to dominate the total energy output of the supernova. So far, the only supernova detected on Earth is SN1987A, which occurred at a distance of about 50 kiloparsecs (kpc)
\cite{Lagage:1987xu}. Neutrinos from SN1987A were observed by detectors such as Kamiokande II, Irvine-Michigan-Brookhaven (IMB), and Baksan, approximately 2.5 hours before the detection of electromagnetic radiation \cite{Kamiokande-II:1987idp, IMB:1987klg, Novoseltseva:2009cr}. In the case of black hole formation, the stellar core may collapse without producing any bright electromagnetic signal; in such scenarios, a burst of neutrinos can play a vital role in identifying the location of black hole formation in the sky.

Neutrino oscillations, the phenomenon in which neutrinos change flavours as they propagate, are now well established by solar, atmospheric, reactor, and accelerator experiments \cite{PhysRevLett.81.1562, PhysRevLett.89.011301, PhysRevD.74.072003}. Precision measurements of the oscillation parameters have been achieved through recent global fits \cite{Esteban:2024eli}. Despite this progress, several fundamental questions remain unresolved, including the neutrino mass ordering, the value of the CP violating phase $\dcp$, and the octant of the mixing angle $\theta_{23}$. Current generation neutrino oscillation experiments have achieved important milestones in precise neutrino detection, energy measurement, and classification of interaction channels, and are therefore expected to observe a large number of supernova neutrino events if a nearby core collapse supernova occurs in the future. As supernova neutrinos propagate through dense stellar matter, vacuum, and the Earth, their flavor evolution is strongly influenced by matter effects and collective oscillations, making the detected signal highly sensitive to intrinsic neutrino properties. Consequently, astrophysical neutrino bursts provide a valuable opportunity to extract information on fundamental neutrino parameters while simultaneously improving our understanding of supernova dynamics.  

Among the current and next generation neutrino detectors, the Deep Underground Neutrino Experiment (DUNE) \cite{DUNE:2020ypp} is uniquely positioned to study neutrinos from a core collapse supernova. Based on large scale liquid argon time projection chamber (LArTPC) technology \cite{Rubbia:1977zz, ICARUS:2004wqc, bagby2021icarus, Acciarri_2017, MicroBooNE:2015bmn, abud2022protodune, abi2020dunephysics, DUNE:2020ypp}, DUNE provides excellent energy and time resolution together with detailed event reconstruction. Most importantly, DUNE is highly sensitive to electron neutrinos through the charged current interaction $\nu_e + {}^{40}\mathrm{Ar} \rightarrow e^- + {}^{40}\mathrm{K}^\ast$, in contrast to water Cherenkov \cite{Hyper-Kamiokande:2018ofw} and liquid scintillator detectors \cite{kamland2003, borexino2014, dayabay2012, lsnd2001, miniboone2018}, which primarily detect $\bar{\nu}_e$. This enhanced $\nu_e$ sensitivity enables DUNE to observe the early neutronization burst of a supernova, a robust and relatively model independent feature whose flavor content and spectral characteristics depend strongly on the neutrino mass ordering. In addition, the large detector mass and low energy threshold allow DUNE to record a high statistics neutrino signal from a Galactic supernova, enabling detailed studies of the time and energy evolution of the burst. Consequently, DUNE offers a powerful and complementary avenue for probing neutrino flavor transformation effects and extracting information on the neutrino mass ordering from a future supernova neutrino event.

Neutrino flavor evolution, which underlies the supernova neutrino signal observable at DUNE, is inherently quantum mechanical in nature. As neutrinos propagate and oscillate, their flavor states exist as coherent superpositions of mass eigenstates, leading to non-classical correlations among different flavor modes. Such correlations are naturally described within the framework of quantum entanglement, a fundamental property of multipartite quantum systems \cite{Horodecki:2009zz}. In recent years, quantum entanglement has emerged as a promising tool in neutrino physics, offering insights beyond conventional analyses based solely on event rates or energy spectra. Several well-established measures from quantum information theory, including the entanglement of formation (EOF) \cite{Bennett:1996gf, Wootters:1997id}, concurrence (CON) \cite{Hill:1997pfa, Wootters:1997id, Rungta:2001zcj}, and negativity (NEG) \cite{Vidal:2002zz, Peres:1996dw, Horodecki:2009zz}, allow these quantum correlations to be quantified directly in terms of neutrino oscillation probabilities. In the extreme matter conditions present during a core collapse supernova, where flavor transformation is strongly affected by matter and collective effects, the resulting entanglement patterns are expected to be sensitive to the underlying neutrino mass ordering. This opens up the possibility of employing entanglement based observables, accessible through supernova neutrino detection at DUNE, as an alternative and complementary approach to investigating the neutrino mass ordering. In this work, we investigate the sensitivity of DUNE to the supernova neutrino mass ordering using quantum entanglement measures, with event simulations carried out using the Supernova Neutrino Observatories with GLoBES (\texttt{SNOwGLoBES}) software \cite{snowglobes}. Studies of supernova neutrinos have been carried out extensively in the literature, with several works employing the \texttt{SNOwGLoBES} framework for detector level simulations \cite{DedinNeto:2023hhp, Dasgupta:2008cd, Wu:2014kaa, Scholberg:2017czd, Linzer:2019swe, Dasgupta:2008my, Dasgupta:2009mg, Gil-Botella:2016sfi, Das:2017iuj, Hyper-Kamiokande:2021frf, Dighe:2003be, Dighe:1999bi, Chiu:2013dya, Brdar:2022vfr, Serpico:2011ir, Choubey:2010up, Brdar:2018zds, Panda:2023rxa, Panda:2024avc}.

The paper is organized as follows. In Section~\ref{sec:framework}, we discuss the three flavor neutrino oscillation framework and the density matrix formalism, along with the supernova neutrino flux and survival probabilities. The details of the quantum entanglement measures, namely the entanglement of formation, concurrence, and negativity, are presented in Section~\ref{sec:entanglement}. Experimental setup and simulation details are described in Section~\ref{sec:expt_sim}. The results and their discussion are presented in Section~\ref{sec:result}. Finally, we summarize our findings and conclude the paper in Section~\ref{sec:conclusion}.
\section{Theoretical Framework}
\label{sec:framework}

\subsection{Three flavor Neutrino Oscillations}
\label{subsec:oscilaltions}

In the relativistic limit, neutrino flavor states can be treated as distinct quantum modes, and their time evolution naturally induces quantum correlations among these modes. Flavor oscillations arise from the coherent superposition of mass eigenstates, and when such correlations cannot be factorized into independent single-mode states, the system exhibits quantum entanglement. In this context, neutrino oscillations can be viewed as a manifestation of mode entanglement among the flavor degrees of freedom.

In the three-flavor framework, the flavor eigenstates $\ket{\nu_\alpha}$ $(\alpha=e,\mu,\tau)$ are related to the mass eigenstates $\ket{\nu_i}$ $(i=1,2,3)$ through the unitary PMNS matrix $U$,
\begin{equation}
\ket{\nu_\alpha} = \sum_{i=1}^{3} U_{\alpha i} \ket{\nu_i},
\end{equation}
which is parametrized in terms of the mixing angles $\theta_{12}$, $\theta_{13}$, $\theta_{23}$ and the CP-violating phase $\delta$ as
\begin{equation}
U = \begin{pmatrix} 
c_{12}c_{13} & s_{12}c_{13} & s_{13}e^{-i\delta} \\ 
-s_{12}c_{23}-c_{12}s_{13}s_{23}e^{i\delta} & c_{12}c_{23}-s_{12}s_{13}s_{23}e^{i\delta} & c_{13}s_{23} \\ 
s_{12}s_{23}-c_{12}s_{13}c_{23}e^{i\delta} & -c_{12}s_{23}-s_{12}s_{13}c_{23}e^{i\delta} & c_{13}c_{23} 
\end{pmatrix}, 
\end{equation}
where $s_{ij}=\sin\theta_{ij}$ and $c_{ij}=\cos\theta_{ij}$. In vacuum, each mass eigenstate evolves according to
\begin{equation}
\ket{\nu_i(t)} = e^{-i E_i t}\ket{\nu_i},
\end{equation}
so that the time-evolved flavor state reads
\begin{equation}
\ket{\nu_\alpha(t)} = \sum_{i=1}^{3} U_{\alpha i} e^{-i E_i t} \ket{\nu_i}.
\end{equation}

To make the entanglement structure explicit, we adopt the occupation-number representation of flavor modes \cite{Blasone:2007vw, Konwar:2024nrd}, where the three flavor states are written as
\begin{align}
\ket{\nu_e} &\equiv \ket{1}_e \ket{0}_\mu \ket{0}_\tau \equiv \ket{100}_{e\mu\tau}, \notag\\
\ket{\nu_\mu} &\equiv \ket{0}_e \ket{1}_\mu \ket{0}_\tau \equiv \ket{010}_{e\mu\tau}, \\
\ket{\nu_\tau} &\equiv \ket{0}_e \ket{0}_\mu \ket{1}_\tau \equiv \ket{001}_{e\mu\tau}. \notag
\end{align}
In this basis, the time-evolved flavor state can be written as a tripartite quantum state,
\begin{equation}
\ket{\nu_\alpha(t)} =
\Bar{U}_{\alpha e}(t)\ket{100}_{e\mu\tau}
+ \Bar{U}_{\alpha \mu}(t)\ket{010}_{e\mu\tau}
+ \Bar{U}_{\alpha \tau}(t)\ket{001}_{e\mu\tau},
\end{equation}
where
\begin{equation}
\Bar{U}(t) = U e^{-i\mathcal{H}_m t} U^\dagger, \qquad
\mathcal{H}_m = \mathrm{diag}(E_1,E_2,E_3).
\end{equation}
This explicitly shows that the evolved flavor state is, in general, an entangled superposition of the three flavor modes.

\subsection{Density Matrix Formalism}
\label{subsec:density}

To quantify the quantum correlations among the flavor modes, we define the density matrix corresponding to the time-evolved flavor state $\ket{\nu_\alpha(t)}$ as \begin{eqnarray}
\rho^{\alpha}_{ABC}(t)&=&\ket{\nu_\alpha(t)}\bra{\nu_\alpha(t)}\,, \nn \\ &=&\begin{pmatrix} 0 & 0 & 0 & 0 & 0 & 0 & 0 & 0 \\ 0 & \rho^{\alpha}_{\tau\tau} & \rho^{\alpha}_{\tau\mu} & 0 & \rho^{\alpha}_{\tau e} & 0 & 0 & 0 \\ 0 & \rho^{\alpha}_{\mu\tau} & \rho^{\alpha}_{\mu\mu} & 0 & \rho^{\alpha}_{\mu e} & 0 & 0 & 0 \\ 0 & 0 & 0 & 0 & 0 & 0 & 0 & 0 \\ 0 & \rho^{\alpha}_{e\tau} & \rho^{\alpha}_{e\mu} & 0 & \rho^{\alpha}_{ee} & 0 & 0 & 0 \\ 0 & 0 & 0 & 0 & 0 & 0 & 0 & 0 \\ 0 & 0 & 0 & 0 & 0 & 0 & 0 & 0 \\ 0 & 0 & 0 & 0 & 0 & 0 & 0 & 0 
\end{pmatrix}, 
\end{eqnarray} 
where the non-vanishing elements are given by
\begin{align} 
\rho^{\alpha}_{ee} &= |\Bar{U}_{\alpha e}(t)|^2, \notag\\ \rho^{\alpha}_{\mu\mu} &= |\Bar{U}_{\alpha \mu}(t)|^2, \notag\\ \rho^{\alpha}_{\tau\tau} &= |\Bar{U}_{\alpha \tau}(t)|^2, \notag\\ \rho^{\alpha}_{e\mu} &= \Bar{U}_{\alpha e}(t)\Bar{U}^*_{\alpha \mu}(t), \notag\\ \rho^{\alpha}_{e\tau} &= \Bar{U}_{\alpha e}(t)\Bar{U}^*_{\alpha \tau}(t), \notag\\ \rho^{\alpha}_{\mu\tau} &= \Bar{U}_{\alpha \mu}(t)\Bar{U}^*_{\alpha \tau}(t), \end{align} 
together with their Hermitian conjugates $\rho^{\alpha}_{ij} = (\rho^{\alpha}_{ji})^*$.

\section{Entanglement Measures}
\label{sec:entanglement}

Quantum entanglement provides a fundamental framework to quantify non-classical correlations in multipartite quantum systems. For a three qubit system, quantum states can be classified as fully separable, biseparable, or genuinely tripartite entangled, depending on whether correlations are absent, restricted to two subsystems, or inseparably shared among all three. This framework is particularly useful in neutrino physics, where the three flavor neutrino state can be mapped onto an effective three qubit system arising from the coherent superposition of mass eigenstates. In this context, entanglement measures provide information beyond conventional oscillation probabilities, enabling a quantitative characterization of the strength and distribution of quantum correlations among neutrino flavor modes. In this work, we employ three widely used entanglement measures entanglement of formation \cite{Bennett:1996gf,Wootters:1997id},
concurrence \cite{Hill:1997pfa,Wootters:1997id},
and negativity \cite{Peres:1996dw,Vidal:2002zz}
 to characterize bipartite and tripartite correlations in neutrino oscillations \cite{Guo:2019mue,Jha:2020dav,Jha:2022yik}.

\subsection{Entanglement of Formation}

Consider a bipartite quantum system described by a density matrix $\rho$, which can be expressed as a convex combination of pure states $\ket{\psi_i}$,
\begin{equation}
    \rho = \sum_i p_i \ket{\psi_i}\bra{\psi_i}, \qquad \sum_i p_i = 1.
\end{equation}
For a pure state $\ket{\psi_i}$, the entanglement of formation is defined as the Von Neumann entropy of either reduced subsystem,
\begin{equation}
    EOF(\ket{\psi_i}) = -\mathrm{Tr}(\rho_A \log_2 \rho_A)
    = -\mathrm{Tr}(\rho_B \log_2 \rho_B),
\end{equation}
where $\rho_A = \mathrm{Tr}_B(\ket{\psi_i}\bra{\psi_i})$ and $\rho_B = \mathrm{Tr}_A(\ket{\psi_i}\bra{\psi_i})$.

For a tripartite pure state $\rho_{ABC}(t)$, EOF can be generalized as \cite{Guo:2019mue}
\begin{equation}
    EOF(\rho_{ABC}(t)) = \frac{1}{3}
    \left[ S(\rho_A) + S(\rho_B) + S(\rho_C) \right],
\end{equation}
where $\rho_A = \mathrm{Tr}_{BC}(\rho_{ABC})$, $\rho_B = \mathrm{Tr}_{AC}(\rho_{ABC})$, $\rho_C = \mathrm{Tr}_{AB}(\rho_{ABC})$, and
$S(\rho) = -\mathrm{Tr}(\rho \log_2 \rho)$.

In the neutrino oscillation framework, EOF can be expressed directly in terms of the survival and transition probabilities $P_{\alpha\beta}$ for detecting flavor $\beta$ given an initial flavor $\alpha$ \cite{Li:2021Entanglement, Konwar:2024nrd},
\begin{equation}\label{EOF}
\begin{aligned}
EOF^\alpha = -\frac{1}{2} \Big[
& P_{\alpha e}\log_2 P_{\alpha e}
+ P_{\alpha \mu}\log_2 P_{\alpha \mu}
+ P_{\alpha \tau}\log_2 P_{\alpha \tau} \\
& + (P_{\alpha e}+P_{\alpha \mu})\log_2 (P_{\alpha e}+P_{\alpha \mu}) \\
& + (P_{\alpha e}+P_{\alpha \tau})\log_2 (P_{\alpha e}+P_{\alpha \tau}) \\
& + (P_{\alpha \mu}+P_{\alpha \tau})\log_2 (P_{\alpha \mu}+P_{\alpha \tau})
\Big].
\end{aligned}
\end{equation}

\subsection{Concurrence}

Concurrence is a widely used bipartite entanglement measure \cite{Streltsov:2009azs}, which can be generalized to tripartite systems as \cite{Guo:2019yyl}
\begin{equation}
    C(\rho_{ABC}) =
    \sqrt{3 - \mathrm{Tr}(\rho_A^2)
              - \mathrm{Tr}(\rho_B^2)
              - \mathrm{Tr}(\rho_C^2)},
\end{equation}
where $\rho_A$, $\rho_B$, and $\rho_C$ denote the reduced density matrices of the respective subsystems.

For neutrino oscillations, concurrence can be written in terms of flavor transition probabilities as \cite{Li:2021Entanglement}
\begin{equation}\label{Concurrence}
    C^\alpha =
    \sqrt{3 - 3 P_S
    - 2 P_{\alpha e}(P_{\alpha \mu}+P_{\alpha \tau})
    - 2 P_{\alpha \mu} P_{\alpha \tau}},
\end{equation}
where $P_S = P_{\alpha e}^2 + P_{\alpha \mu}^2 + P_{\alpha \tau}^2$.

\subsection{Negativity}

Negativity is an alternative entanglement measure particularly suited for multipartite systems. For a three-qubit state, it is defined as \cite{Sabin:2008naa,Vidal:2002zz}
\begin{equation}
    N = \left(
    N_{A-BC}\,
    N_{B-AC}\,
    N_{C-AB}
    \right)^{1/3},
\end{equation}
where $N_{A-BC} = \sum_i |\lambda_i^A| - 1$, and $\lambda_i^A$ are the negative eigenvalues of the partial transpose of $\rho_{ABC}$ with respect to subsystem $A$ (and similarly for $B$ and $C$).

For neutrino flavor oscillations in vacuum, negativity can be approximately expressed in terms of survival and transition probabilities as \cite{Li:2021Entanglement}
\begin{equation} \label{Negativity}
\begin{aligned}
N^\alpha &=
\Big[
\sqrt{P_{\alpha e}(P_{\alpha \mu}+P_{\alpha \tau})}
\sqrt{P_{\alpha \mu}(P_{\alpha e}+P_{\alpha \tau})} \\
& \times
\sqrt{P_{\alpha \tau}(P_{\alpha e}+P_{\alpha \mu})}
\Big]^{1/3}.
\end{aligned}
\end{equation}

\begin{figure*}[htb]
    \centering
    \includegraphics[width=0.47\linewidth]{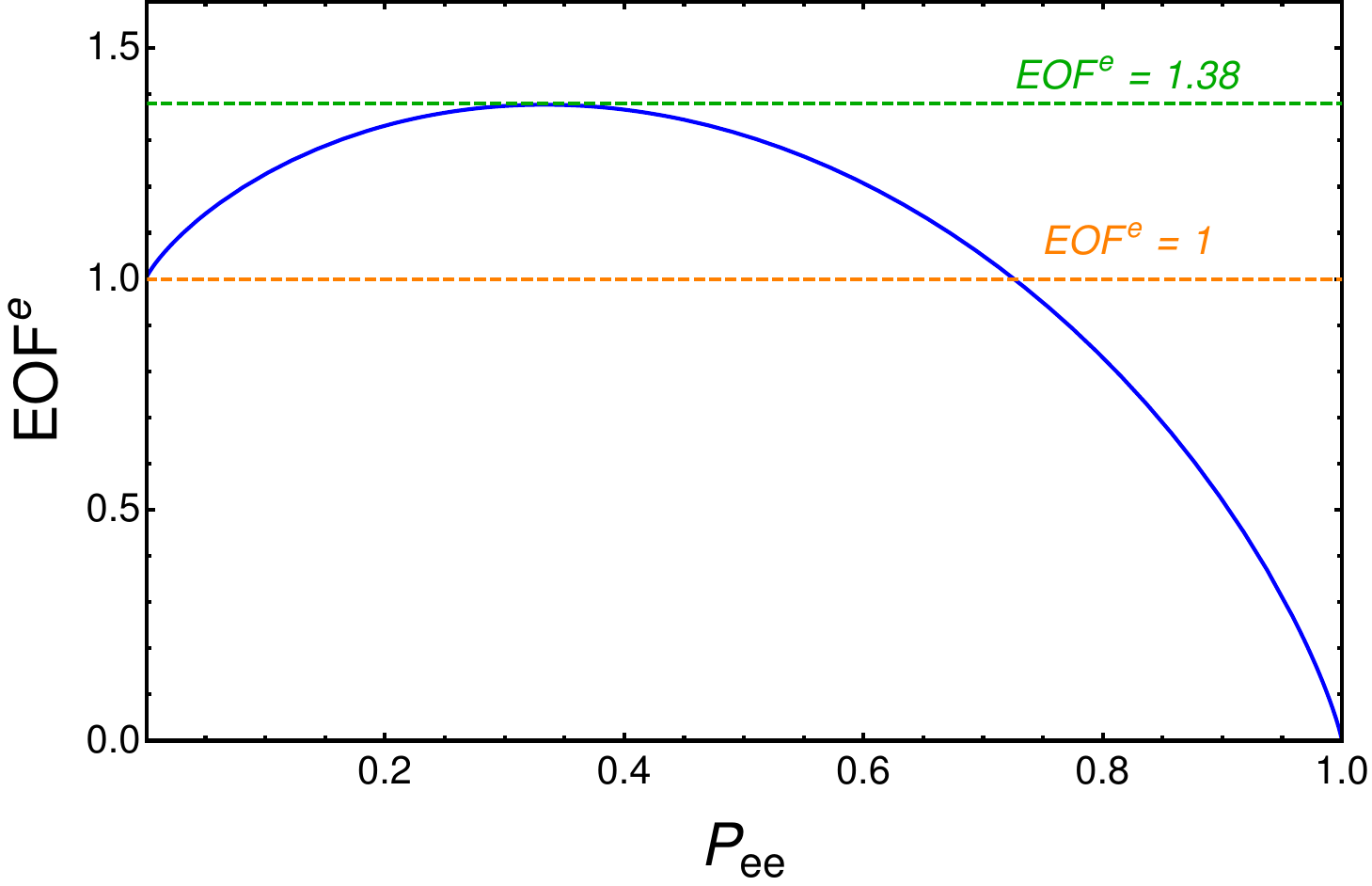}\quad
    \includegraphics[width=0.47\linewidth]{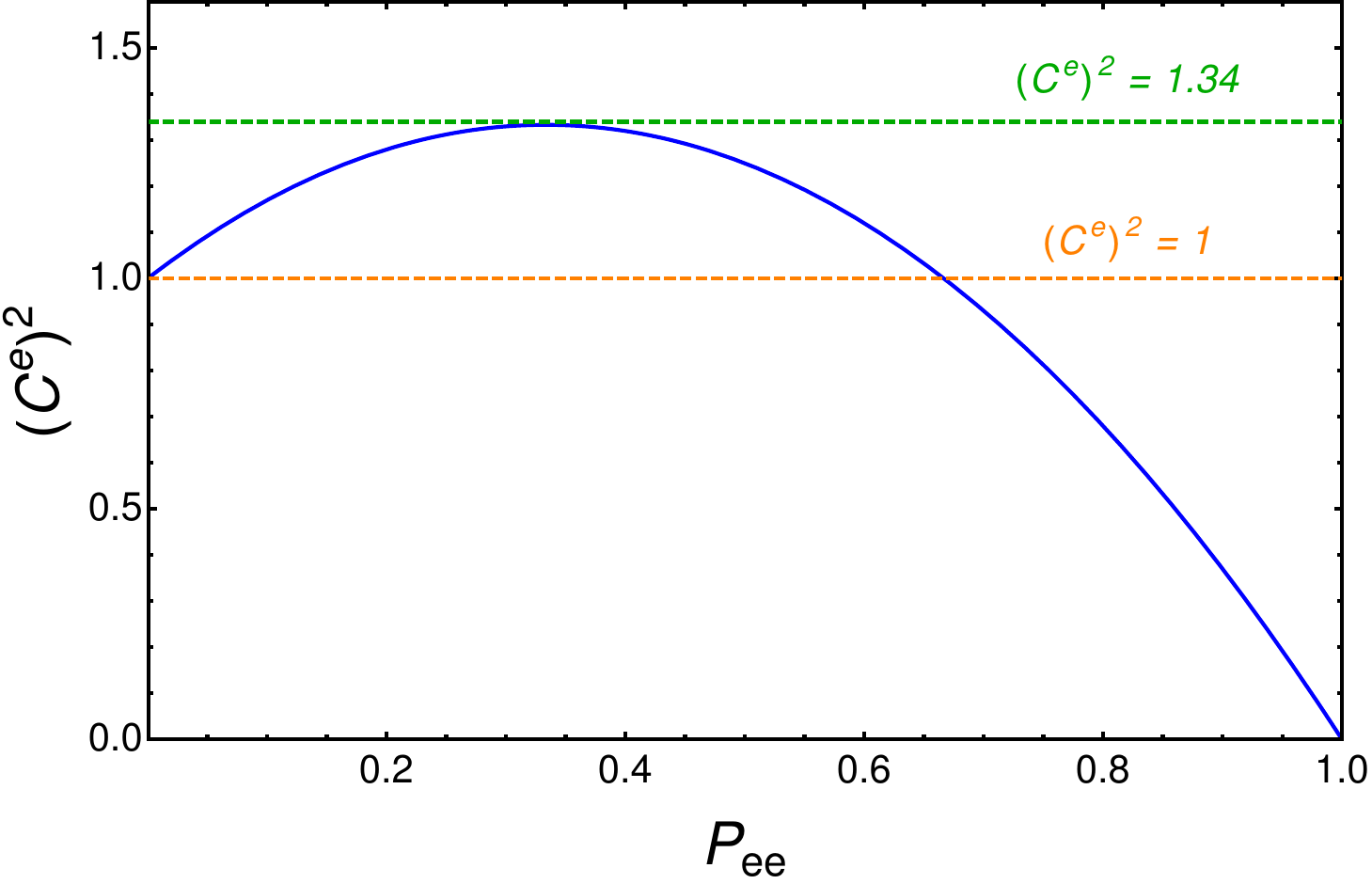}\quad
    \includegraphics[width=0.47\linewidth]{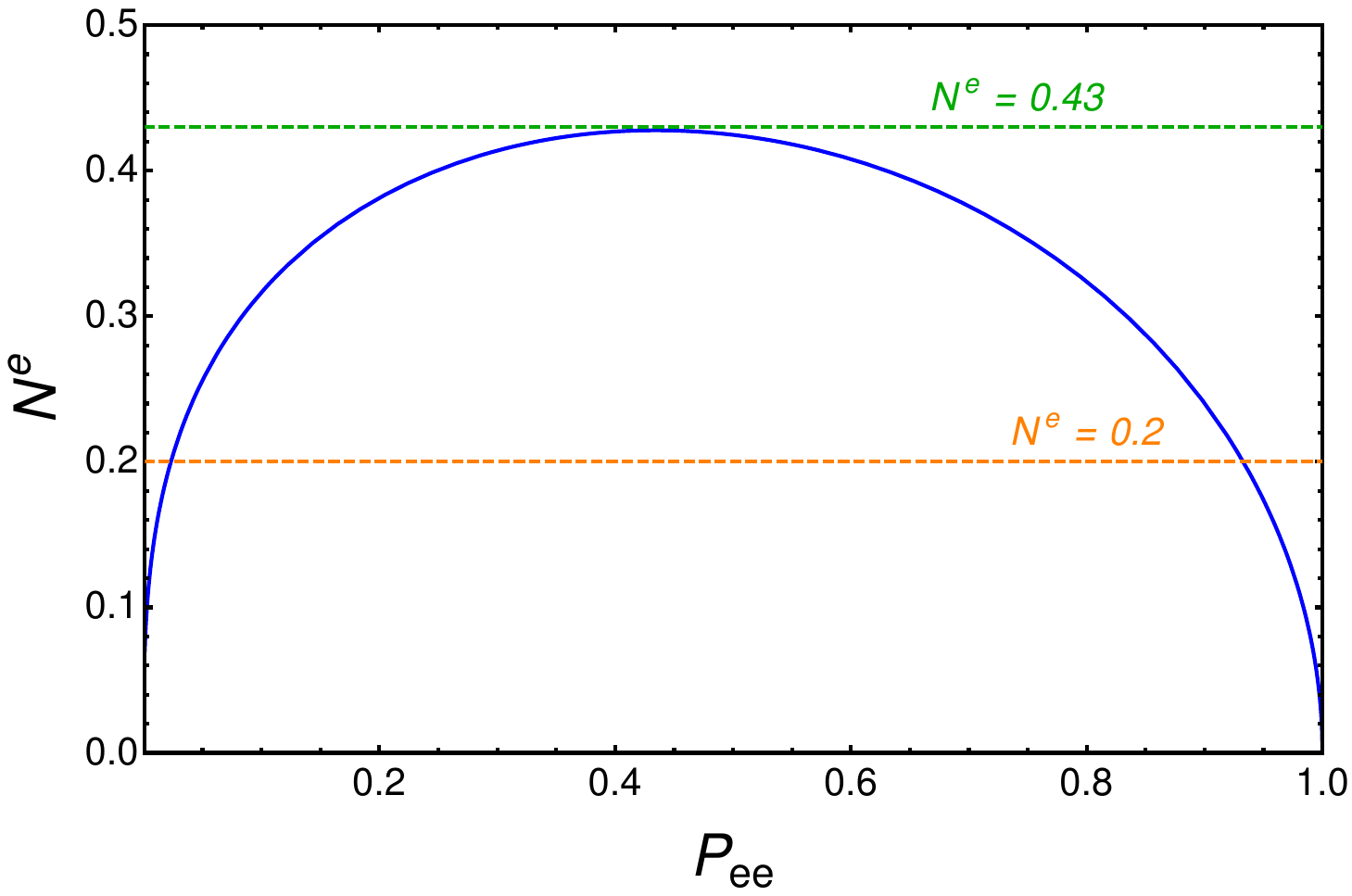}
    \caption{Entanglement measures as functions of the electron neutrino survival probability $P_{ee}$: Entanglement of formation EOF$^{e}$ (top left), squared concurrence $(C^{e})^{2}$ (top right), and negativity $N^{e}$ (bottom).}
    \label{fig:entanglement}
\end{figure*}

Figure~\ref{fig:entanglement} shows the behavior of the entanglement measures EOF$^{e}$ (upper left panel), squared concurrence $(C^{e})^{2}$ (upper right panel), and negativity $N^{e}$ (bottom panel) as functions of the electron neutrino survival probability $P_{ee}$. All three measures increase with $P_{ee}$ up to an intermediate value and subsequently decrease as $P_{ee}$ approaches unity, indicating the presence of non-classical correlations among the neutrino flavor modes during oscillations. The horizontal dashed lines in each panel indicate two representative values of the corresponding entanglement measures: for EOF$^e$ and $(C^e)^2$, the values are 1 and their respective maxima, while for negativity $N^e$, the values are 0.2 and its maximum. The benchmark values of the survival probability $P_{ee}$ associated with EOF$^{e}$ in the range $1.0$--$1.38$, $(C^{e})^{2}$ in the range $1.0$--$1.34$, and $N^{e}$ in the range $0.20$--$0.43$ are summarized in Table~\ref{tab:pee_entanglement}.


\begin{table}[htb]
\centering
\begin{tabular}{|c|c|c|c|c|}
\hline
Entanglement measure &
\multicolumn{4}{c|}{Benchmark values of $P_{ee}$ ($\Delta p$)} \\
\hline
EOF$^e \in [1.0, 1.38]$ & ~0.698~ & ~0.606~ & ~0.333~ & ~0.082~ \\
\hline
$(C^e)^2 \in [1.0, 1.34]$ & ~0.667~ & ~0.544~ & ~0.333~ & ~0.123~ \\
\hline
$N^e \in [0.20, 0.43]$ & ~0.698~ & ~0.434~ & ~0.083~ & ~0.023~ \\
\hline
\end{tabular}
\caption{Benchmark values of the electron neutrino survival probability $P_{ee}$ (denoted by $\Delta p$) corresponding to representative ranges of the entanglement measures EOF$^e$, $(C^e)^2$, and $N^e$, as indicated by the horizontal dashed lines in Fig.~\ref{fig:entanglement}.}
\label{tab:pee_entanglement}
\end{table}

\section{Supernova Neutrino Flux}
\label{sec:supernova}

The energy spectrum of neutrinos emitted from a supernova is modeled using a pinched thermal distribution \cite{Scholberg:2012id, Das:2017iuj, Keil:2002in, Dighe:1999bi}, 
\begin{equation}
\phi_\nu(E_\nu)=
\frac{(\alpha+1)^{\alpha+1}}{\langle E_\nu\rangle\,\Gamma(\alpha+1)}
\left(\frac{E_\nu}{\langle E_\nu\rangle}\right)^{\alpha}
\exp\!\left[-(\alpha+1)\frac{E_\nu}{\langle E_\nu\rangle}\right],
\end{equation}
where $\langle E_\nu\rangle$ is the average neutrino energy and $\alpha$ characterizes the degree of spectral pinching. The corresponding time-integrated flux at Earth is given by
\begin{equation}
F_\nu(E_\nu)=\frac{L_\nu}{4\pi D^2\,\langle E_\nu\rangle}\,\phi_\nu(E_\nu),
\end{equation}
where $L_\nu$ is the total energy emitted in neutrinos of a given flavor and $D$ is the distance to the supernova.

As neutrinos propagate through the dense supernova envelope, their flavor composition is modified by the Mikheyev–Smirnov–Wolfenstein (MSW) effect. Neglecting collective neutrino–neutrino interactions and Earth matter effects \cite{Seadrow:2018ftp}, the electron neutrino and antineutrino fluxes arriving at Earth can be expressed in terms of the primary fluxes and neutrino mixing angles \cite{Dighe:1999bi}. To account for the impact of quantum correlations among neutrino flavors, the modified electron (anti)neutrino fluxes, including the entanglement contribution $\Delta p$, can be written for both normal hierarchy (NH) and inverted hierarchy (IH) as
\begin{equation}
\begin{aligned}
F_{\nu_e}^{\rm NH} &= (\sin^2\theta_{13} + \Delta p)\,F^0_{\nu_e} + (\cos^2\theta_{13} + \Delta p)\,F^0_{\nu_x},\\
F_{\bar\nu_e}^{\rm NH} &= (\cos^2\theta_{12}\cos^2\theta_{13} + \Delta p)\,F^0_{\bar\nu_e} 
\\& + \left[1 - \cos^2\theta_{12}\cos^2\theta_{13} + \Delta p \right]\,F^0_{\bar\nu_x},\\
F_{\nu_e}^{\rm IH} &= (\cos^2\theta_{12}\cos^2\theta_{13} + \Delta p)\,F^0_{\nu_e} 
\\ &+ \left[1 - \cos^2\theta_{12}\cos^2\theta_{13} + \Delta p \right]\,F^0_{\nu_x},\\
F_{\bar\nu_e}^{\rm IH} &= (\sin^2\theta_{13} + \Delta p)\,F^0_{\bar\nu_e} + (\cos^2\theta_{13} + \Delta p)\,F^0_{\bar\nu_x},
\end{aligned}
\end{equation}
where $F^0_{\nu_\alpha}$ denotes the primary flux of flavor $\alpha = e,\mu,\tau$ at the supernova source, and $F^0_{\nu_x}$ represents the flux of a non-electron neutrino flavor (either $\nu_\mu$ or $\nu_\tau$).

\section{Experimental Setup and Simulation Details}
\label{sec:expt_sim}
\subsection{Supernova Neutrino Interaction Channels}
\label{subsec:channels}

The detection of neutrinos from a core collapse supernova at DUNE is dominated by a few
interaction channels in liquid argon. In the energy range relevant for supernova neutrinos,
$\mathcal{O}(10)$~MeV, these channels provide complementary sensitivity to neutrino flavor,
energy, and temporal structure \cite{Abi:2020evt}. The three dominant interaction modes
considered in this work are
\begin{itemize}
    \item \textbf{Channel A: Electron neutrino charged current (CC) interactions on argon:}
    \[
    \nu_e + {}^{40}\mathrm{Ar} \rightarrow e^- + {}^{40}\mathrm{K}^*,
    \]
    which provide direct sensitivity to the $\nu_e$ flux and dominate the signal during
    the neutronization burst phase.

    \item \textbf{Channel B: Electron antineutrino charged current  interactions on argon:}
    \[
    \bar{\nu}_e + {}^{40}\mathrm{Ar} \rightarrow e^+ + {}^{40}\mathrm{Cl}^*,
    \]
    which probe the $\bar{\nu}_e$ flux primarily during the accretion and cooling phases.

    \item \textbf{Channel C: Elastic scattering on electrons:}
    \[
    \nu_\alpha + e^- \rightarrow \nu_\alpha + e^-,
    \]
    which is sensitive to all neutrino flavors, with an enhanced contribution from
    electron neutrinos due to their larger weak-interaction couplings.
\end{itemize}

\begin{figure*}[htb]
    \centering
    \includegraphics[width=0.48\linewidth]{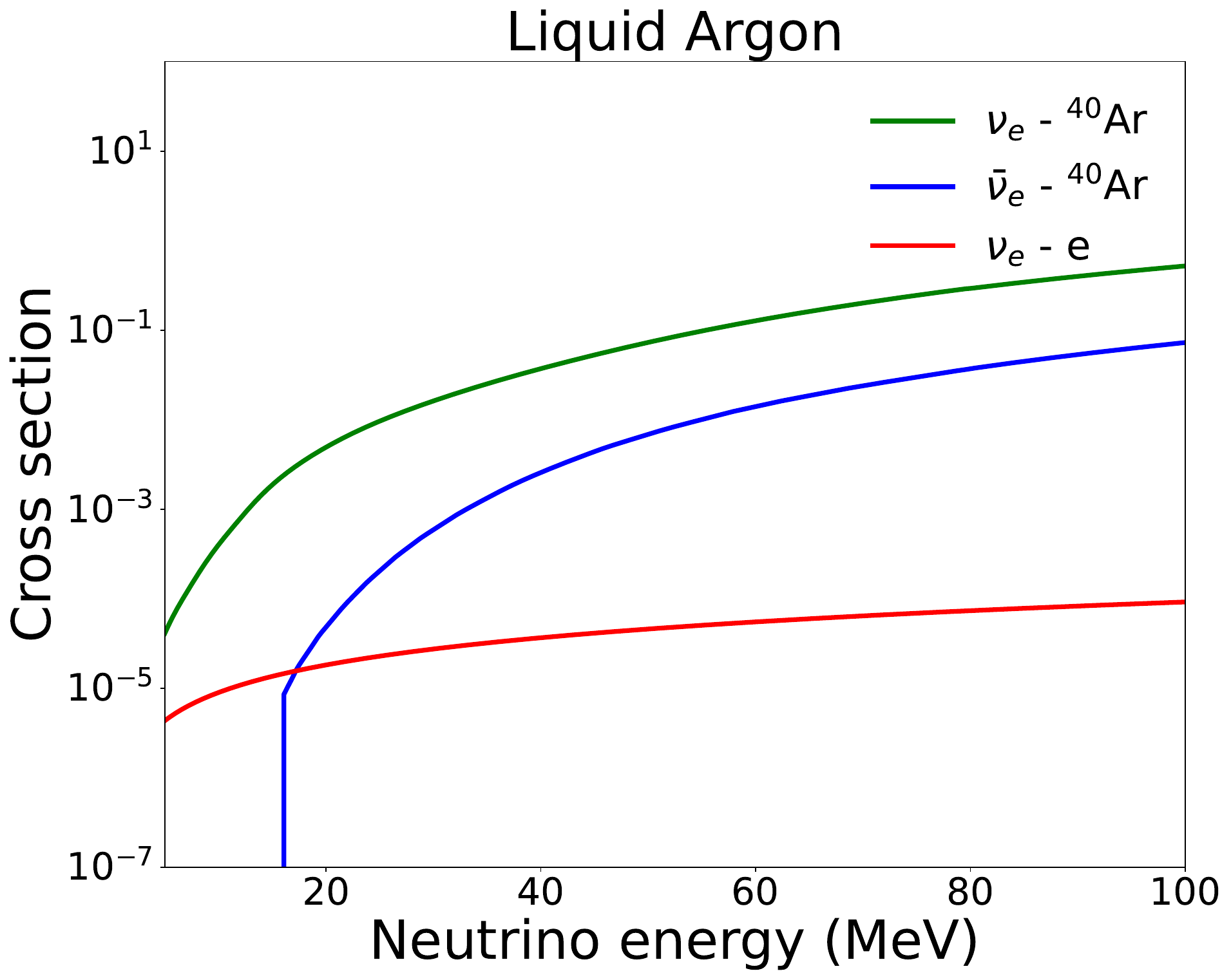}
    \caption{Energy-dependent cross sections for the dominant supernova neutrino interaction
    channels in liquid argon.}
    \label{fig:cross-section}
\end{figure*}

The energy dependent cross sections used in our simulations are shown in
Fig.~\ref{fig:cross-section}. Among these channels, the $\nu_e$ charged current interaction on argon exhibits the largest cross section throughout the supernova neutrino energy range, making DUNE uniquely sensitive to electron neutrinos compared to other detector technologies. This enhancement arises from the large nuclear charge of argon and the coherent contribution of nuclear transitions in CC interactions. The elastic scattering channel, although sensitive to all neutrino flavors, has a much
smaller cross section than the charged current interactions on argon. This suppression is due to the absence of nuclear enhancement, the light electron target, and kinematic
limitations of the neutral current interaction. At low neutrino energies
($E_\nu \lesssim 20~\mathrm{MeV}$), elastic scattering exhibits a larger cross section than the $\bar{\nu}_e$ charged current channel; however, at higher energies, the
$\bar{\nu}_e$--argon interaction becomes dominant. Despite its smaller rate, the elastic
scattering channel provides valuable directional information and is therefore included in
our analysis.

\subsection{Supernova Neutrino Detection at DUNE}
\label{subsec:dune}

The DUNE consists of a modular liquid argon time projection chamber detector with a total fiducial mass of 40~kton, located at the Sanford Underground Research Facility (SURF) in South Dakota \cite{Abi:2020evt,DUNE:2020ypp}. The far detector comprises four independent modules, each with a fiducial mass of approximately 10~kton, allowing operational flexibility and enabling each module to function as an independent supernova neutrino observatory. For the computation of supernova neutrino event rates in this work, we assume the full 40~kton fiducial mass, corresponding to the combined exposure of all four modules. The neutrino signal is assumed to be isotropic across the detector volume. DUNE’s excellent spatial resolution, calorimetric energy reconstruction, and particle identification capabilities make it ideally suited for detecting low energy neutrinos from a Galactic core collapse supernova. In particular, the liquid argon target provides unique sensitivity to electron neutrinos via charged current interactions, enabling detailed studies of the neutronization burst and early time supernova dynamics. 

The primary supernova neutrino detection channels are described in subsection~\ref{subsec:channels}. The $\nu_e$ charged-current interaction on argon dominates the event rate during the neutronization burst, while the $\bar{\nu}_e$ charged current channel is most relevant during the accretion and cooling phases. Although elastic scattering events contribute fewer events, they provide valuable directional information that can aid in supernova localization. All three channels are included in our analysis to compute the expected event rates and flavor dependent probabilities, which serve as inputs for the subsequent entanglement study.

\subsection{Simulation Details Using \texttt{SNOwGLoBES}}
\label{subsec:simulation}

To simulate the detection of supernova neutrinos at DUNE, we employ the \texttt{SNOwGLoBES} framework \cite{snowglobes}, which provides realistic predictions of event rates by folding supernova neutrino fluences with interaction cross sections and detector response functions. We adopt the Garching supernova model, in which the time integrated fluence of each flavor is characterized by the total emitted energy, the average neutrino energy, and a pinching parameter that accounts for deviations from a pure thermal spectrum \cite{Mirizzi:2015eza}.

Neutrino oscillation effects are incorporated through flavor survival probabilities determined by the neutrino mass ordering, as described in Sec.~\ref{subsec:oscilaltions}. We consider a benchmark Galactic supernova at a distance of $D = 10$~kpc. The reconstructed neutrino energy window is chosen to be $1$--$100$~MeV, covering the full supernova neutrino spectrum relevant for DUNE. Oscillation parameters are taken from recent global fits \cite{Esteban:2024eli}.

To quantify the sensitivity to the neutrino mass ordering we employ a Poissonian
log-likelihood \(\chi^2\) (statistical part), comparing the predicted event spectrum
for the test hierarchy with the true-hierarchy spectrum:
\begin{equation}\label{eq:chi2_stat}
\chi^2_{\rm stat}
= 2\sum_{i=1}^{n}\left[ N^{\rm test}_i - N^{\rm true}_i 
- N^{\rm true}_i \ln\!\left(\frac{N^{\rm test}_i}{N^{\rm true}_i}\right)\right],
\end{equation}
where \(N^{\rm true}_i\) and \(N^{\rm test}_i\) are the predicted event counts in the
\(i\)-th reconstructed-energy bin for the true and test ordering, respectively, and
the sum runs over all bins. 

To take into account systematic uncertainties, we include normalization and energy calibration errors using the pull-method approach \cite{Fogli:2002pt, Huber:2002mx}. In the presence of energy calibration uncertainties, the expected number of events for the test mass ordering in the $i$-th energy bin is modified as
\begin{equation}
N^{\rm test}_i \;\to\; N^{\rm test}_i
\left[
(1 + x_1 \zeta_1)
+ x_2 \zeta_2 \,
\frac{(E'_i - \bar{E})}{(E'_{\rm max} - E'_{\rm min})}
\right],
\end{equation}
where $\zeta_1$ and $\zeta_2$ are the pull variables corresponding to the normalization and energy calibration uncertainties, respectively. The parameters $x_1$ and $x_2$ denote the fractional uncertainties associated with the normalization and energy calibration errors, respectively. For this study we take $x_1 = x_2 = 5\%$. Here, $E'_i$ denotes the reconstructed energy of the $i$-th bin, while $E'_{\rm max}$ and $E'_{\rm min}$ represent the maximum and minimum reconstructed energies considered in the analysis. The quantity $\bar{E} = (E'_{\rm max} + E'_{\rm min})/2$ is the midpoint of the reconstructed energy range.

The total $\chi^2$ function including systematic uncertainties is then given by
\begin{equation}
\chi^2_{\rm stat+sys} = \chi^2_{\rm stat} + \zeta_1^2 + \zeta_2^2 .
\end{equation}

Unless explicitly stated otherwise, our analysis is performed using the statistical contribution $\chi^2_{\rm stat}$ only, neglecting systematic uncertainties. For Figs.~\ref{fig:MH-EOF-sys}, systematic uncertainties of $5\%$ are assumed for both normalization and energy calibration errors.

\section{Result and Discussion}
\label{sec:result}
This section presents our results incorporating quantum entanglement among supernova neutrinos. We first discuss the neutrino fluence, followed by event rates, cross sections, and finally the sensitivity to neutrino mass ordering, as detailed in subsections~\ref{Fluence}, \ref{Event rate}, and \ref{MO}, respectively.

\subsection{Fluence}
\label{Fluence}

\begin{figure*}[htbp]
    \centering
    \includegraphics[width=0.48\linewidth]{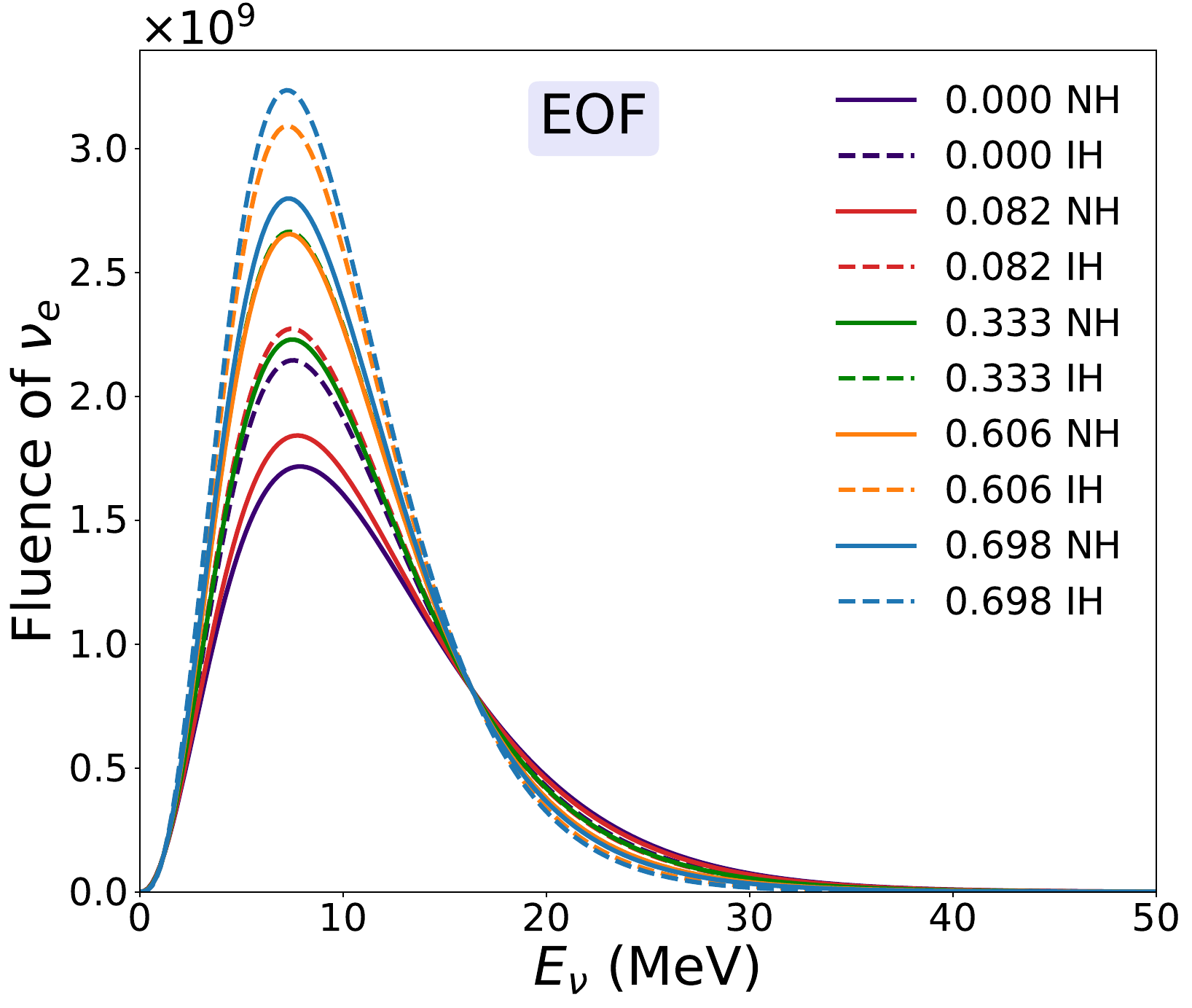}\quad
    \includegraphics[width=0.48\linewidth]{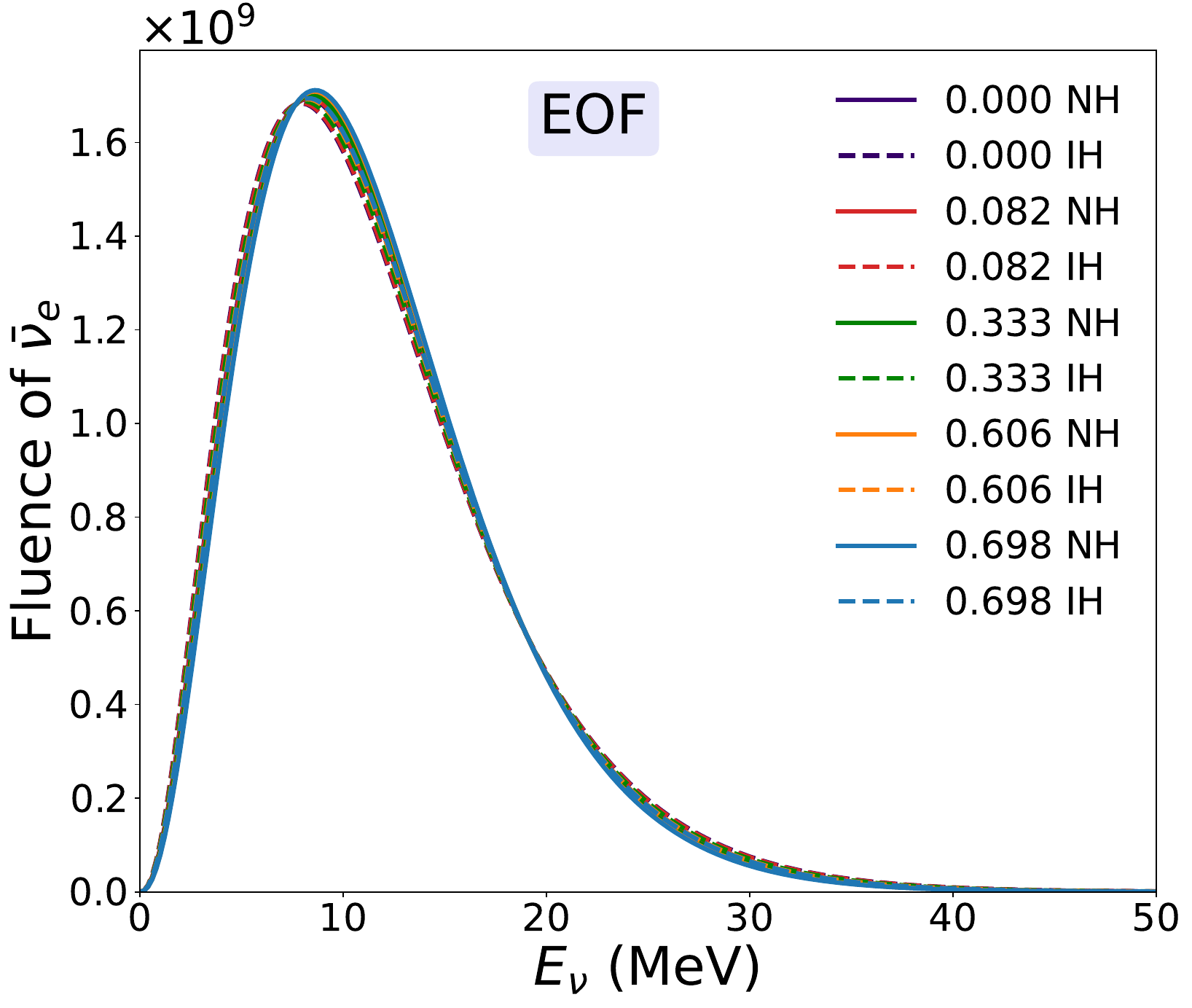}\\
     \includegraphics[width=0.48\linewidth]{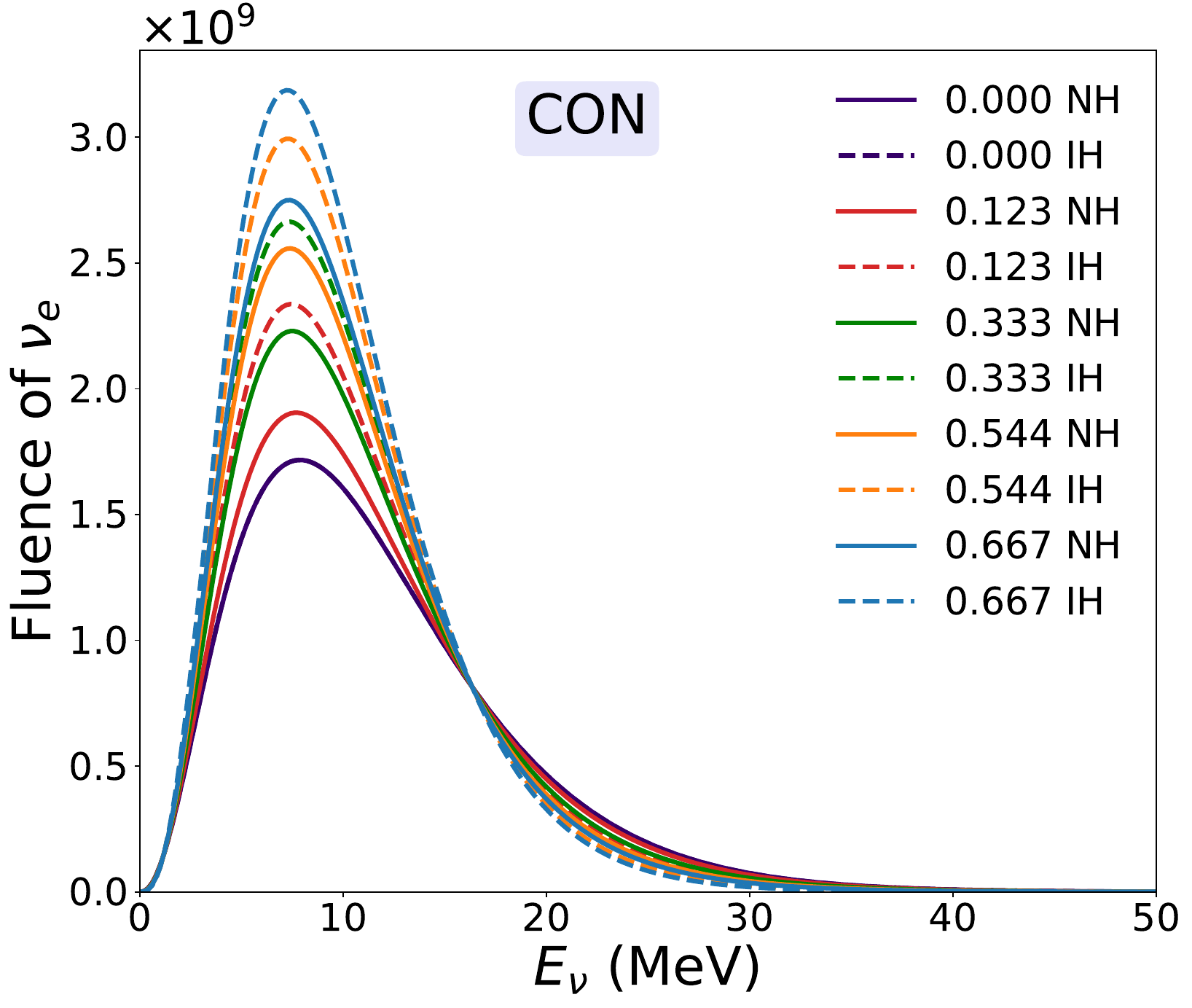}\quad
     \includegraphics[width=0.48\linewidth]{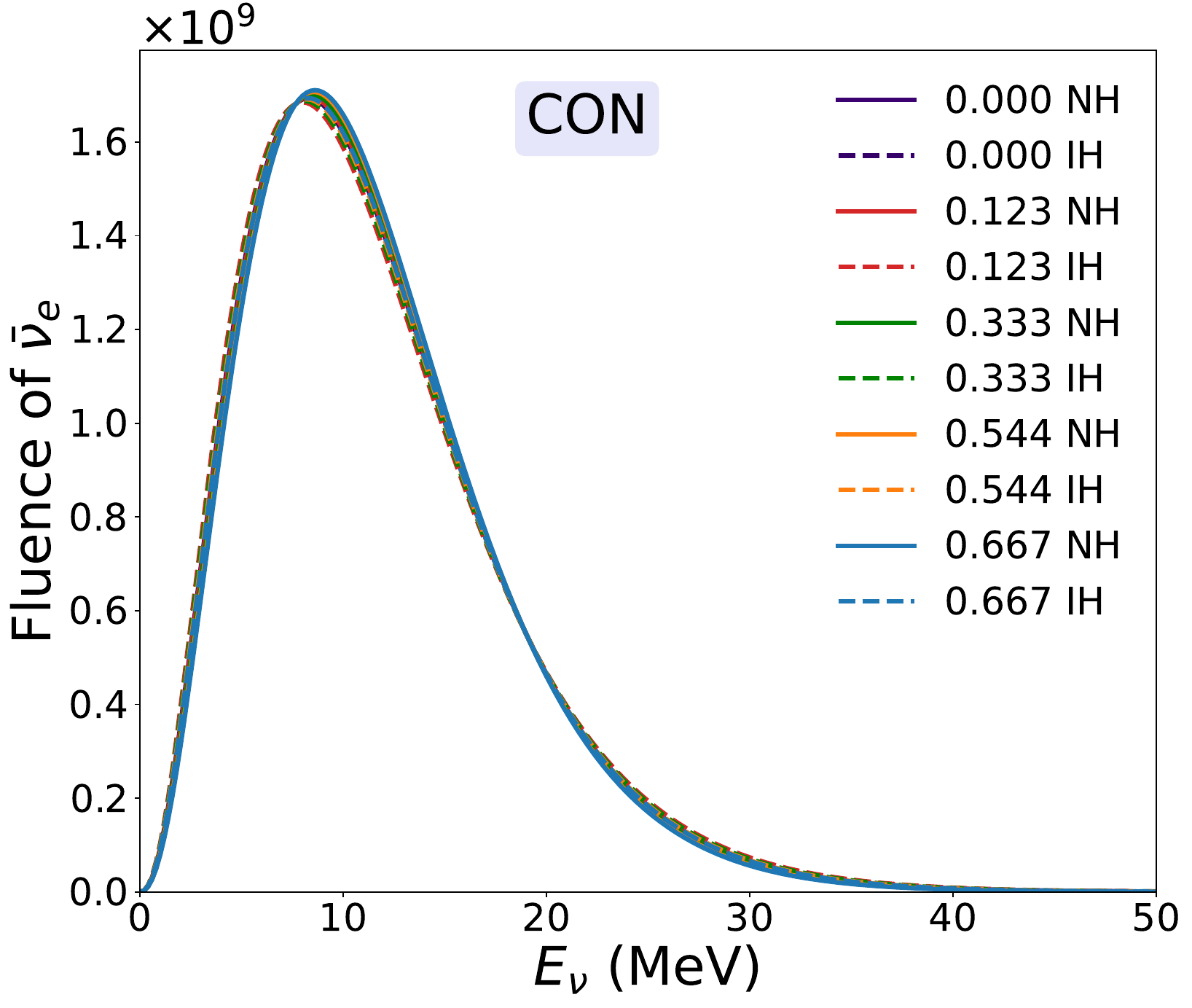}\\
     \includegraphics[width=0.48\linewidth]{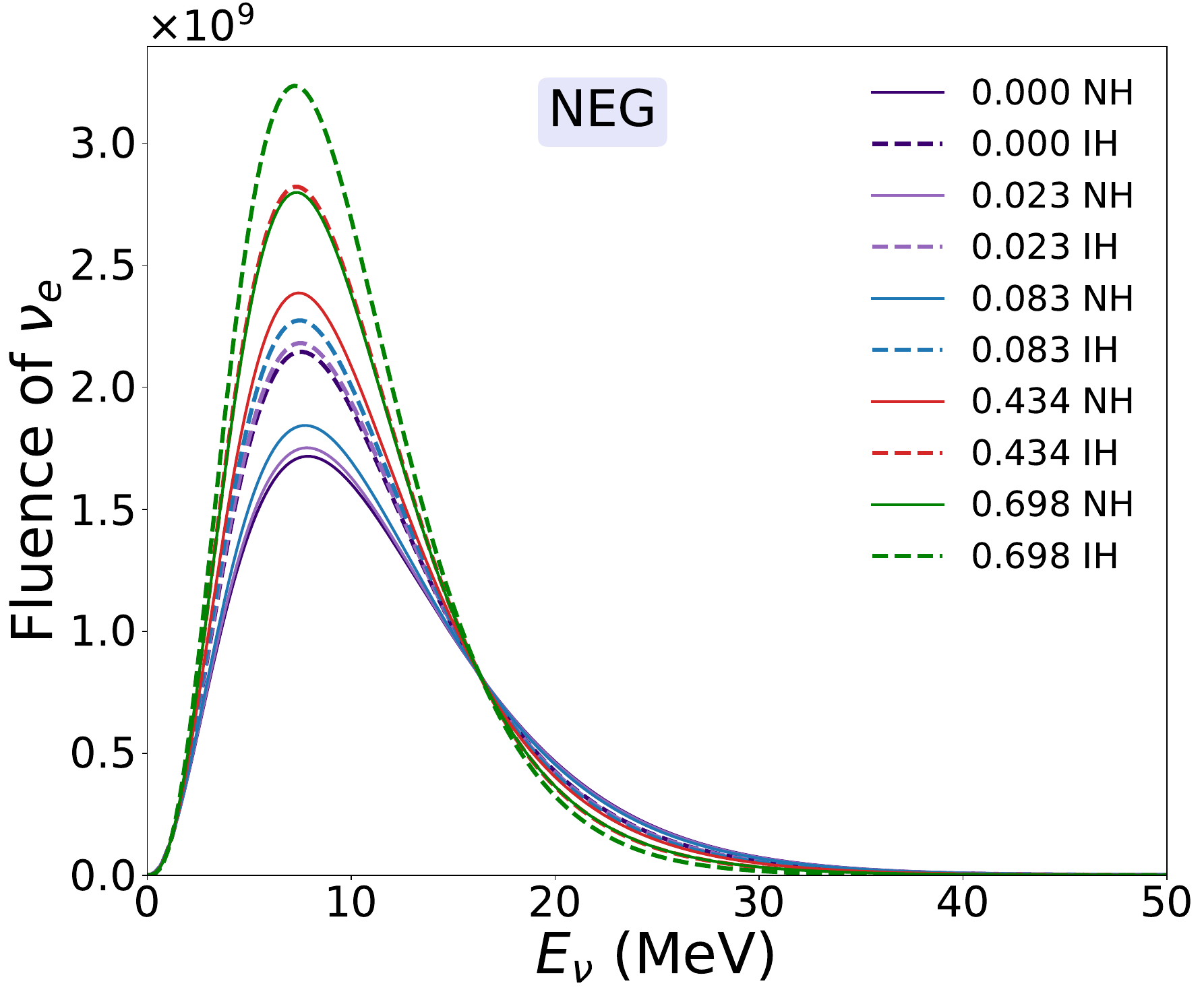}
     \includegraphics[width=0.48\linewidth]{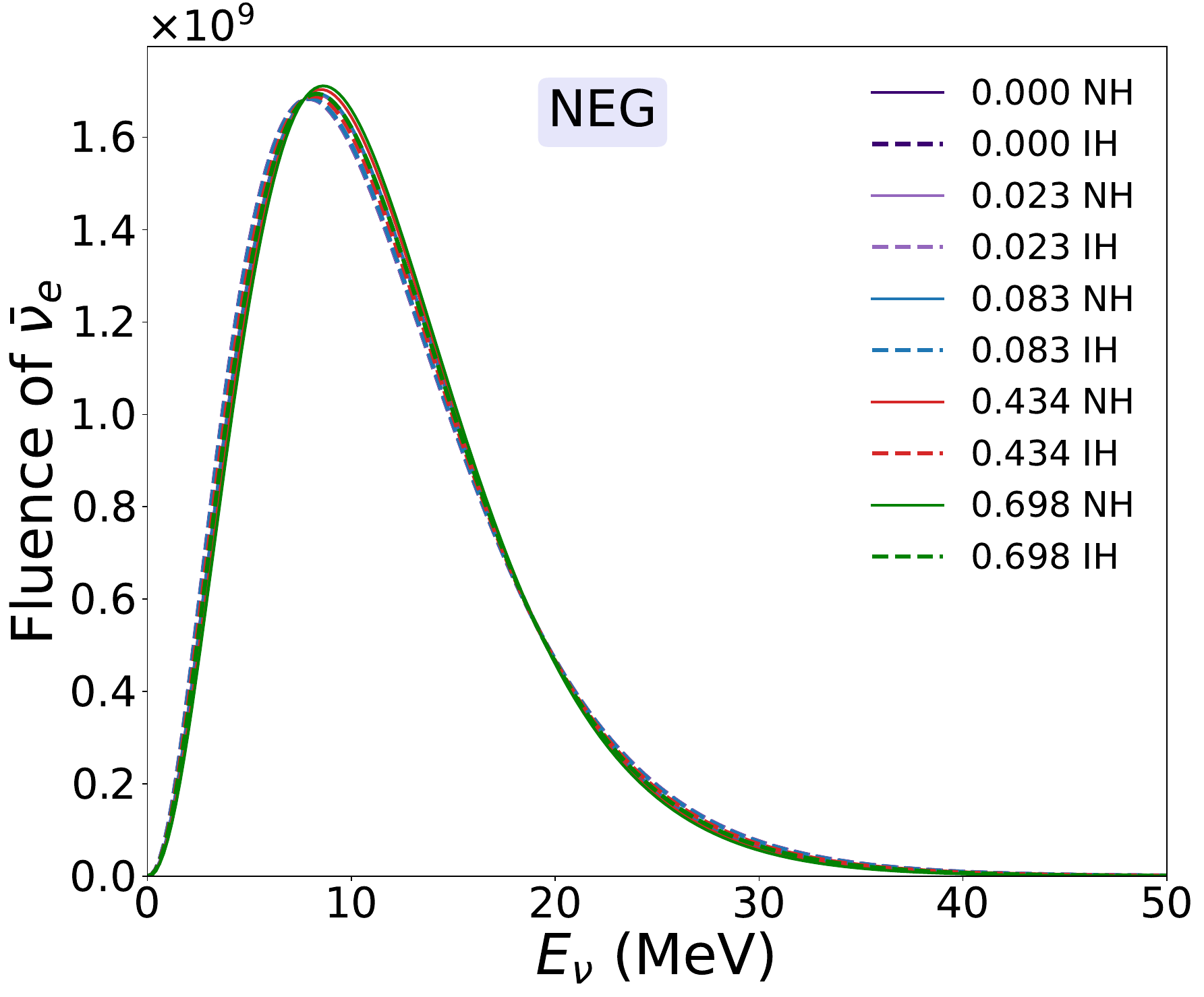}
    \caption{Fluence of $\nu_e$ (left panels) and $\bar{\nu}_e$ (right panels) as functions of neutrino energy for a supernova at a distance of 10~kpc. The upper, middle, and bottom panels correspond to the entanglement of formation, concurrence, and negativity cases, respectively. Solid lines denote the fluence for NH, while dashed lines represent the IH. Different colors correspond to different benchmark values of the survival probability ($\Delta p)$.}
    \label{fig:fluence}
\end{figure*}
Fig.~\ref{fig:fluence} shows the fluence of $\nu_e$ (left panels) and $\bar{\nu}_e$ (right panels) as a function of energy, assuming a supernova distance of 10 kpc. The upper, middle, and bottom panels correspond to the EOF, CON, and NEG measures of entanglement, respectively. In each panel, the fluence is shown for different values of $\Delta p$ for both the NH and IH cases, denoted by solid and dashed curves, respectively. The different values of $\Delta p$ represent different strengths of entanglement.

The fluence peaks at energies below $\sim 10$ MeV, with the most significant variations between different fluence scenarios occurring around this peak. For $\nu_e$, non-zero $\Delta p$ (accounting for quantum entanglement) leads to noticeable deviations relative to the $\Delta p = 0$ case, across all entanglement measures (EOF, CON, and NEG). These deviations are less significant for $\bar{\nu}_e$. For all three entanglement measures, the fluence generally increases with increasing $\Delta p$ for both normal and inverted hierarchy. For $\nu_e$, the fluence in IH is consistently higher than in NH, enhancing the distinction between the two hierarchies.

An interesting observation is that, within the EOF measure, the NH scenario at $\Delta p = 0.606$ closely matches the IH scenario at $\Delta p = 0.333$. A similar correspondence is seen for the NEG measure, where the NH fluence at $\Delta p = 0.698$ nearly coincides with the IH fluence at $\Delta p = 0.434$. This indicates that, for certain $\Delta p$ values, the NH and IH scenarios can mimic each other within EOF, CON, and NEG, potentially reducing hierarchy discrimination. For $\bar{\nu}_e$, however, the differences are much smaller, and it becomes difficult to distinguish the fluence with non-zero $\Delta p$ from the standard case ($\Delta p =0$).

\subsection{Event rate}
\label{Event rate}

\begin{figure*}[htbp]
    \centering
     \includegraphics[scale=0.25]{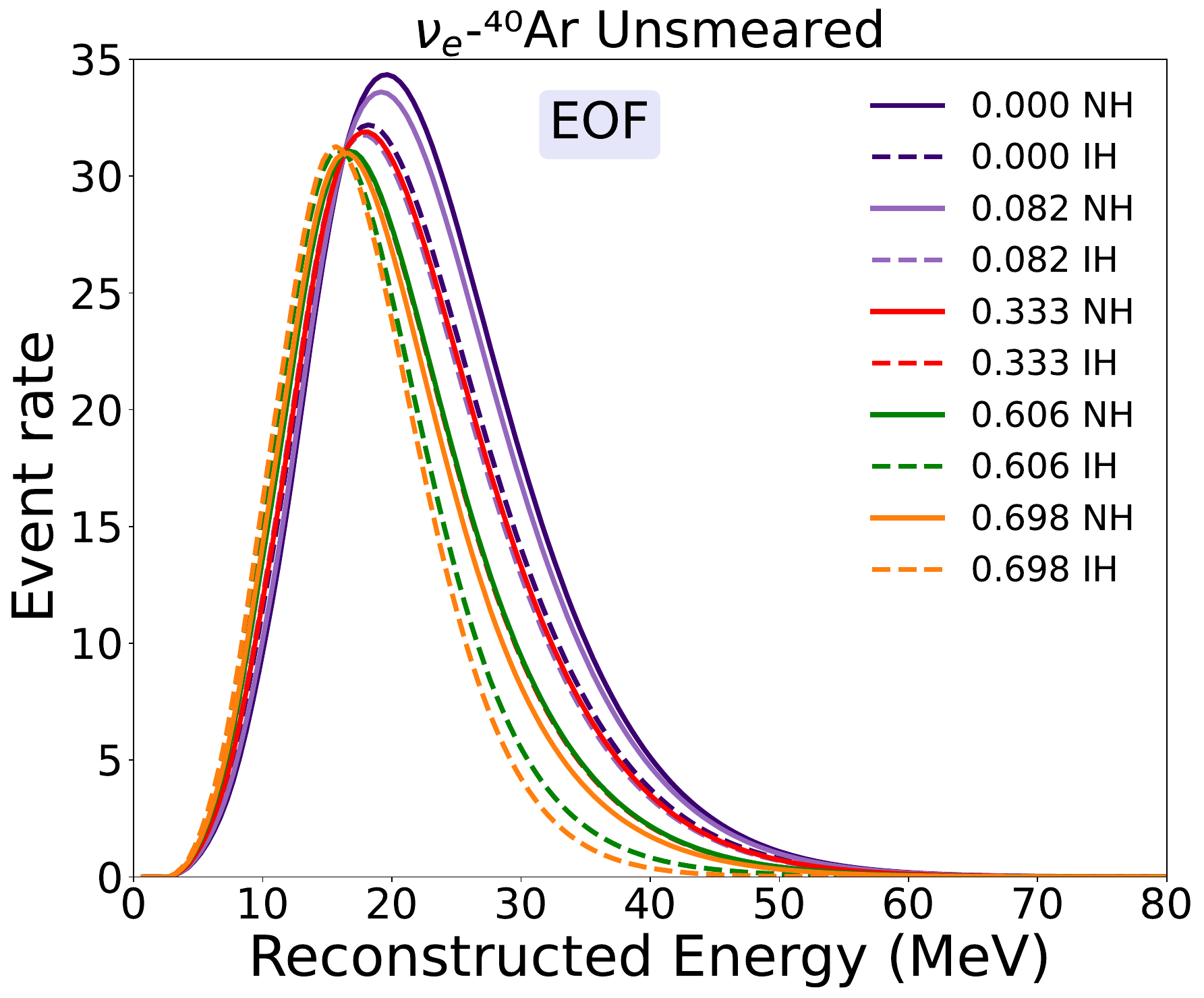}\quad
     \includegraphics[scale=0.25]{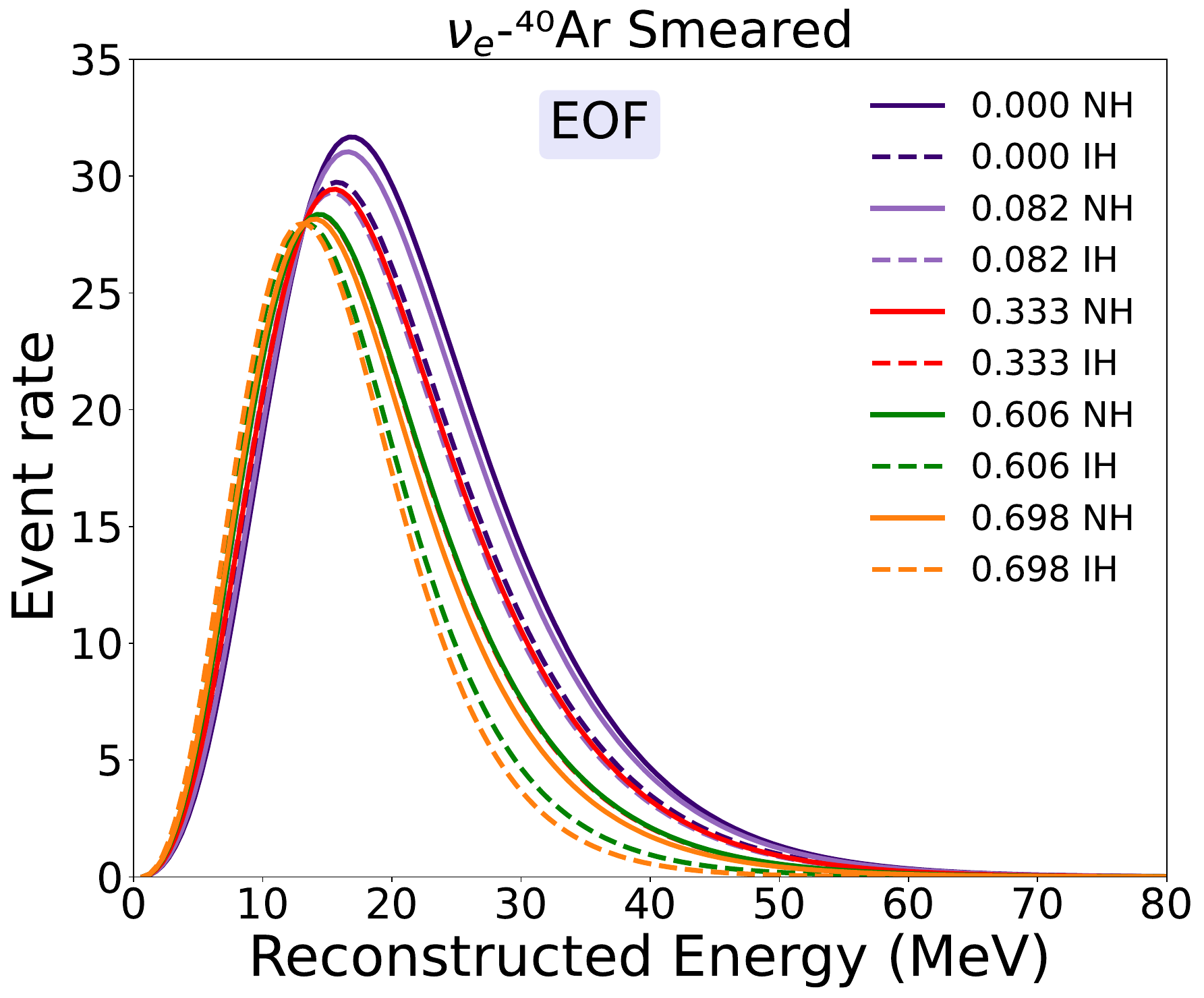}\\
     \includegraphics[scale=0.25]{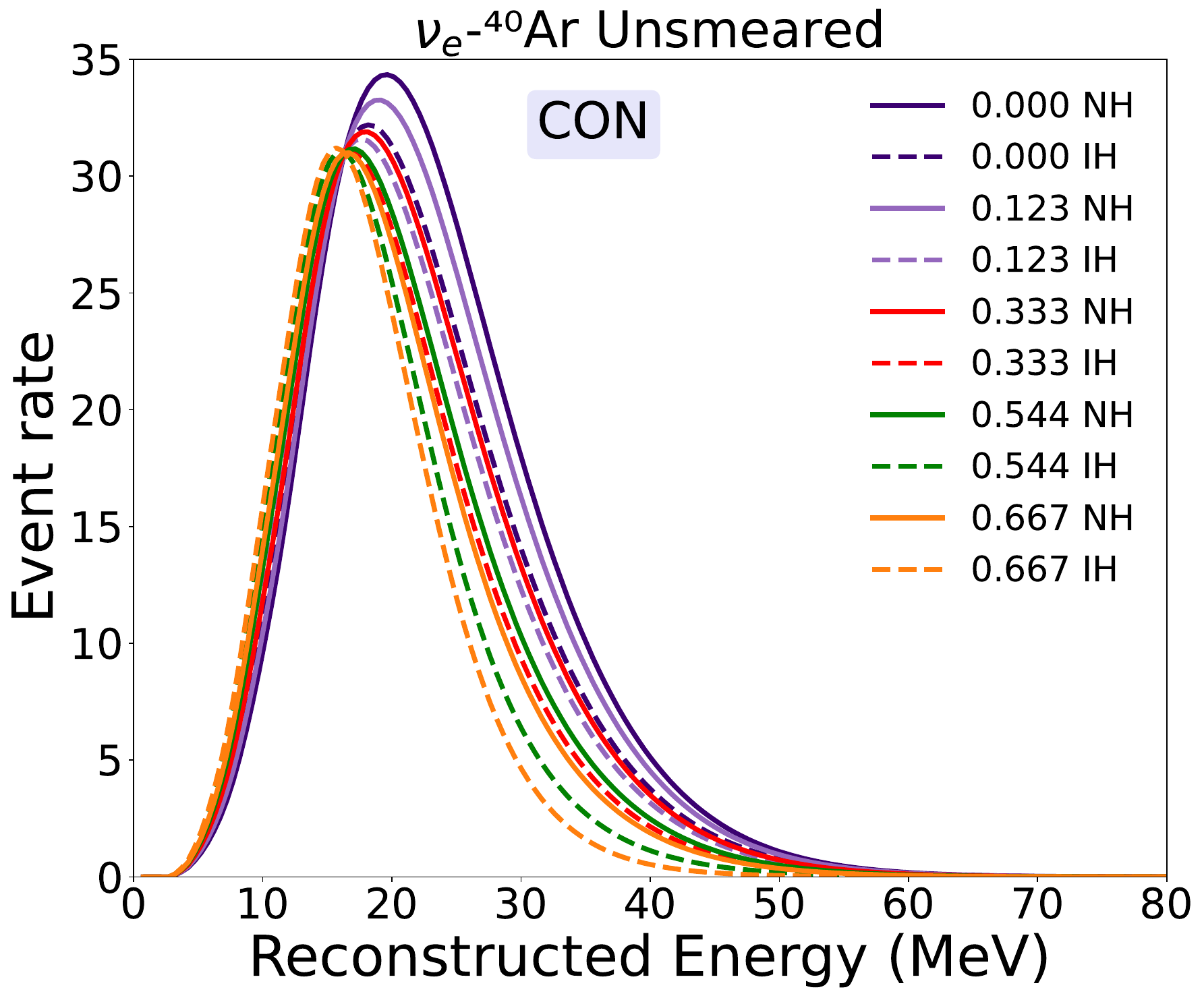}\quad
     \includegraphics[scale=0.25]{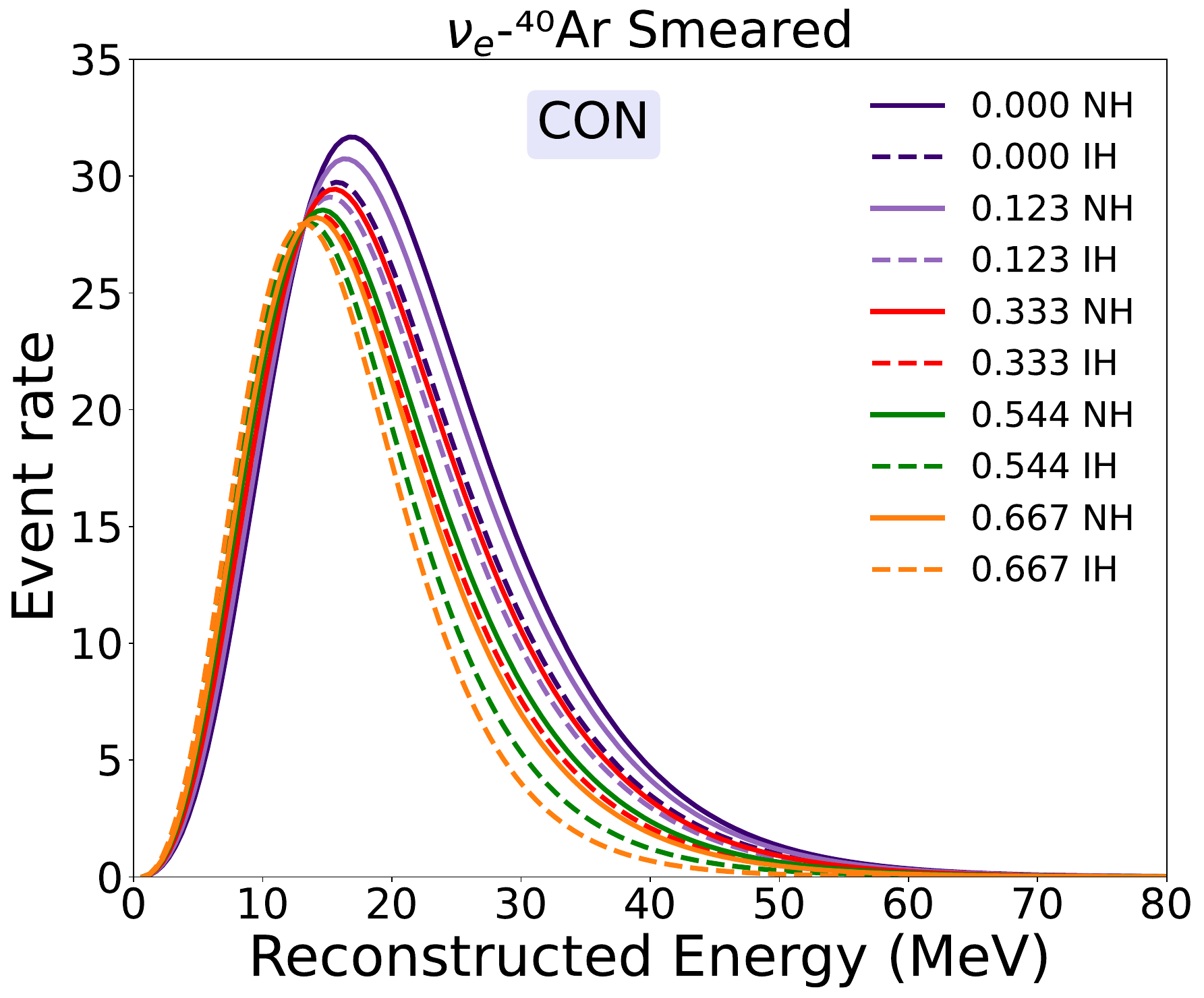}\\
     \includegraphics[scale=0.25]{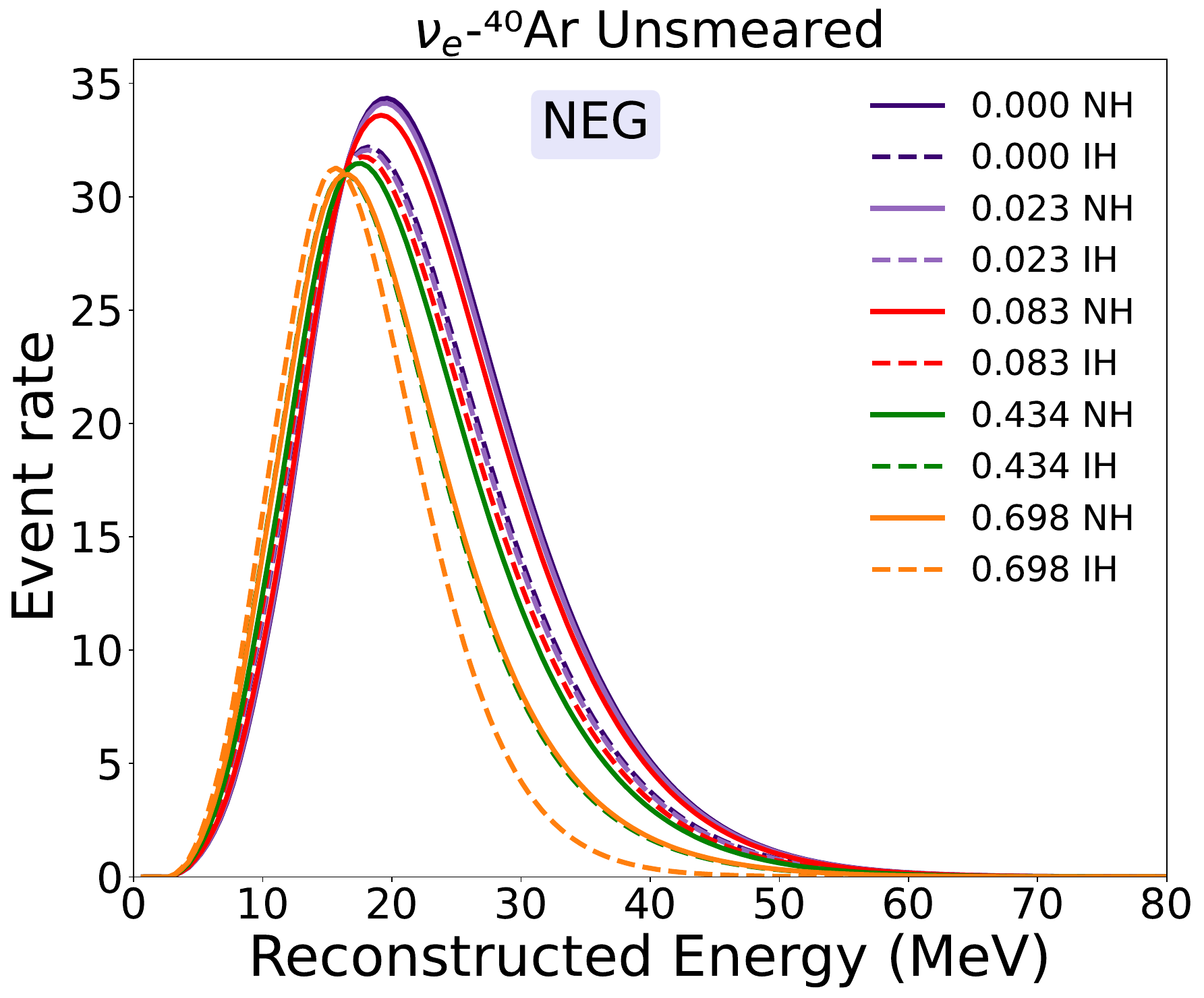}  \quad
     \includegraphics[scale=0.25]{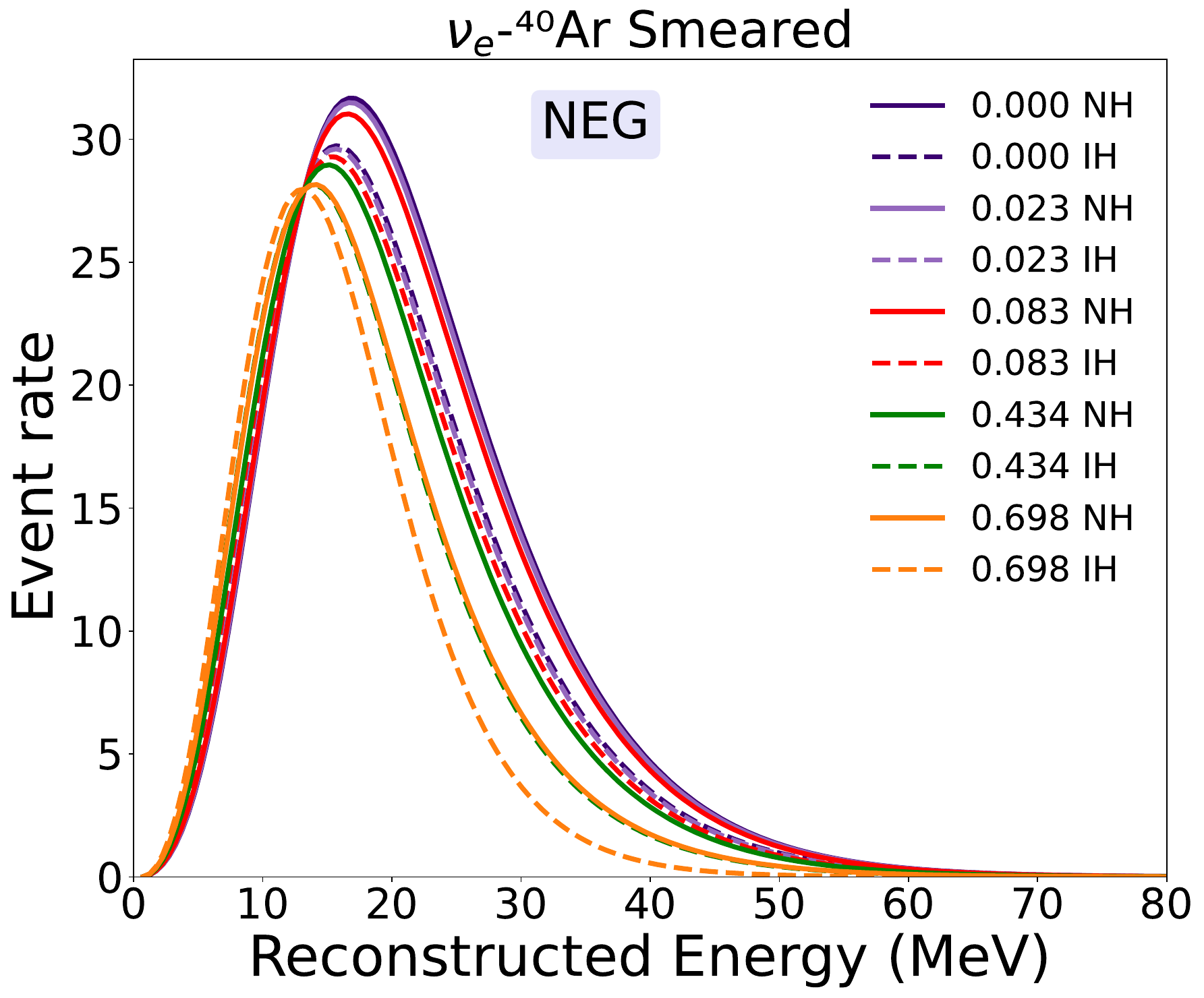}
    \caption{Event rates of $\nu_e$ as functions of reconstructed energy for the entanglement of formation (upper panels), concurrence (middle panels), and negativity (bottom panels). The left and right panels show the unsmeared and smeared event rates, respectively. All results correspond to Channel~A, i.e., the $\nu_e$ charged-current interaction on argon at DUNE.}
    \label{fig:eventrates-channel-A}
\end{figure*}

\begin{figure*}[htbp]
    \centering
    \includegraphics[scale=0.25]{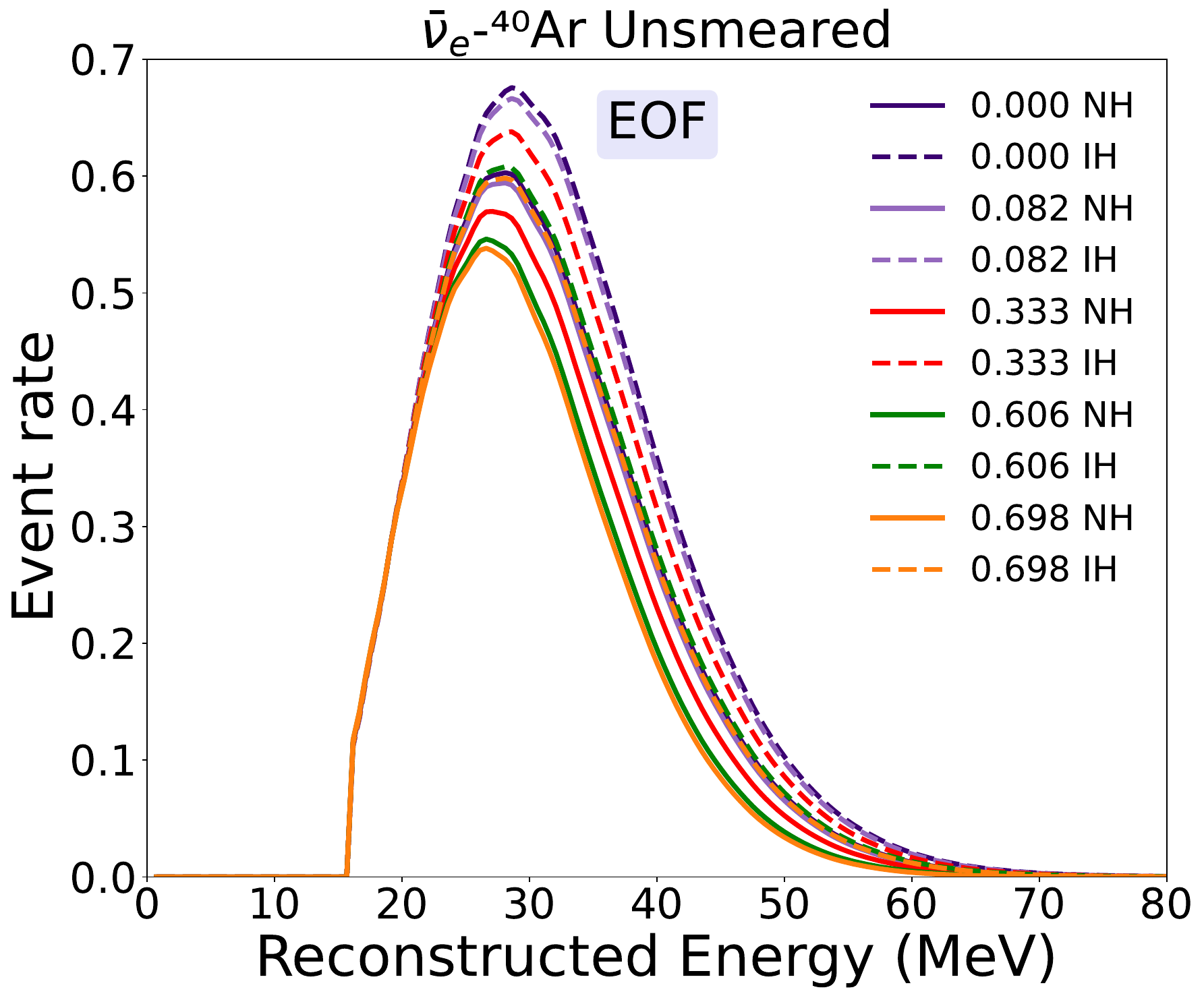}\quad
    \includegraphics[scale=0.25]{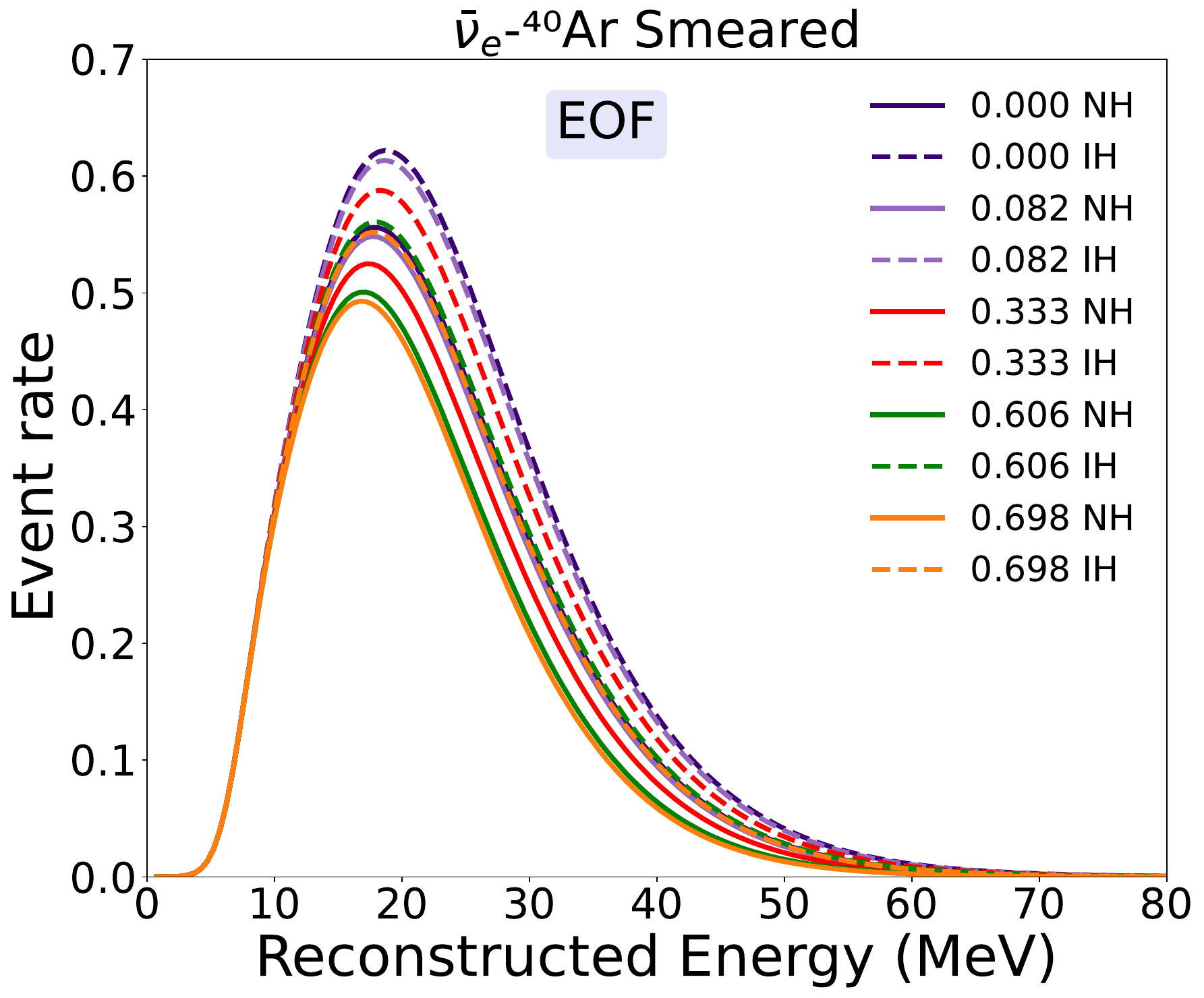}\\
    \includegraphics[scale=0.25]{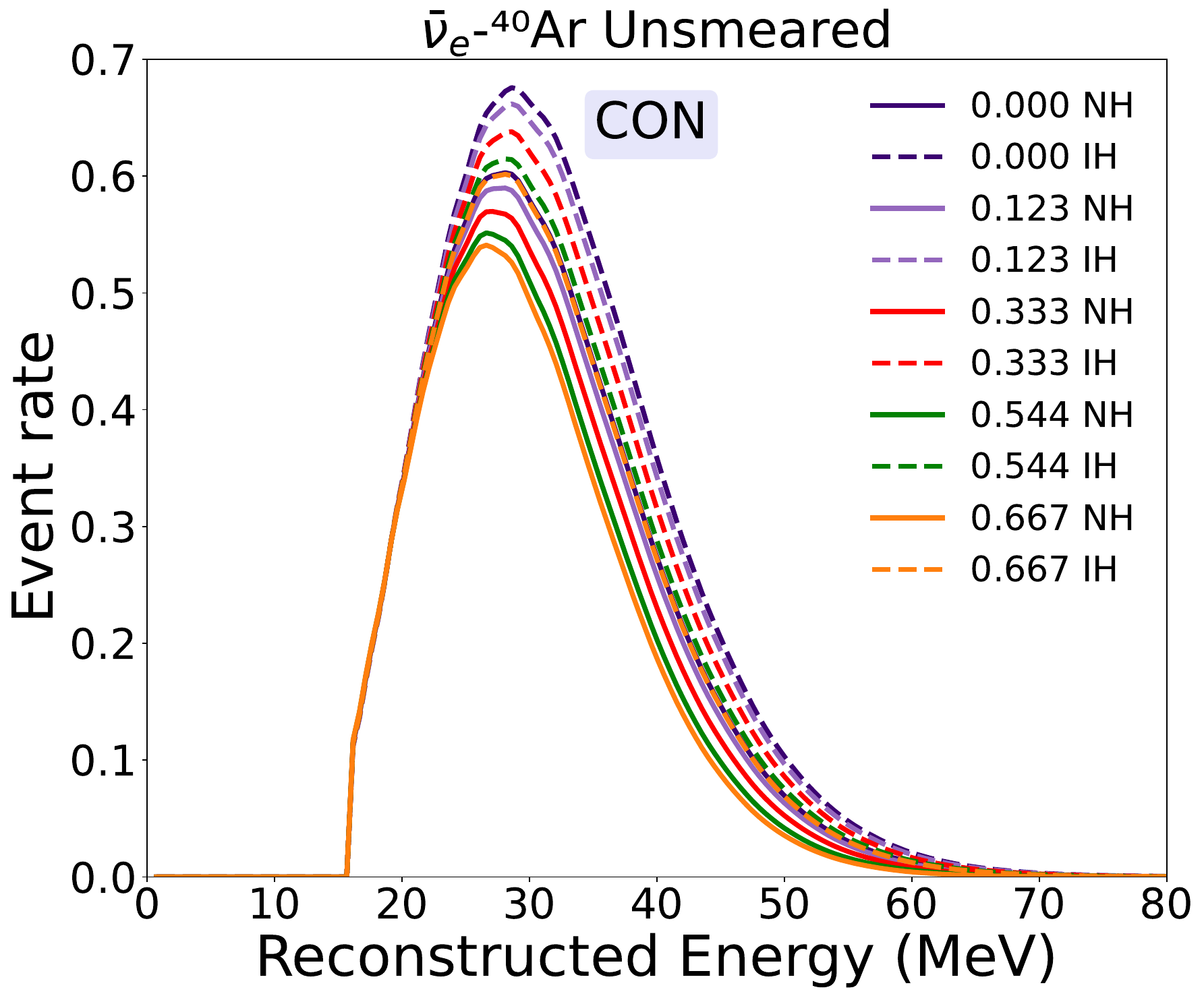}  \quad
     \includegraphics[scale=0.25]{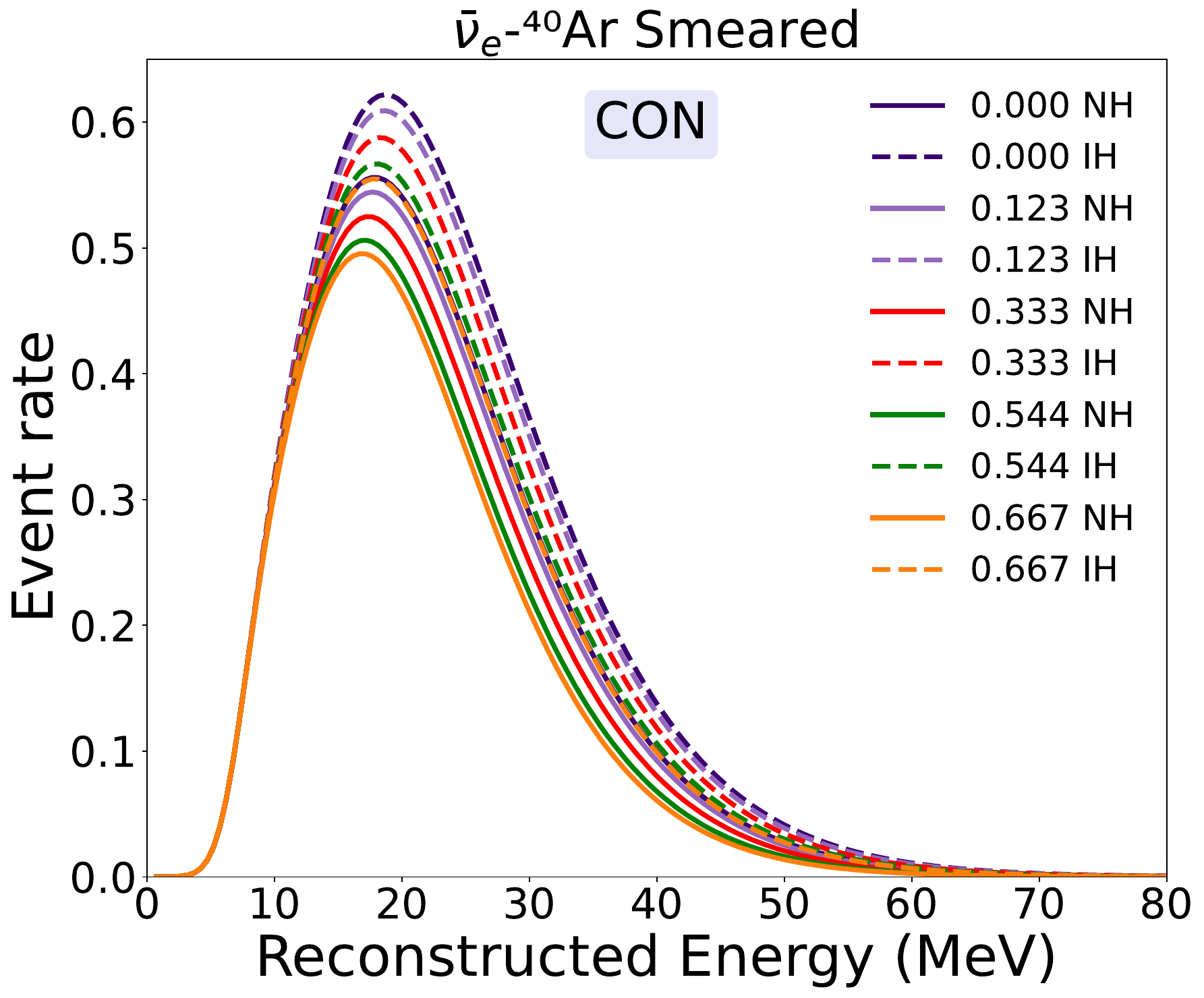} \\
         \includegraphics[scale=0.25]{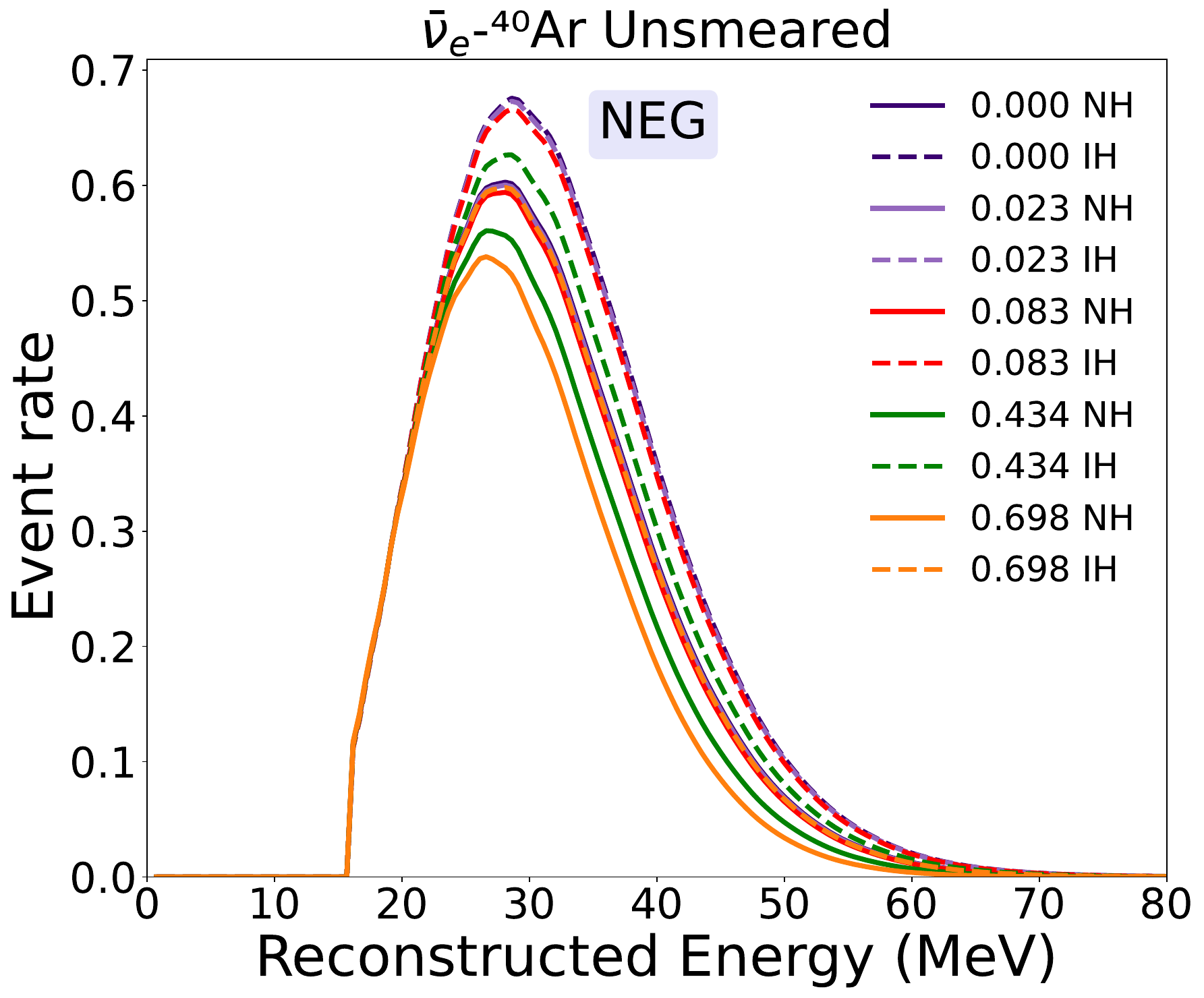}  \quad
     \includegraphics[scale=0.25]{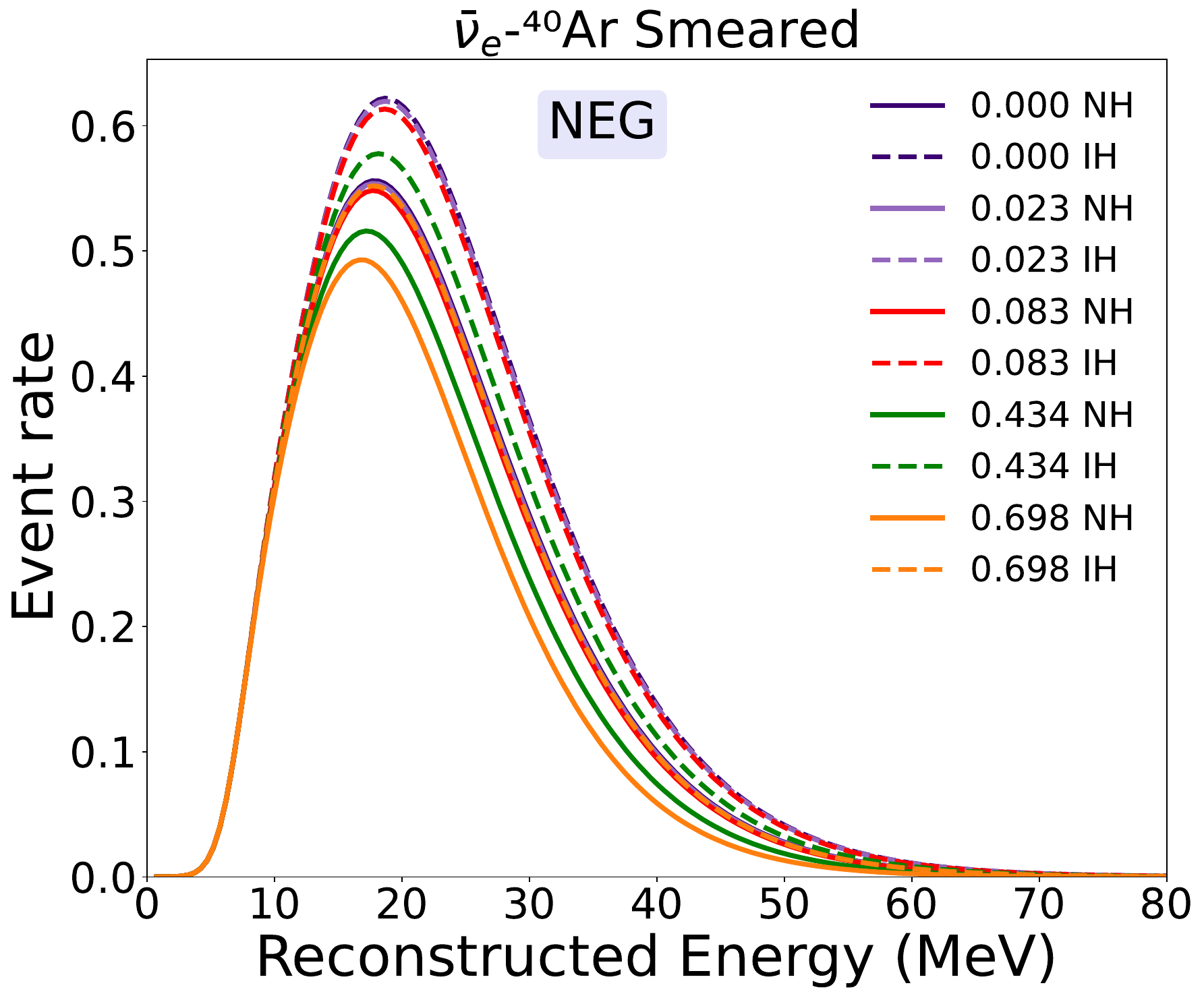}
    \caption{Same as Fig.~\ref{fig:eventrates-channel-A}, but for the event rates of $\bar{\nu}_e$ corresponding to Channel~B, i.e., the $\bar{\nu}_e$ charged-current interaction on argon at DUNE.}
    \label{fig:eventrates-channel-B}
\end{figure*}

\begin{figure*}[htbp!]
    \centering
    \includegraphics[scale=0.25]{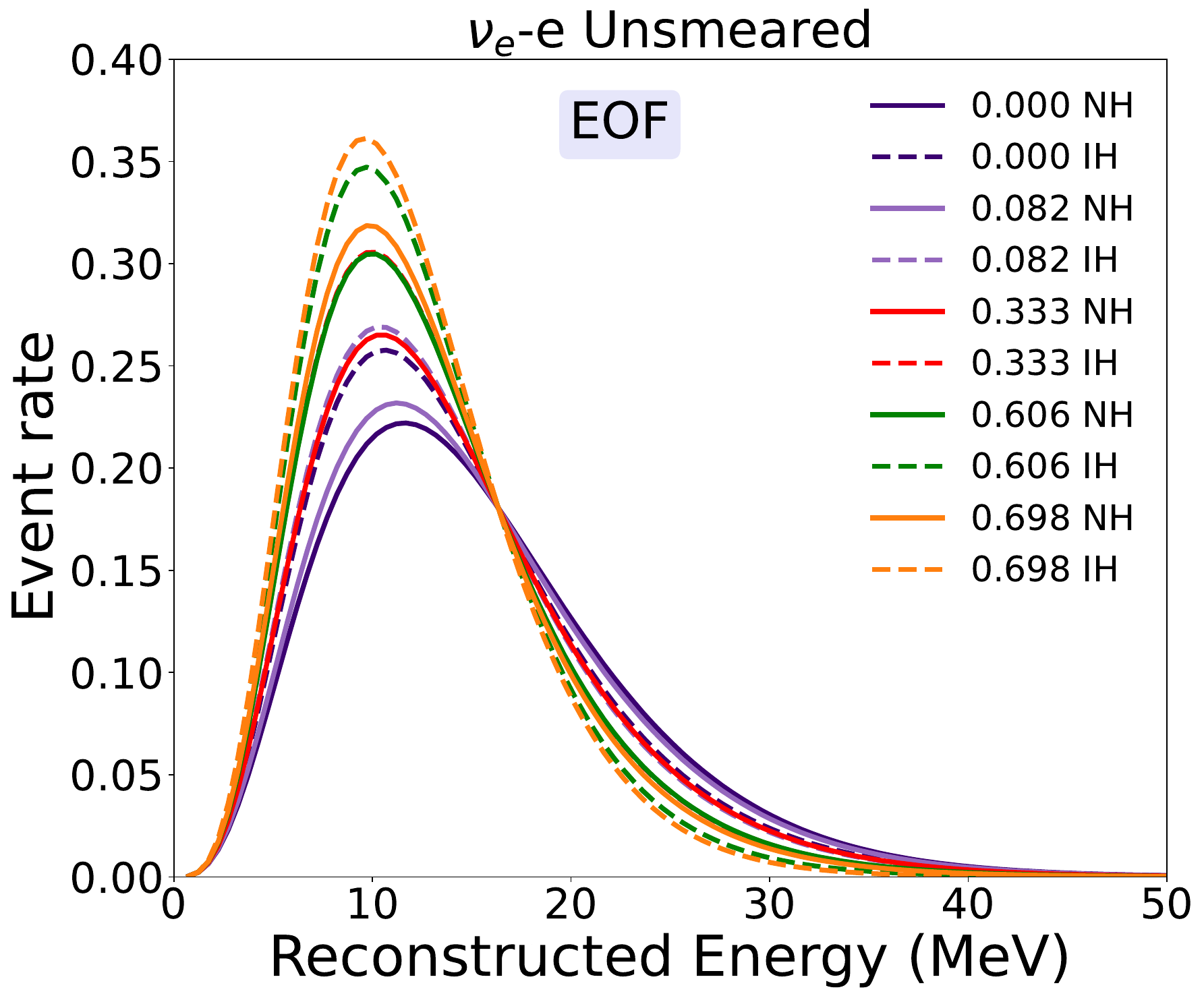}\quad
    \includegraphics[scale=0.25]{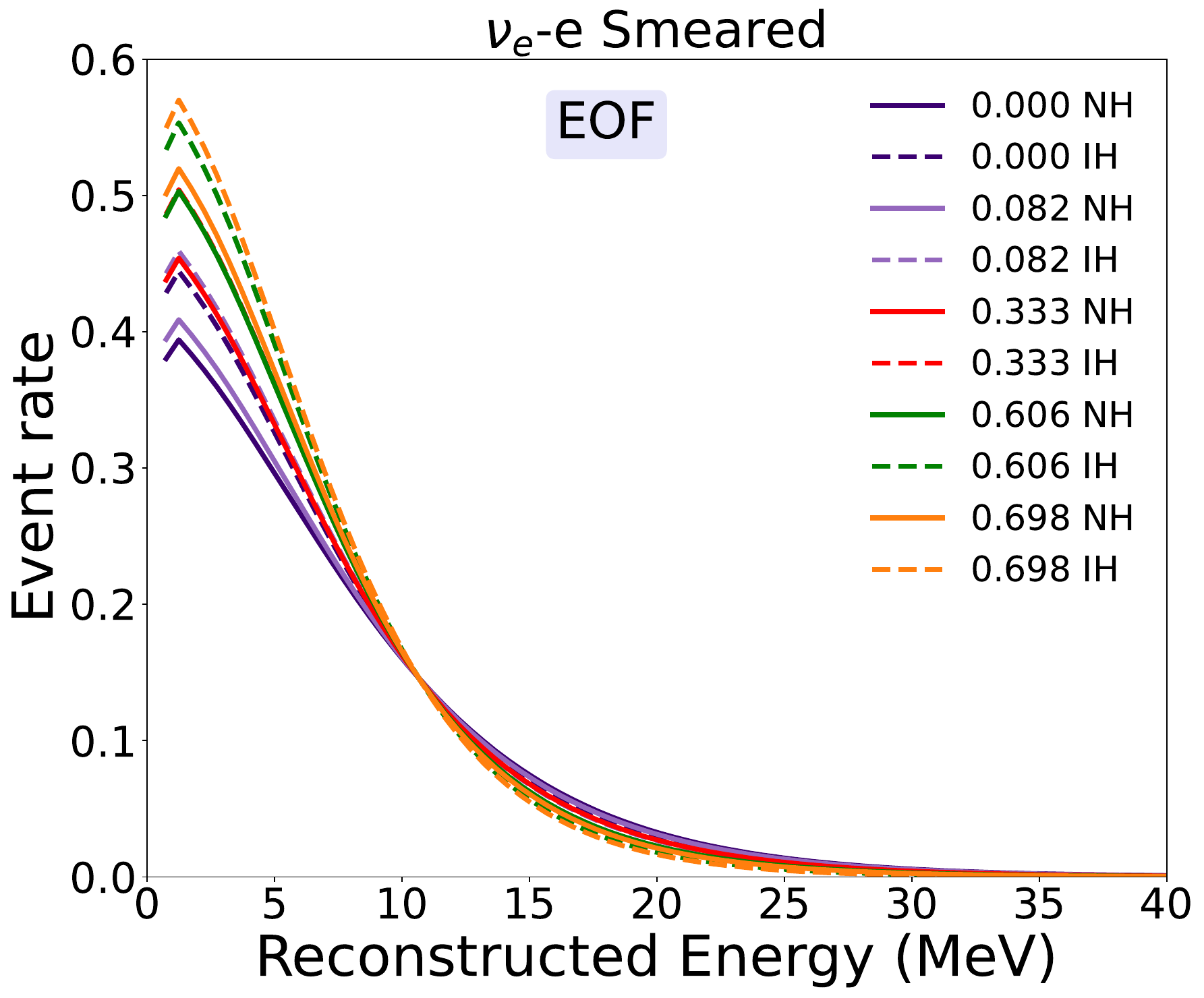}\\
    \includegraphics[scale=0.25]{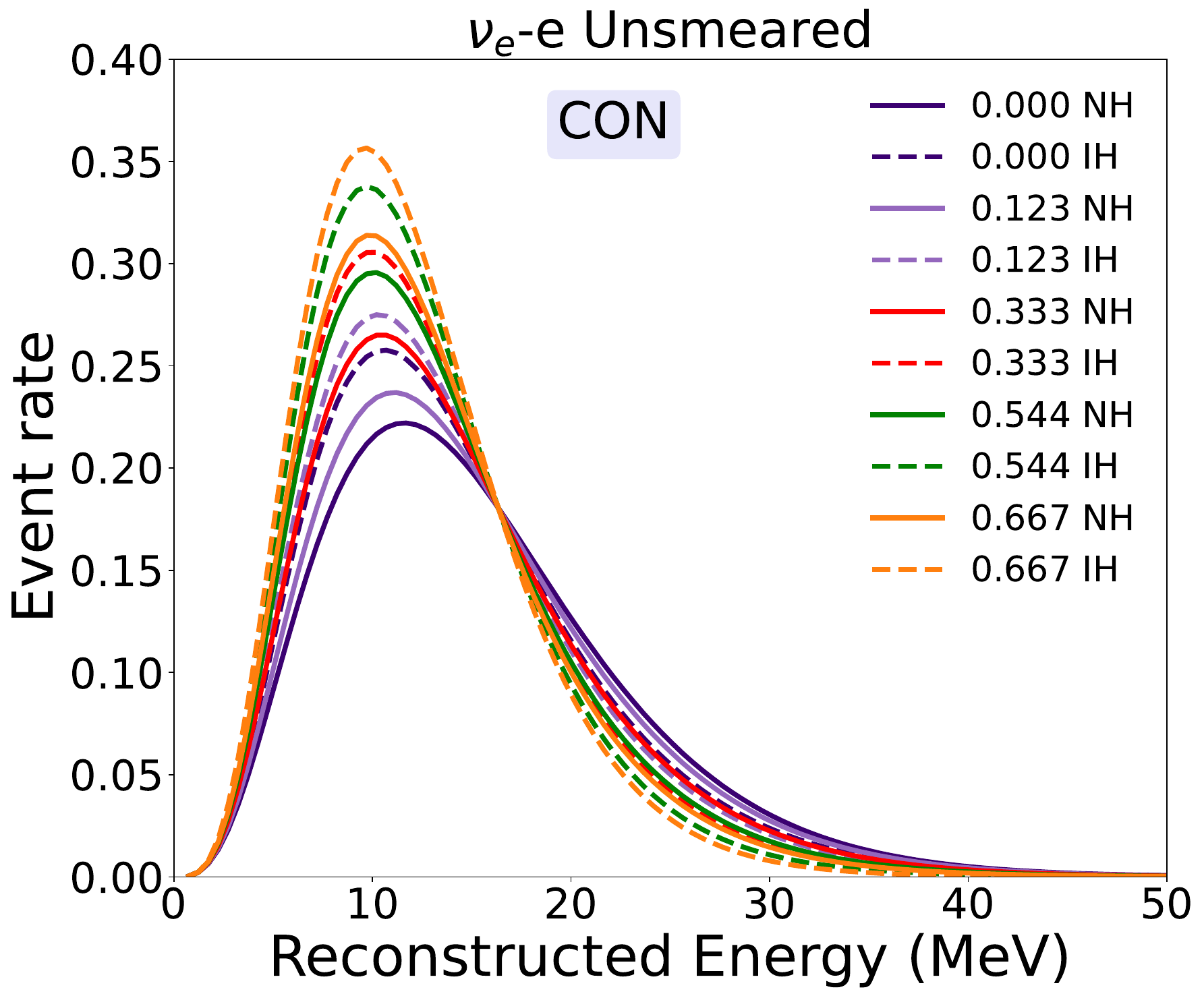}  \quad
     \includegraphics[scale=0.25]{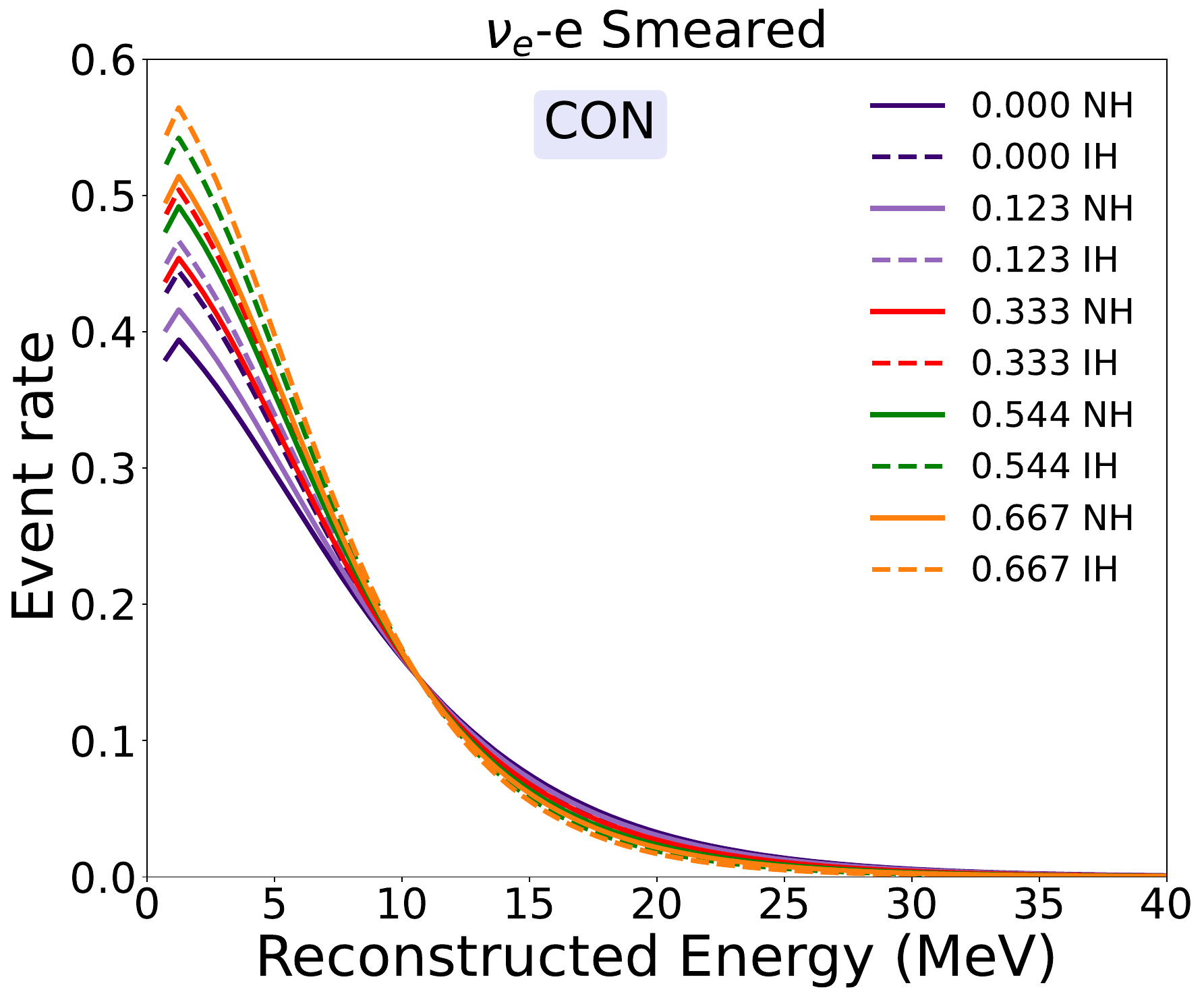}\\
      \includegraphics[scale=0.25]{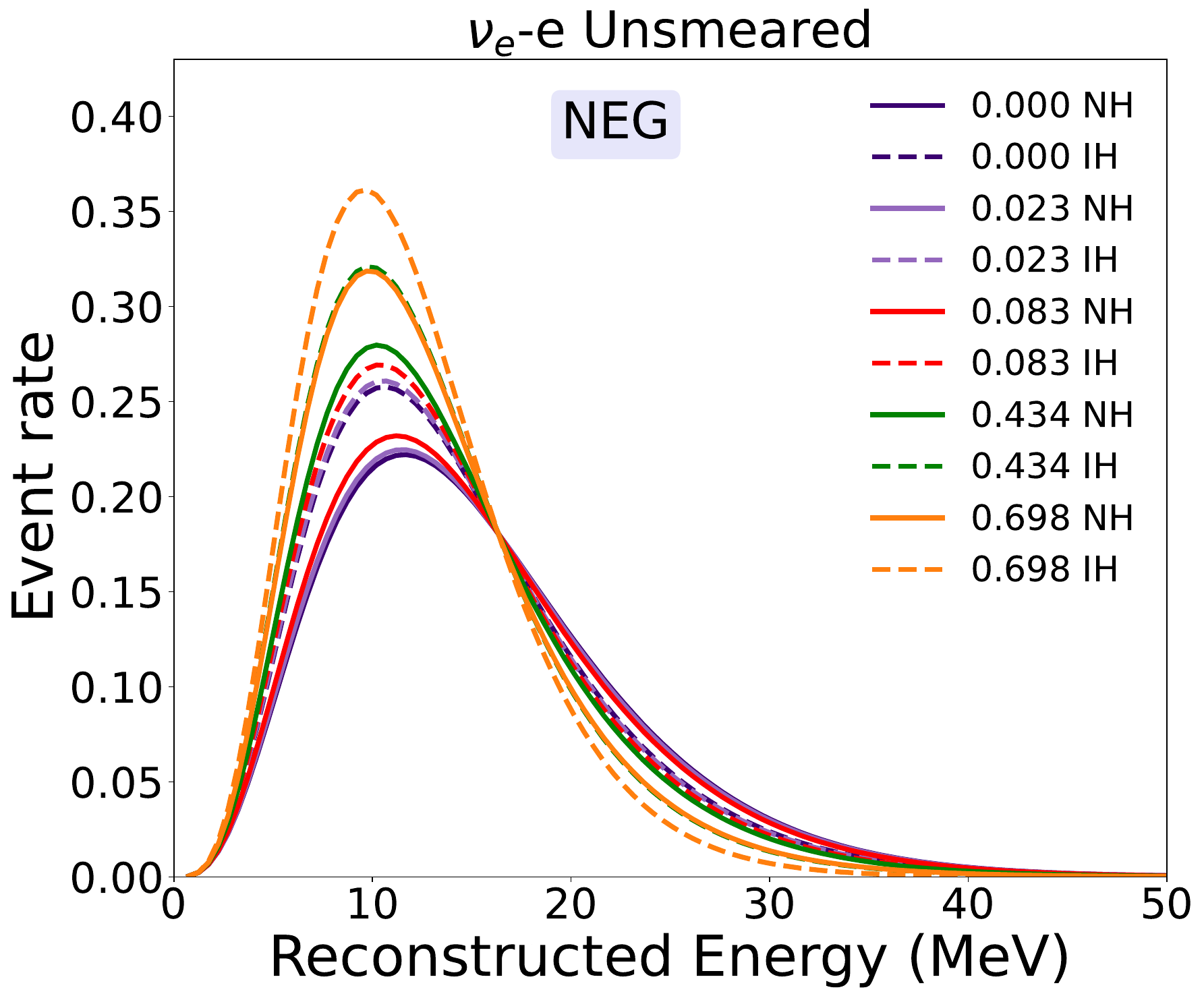}  \quad
     \includegraphics[scale=0.25]{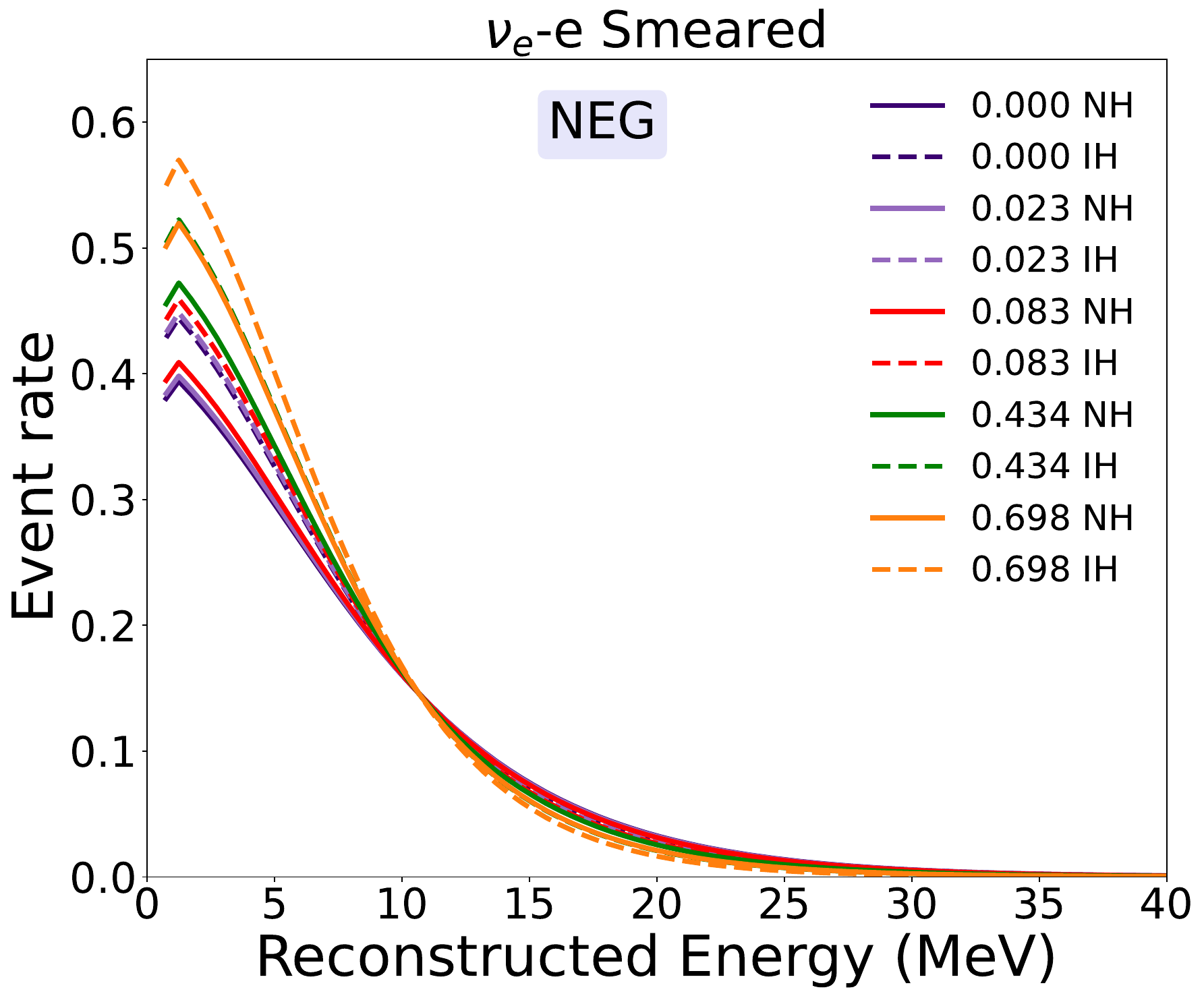}
    \caption{Same as Fig.~\ref{fig:eventrates-channel-A}, but for Channel~C, corresponding to elastic scattering of neutrinos on electrons, $\nu_\alpha + e^- \rightarrow \nu_\alpha + e^-$, at DUNE.}
    \label{fig:eventrates-channel-C}
\end{figure*}

\begin{figure*}[htbp]
    \centering
    \includegraphics[scale=0.25]{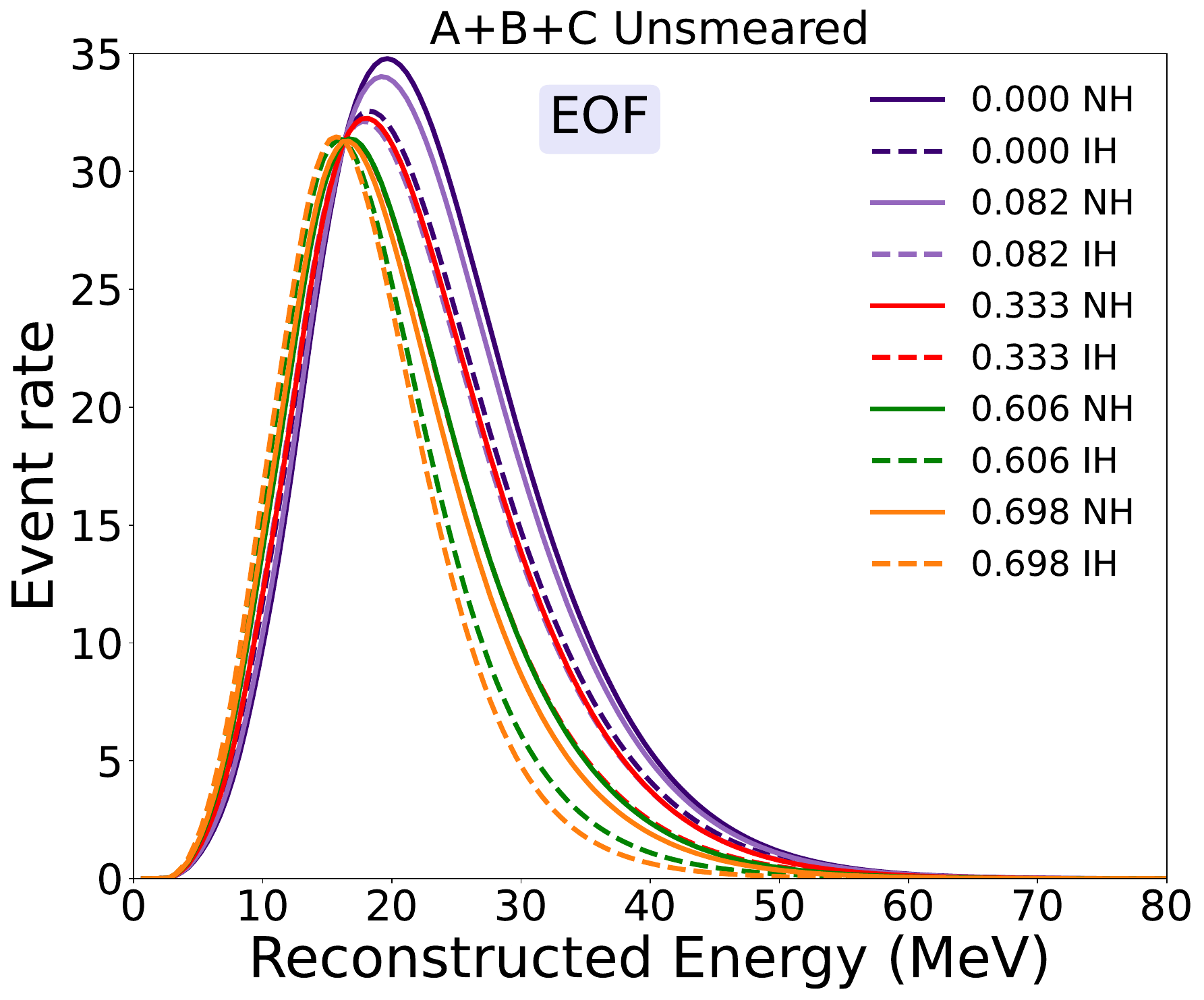}\quad
    \includegraphics[scale=0.25]{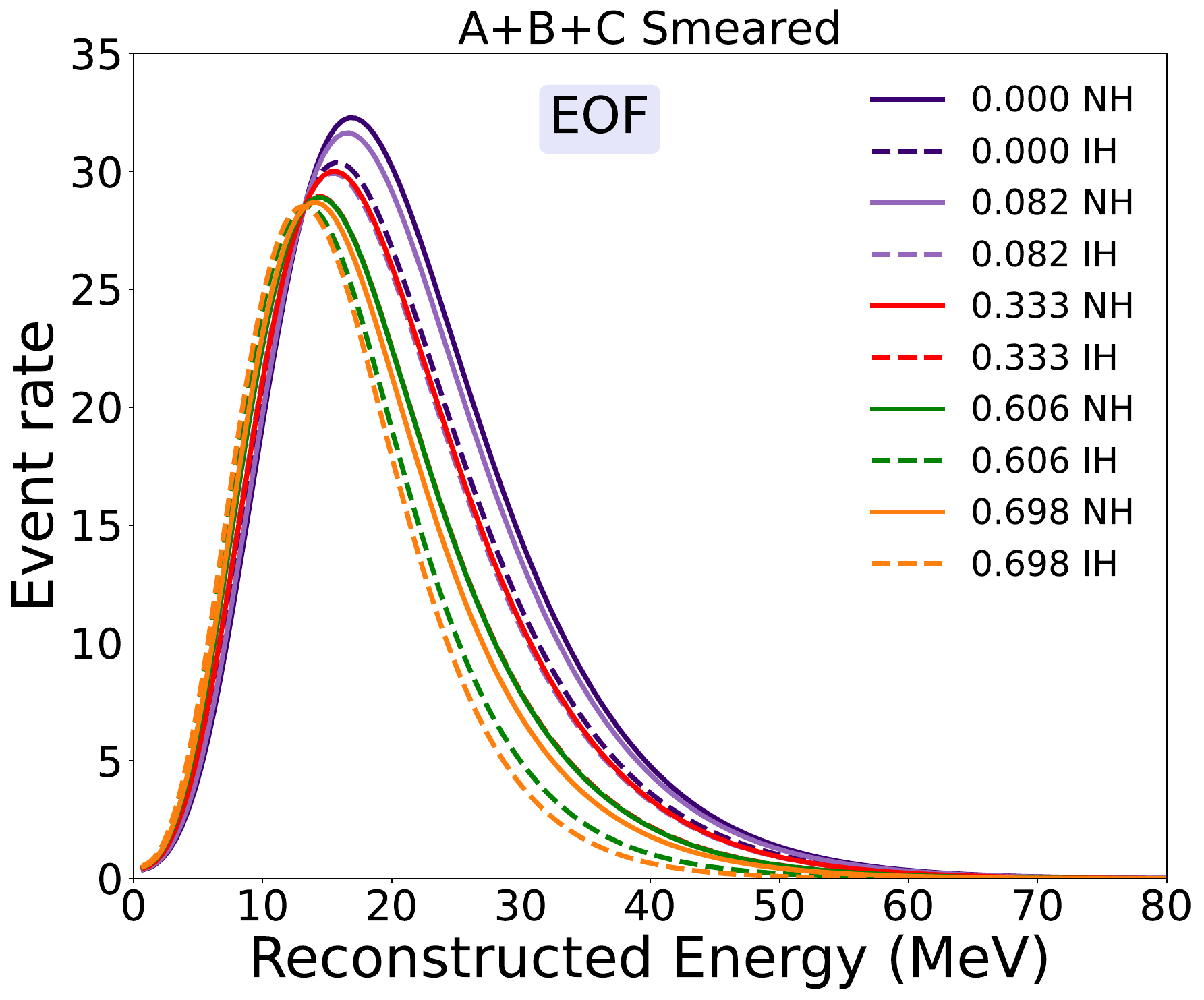}\quad
        \includegraphics[scale=0.25]{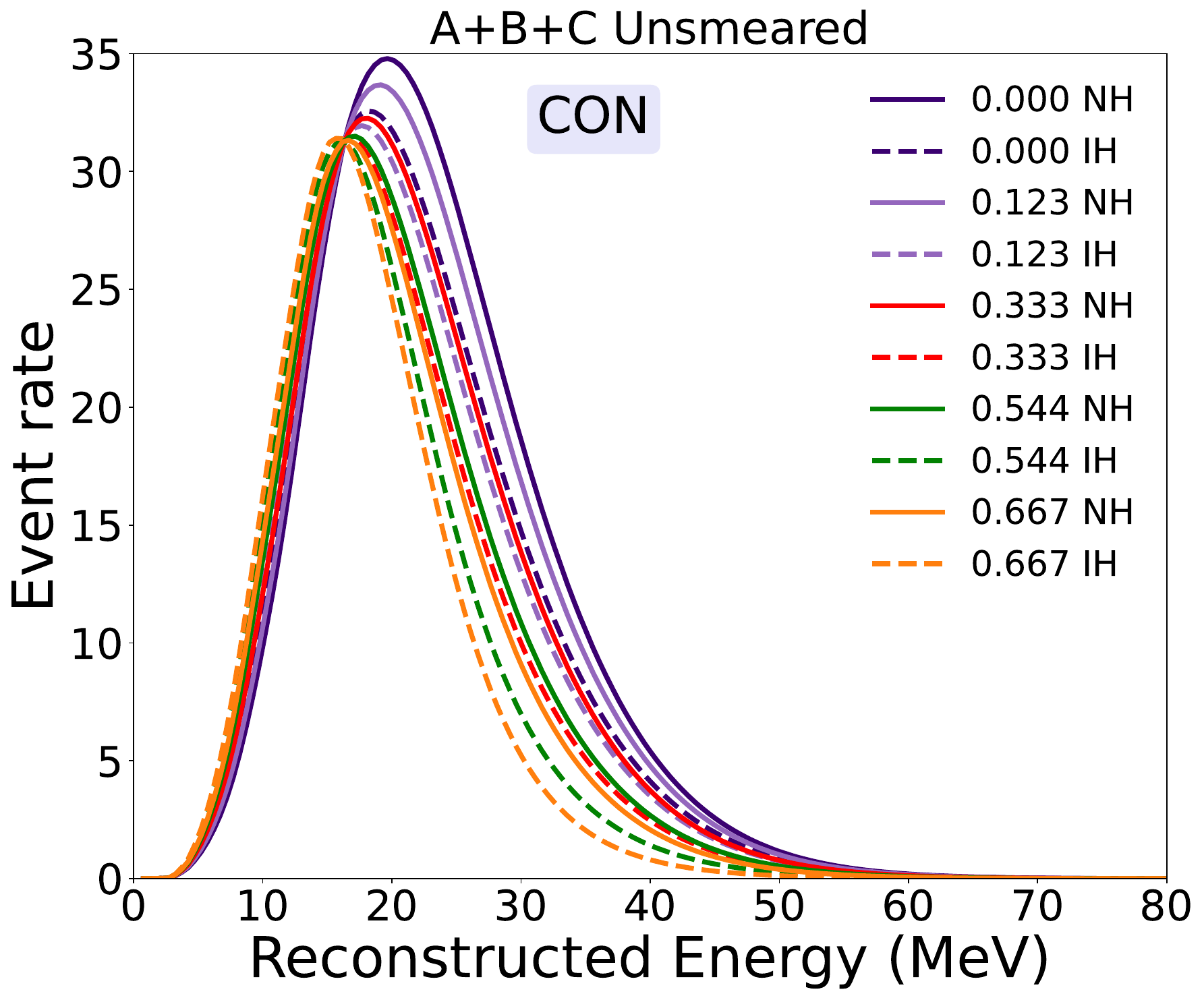}\quad
            \includegraphics[scale=0.25]{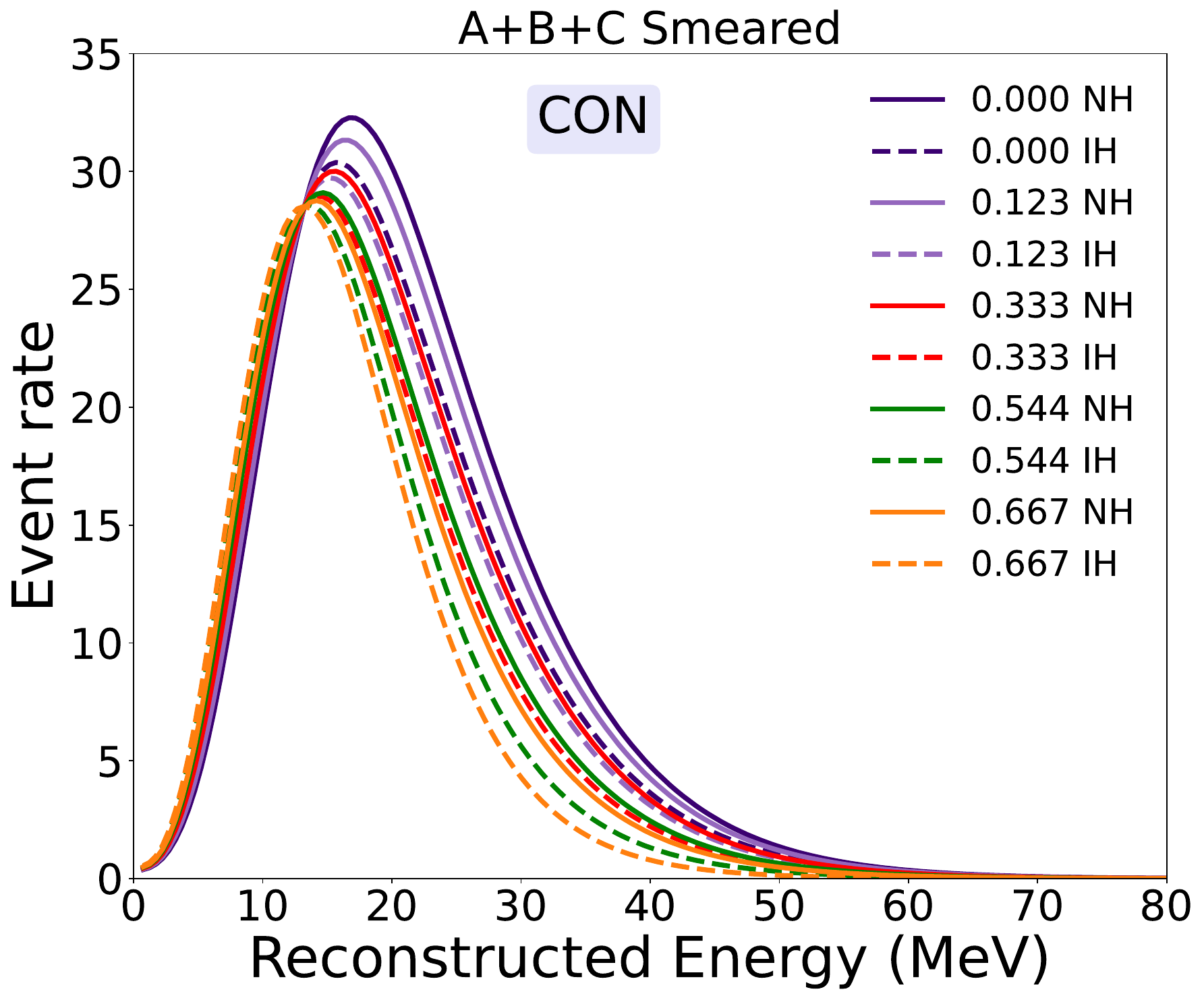}\quad
    \includegraphics[scale=0.25]{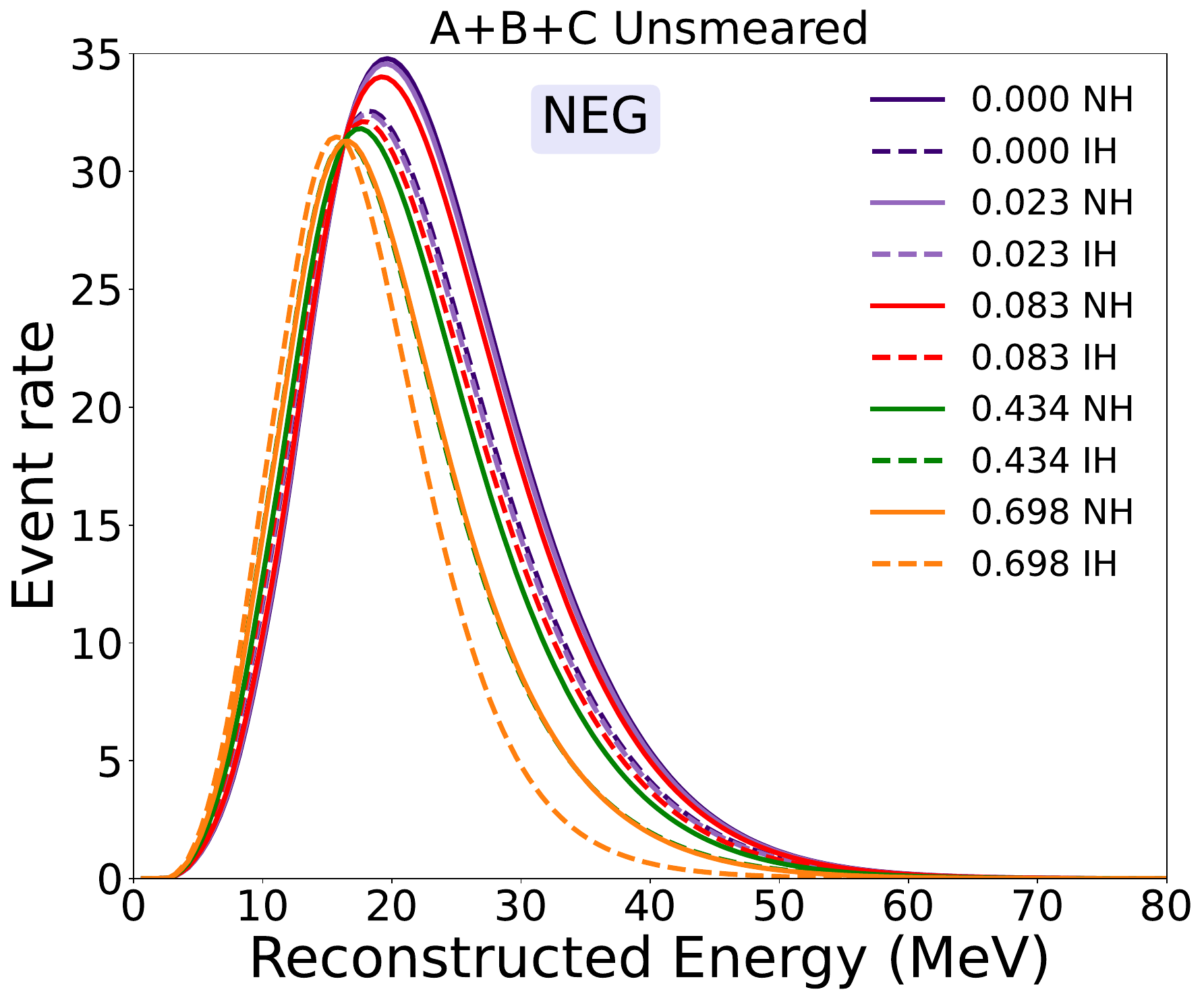}\quad
    \includegraphics[scale=0.25]{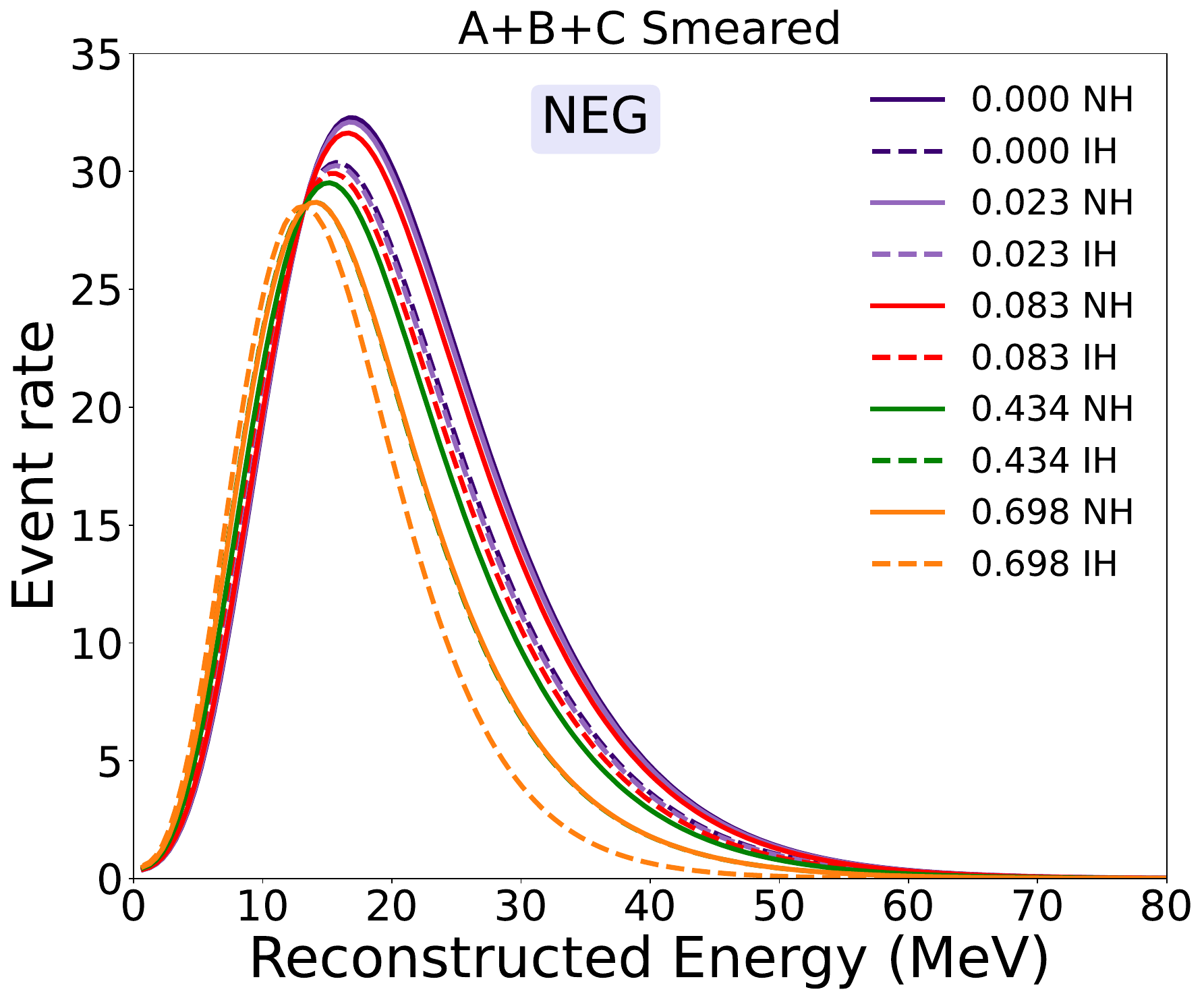}\\
    \caption{Same as Fig.~\ref{fig:eventrates-channel-A}, but for the combined event rates from Channel~A ($\nu_e$ CC on Ar), Channel~B ($\bar{\nu}_e$ CC on Ar), and Channel~C (elastic scattering on electrons) at DUNE.}
    \label{fig:eventrates-channel-ABC}
\end{figure*}

In Fig.~\ref{fig:eventrates-channel-A}, we show the event rate as a function of neutrino energy for EOF, CON and NEG in the upper, middle and lower panels, respectively, assuming a supernova distance of 10 kpc. Here, the left and right panels correspond to the unsmeared and smeared event rates for Channel~A ($\nu_e -{}^{40}{\rm Ar}$). In each panel, the solid and dashed curves denote the NH and IH scenarios, respectively, and different colors represent different benchmark values of the parameter $\Delta p$. The event rate increases with energy in the low energy region and reaches a maximum around 20~MeV. Both the unsmeared and smeared event rates are largest for $\Delta p = 0$ and decrease as $\Delta p$ increases. For the standard case, the event rate peaks near 20~MeV, whereas increasing $\Delta p$ shifts the peak toward lower energies for all three measures (EOF, CON, and NEG). In the NH scenario event rate peaks at higher energy compare to the IH scenario.  At higher energies, the event rate decreases sharply due to the suppression of the fluence.  For all values of $\Delta p$, the event rate for NH remains higher than that for IH, which is opposite to the behavior observed in the fluence curves (Fig.~\ref{fig:fluence}). This behavior arises because the event rate depends on the convolution of the flux with the energy dependent cross section. Since the event rate for nonzero $\Delta p$ shifts toward lower energies relative to the NH case, the integrated contribution from the cross section is reduced, leading to a smaller total event rate compared to NH. The total event rates for both NH and IH scenarios, including the effects of the three entanglement measures, are presented in Table~\ref{Tab:eventrate}.

Similar to Fig.~\ref{fig:eventrates-channel-A}, Fig.~\ref{fig:eventrates-channel-B} shows the event rate as a function of energy for Channel B ($\bar{\nu}_e -{}^{40}{\rm Ar}$) for a supernova at a distance of 10 kpc. The upper, middle, and lower panels correspond to the EOF, CON, and NEG cases, respectively, while the left and right panels show the unsmeared and smeared results. In this channel, the unsmeared event rate peaks at a neutrino energy of approximately 30 MeV, whereas after smearing the reconstructed energy peak shifts to about 20 MeV. Overall, the qualitative behavior of the curves is similar to that in Fig.~\ref{fig:eventrates-channel-A}, except for the hierarchy dependence. In this channel, the IH event rates are higher than the NH rates, which is opposite to the trend observed in Fig.~\ref{fig:eventrates-channel-A} and consistent with the behavior shown in Fig.~\ref{fig:fluence}. This reversal originates from the combined energy dependence of the flux and the cross section in the convolution that determines the event rate.  This reflects the hierarchy dependence of the $\bar{\nu}_e$ flux due to MSW flavor conversions inside the supernova, while the cross section remains hierarchy independent. Table~\ref{Tab:eventrate} summarizes the total event rates for Channel B at different values of $\Delta p$.

The event rates at supernova distance of 10 kpc for Channel C (Elastic Scattering on $e^-$) for EOF, CON and NEG as a function of energy is depicted  in Fig.\ref{fig:eventrates-channel-C}. Both the unsmeared and smeared event rates have shown for different benchmark values of $\Delta p$. In this channel, the event rates are significantly lower than those of Channels A and B. This is primarily due to the smaller interaction cross section and the lower recoil energy of the electron during neutrino scattering. The unsmeared event rates peak around 12 MeV, while the smeared event rates peak at a few MeV of reconstructed energy. Despite the lower statistics, this channel is particularly valuable for supernova localization, as the direction of the scattered electrons provides information about the incoming neutrino \cite{Scholberg:2012id, Super-Kamiokande:2008mmn}.

Figure~\ref{fig:eventrates-channel-ABC} shows the event rate as a function of reconstructed energy for the combined contributions of Channels A, B, and C. The overall behavior is dominated by Channel A, whose larger contribution largely determines the shape of the total event rate. This dominance arises from the comparatively larger charged current $\nu_e$ cross section on ${}^{40}\mathrm{Ar}$, which increases rapidly with energy. In contrast, the elastic scattering channel (Channel C) has a much smaller cross section, while the Channel B event rate is influenced by the mass ordering through the $\bar{\nu}_e$ flux resulting from MSW flavor conversions in the supernova.


\begin{table*}[htbp]
\centering
\begin{tabular}{|l|
c@{\hspace{5pt}}c| 
c@{\hspace{5pt}}c| 
c@{\hspace{5pt}}c| 
c@{\hspace{5pt}}c| 
c@{\hspace{5pt}}c|}
\hline
\textbf{Channels} 
& \textbf{NH} & \textbf{IH} & \textbf{NH} & \textbf{IH} & \textbf{NH} & \textbf{IH} & \textbf{NH} & \textbf{IH} & \textbf{NH} & \textbf{IH} \\
\hline

\multicolumn{11}{|c|}{\textbf{Entanglement of Formation}} \\
\hline
& \multicolumn{2}{c|}{$\boldsymbol{\Delta p=0.000}$} 
& \multicolumn{2}{c|}{$\boldsymbol{\Delta p=0.082}$} 
& \multicolumn{2}{c|}{$\boldsymbol{\Delta p=0.333}$} 
& \multicolumn{2}{c|}{$\boldsymbol{\Delta p=0.606}$} 
& \multicolumn{2}{c|}{$\boldsymbol{\Delta p=0.698}$} \\ \hline
$\nu_e + {}^{40}\mathrm{Ar}$           &1380.93 &1244.4 & 1340.79&1204.26 &1217.91 &1081.38 &1084.26 &947.73 &1039.22 &902.69 \\
$\bar{\nu}_e + {}^{40}\mathrm{Ar}$     & 25.10& 29.46&24.55 & 28.92&22.88 & 27.24& 21.06&25.42 &20.45 & 24.81\\
$\nu_\alpha + e^-$                     &7.49 &7.87 &7.61 & 7.98& 7.95&8.32 &8.31 &8.69  &8.44 &8.81  \\
A+B+C                     &1413.52 &1281.73 &1412.95 &1241.16 & 1248.74&1116.94 &1113.63 &981.84  &1068.11 & 936.31 \\
\hline

\multicolumn{11}{|c|}{\textbf{Concurrence}} \\
\hline
& \multicolumn{2}{c|}{$\boldsymbol{\Delta p=0.000}$} 
& \multicolumn{2}{c|}{$\boldsymbol{\Delta p=0.123}$} 
& \multicolumn{2}{c|}{$\boldsymbol{\Delta p=0.333}$} 
& \multicolumn{2}{c|}{$\boldsymbol{\Delta p=0.554}$} 
& \multicolumn{2}{c|}{$\boldsymbol{\Delta p=0.667}$} \\ \hline
$\nu_e + {}^{40}\mathrm{Ar}$           &1380.93 &1244.4 & 1320.71& 1184.19&1217.91 &1081.38 & 1114.61&978.08 & 1054.39&917.87 \\
$\bar{\nu}_e + {}^{40}\mathrm{Ar}$     &25.10 &29.46 &24.28 & 28.64&22.88 & 27.24& 21.48& 25.84& 20.66& 25.02\\
$\nu_\alpha + e^-$                     & 7.49& 7.87& 7.66& 8.04& 7.95&8.32 &8.23 & 8.60&8.39 & 8.77\\
A+B+C                     & 1413.52& 1281.73&1352.65 & 1220.87&1248.74 & 1116.91& 1144.32& 1012.52 &1083.44 & 951.66 \\
\hline

\multicolumn{11}{|c|}{\textbf{Negativity}} \\
\hline
& \multicolumn{2}{c|}{$\boldsymbol{\Delta p=0.000}$} 
& \multicolumn{2}{c|}{$\boldsymbol{\Delta p=0.023}$} 
& \multicolumn{2}{c|}{$\boldsymbol{\Delta p=0.083}$} 
& \multicolumn{2}{c|}{$\boldsymbol{\Delta p=0.434}$} 
& \multicolumn{2}{c|}{$\boldsymbol{\Delta p=0.698}$} \\ \hline
$\nu_e + {}^{40}\mathrm{Ar}$           &1380.93 &1244.4 &1369.67 &1233.14 &1340.3 &1203.77 & 1168.46&1031.93 & 1039.22& 902.69\\
$\bar{\nu}_e + {}^{40}\mathrm{Ar}$     &25.10 &29.46 & 24.95&29.31 & 24.55& 28.91&22.21 &26.57 & 20.45&24.81 \\
$\nu_\alpha + e^-$                     &7.49 &7.87 & 7.53&7.90 & 7.61&7.98 &8.08 &8.46 &8.44 &8.81 \\
A+B+C                     &1413.52 &1281.73 & 1402.15&1270.35 &1372.46 &1240.66 &1198.75 &  1066.96& 1068.11& 936.31 \\
\hline

\end{tabular}
\caption{Total event rates at DUNE for a Galactic supernova at a distance of 10~kpc. 
Event rates are shown for the individual detection channels (Channel~A, Channel~B, and Channel~C),  as well as for their combined contribution (A+B+C). Results are presented for different benchmark values of the survival probability variation $\Delta p$, 
corresponding to the entanglement of formation, concurrence, and negativity, for both normal and inverted hierarchy.}
\label{Tab:eventrate}
\end{table*}




\subsection{Mass Ordering Sensitivity}
\label{MO}

In this subsection, we study the impact of quantum entanglement on the mass hierarchy (MH) sensitivity of the experiment. The MH sensitivity quantifies the experiment’s ability to distinguish between the true neutrino mass ordering and the wrong one. Throughout this analysis, the normal hierarchy  is assumed as the true hypothesis, while the inverted hierarchy  is taken as the test hypothesis.

\begin{figure*}[htbp]
    \centering
    \includegraphics[width=0.48\linewidth]{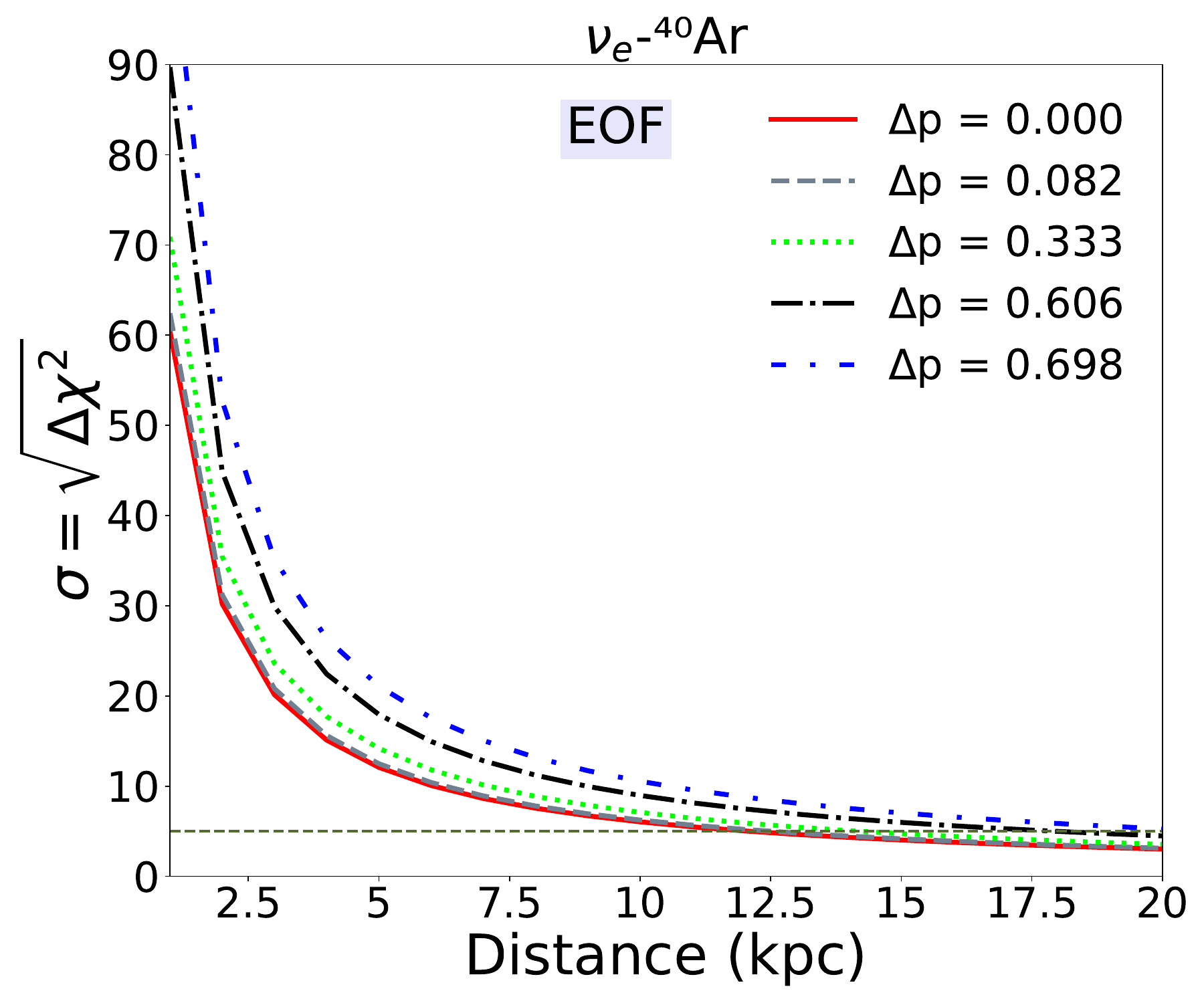}\quad
    \includegraphics[width=0.48\linewidth]{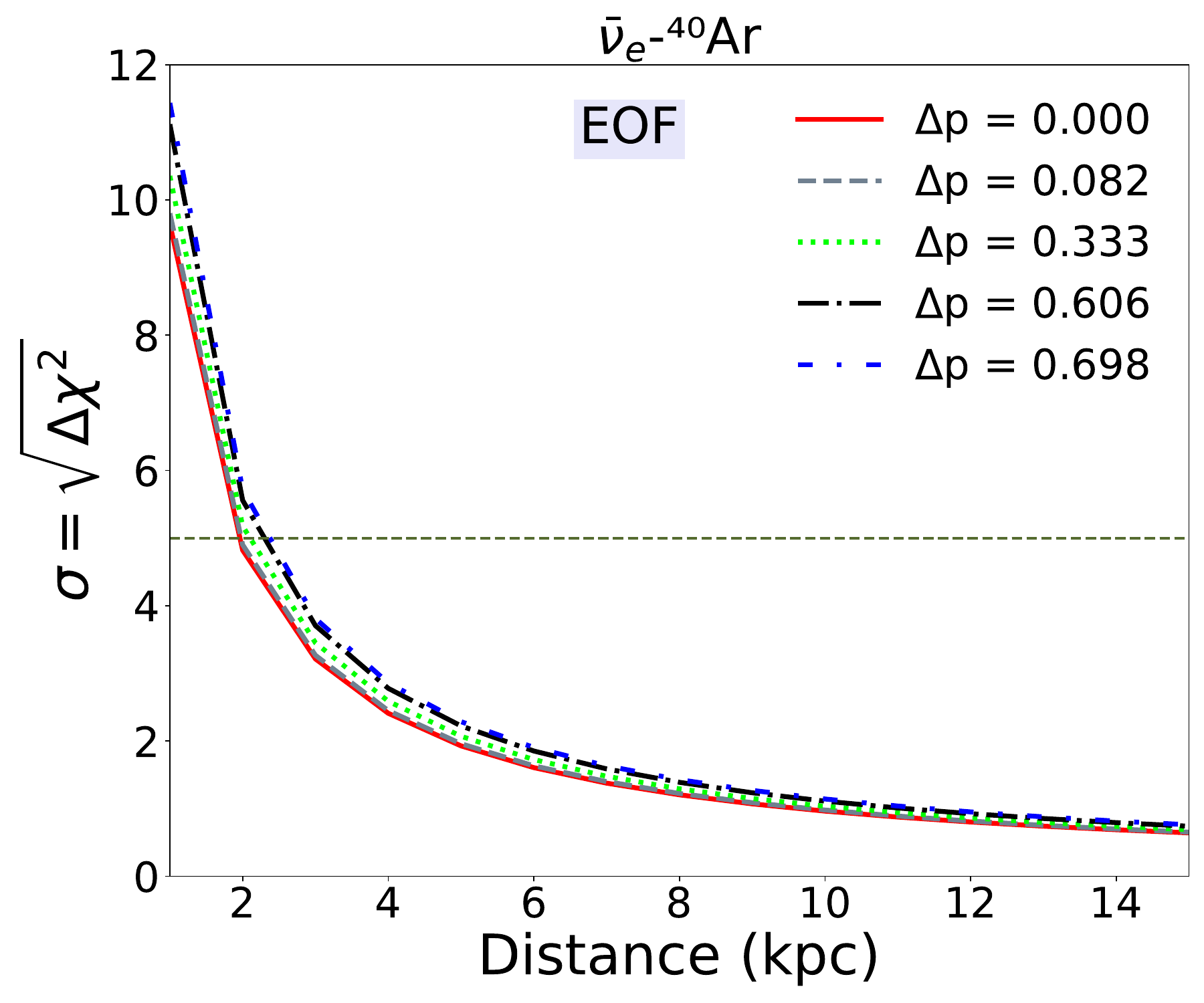}\\
    \includegraphics[width=0.48\linewidth]{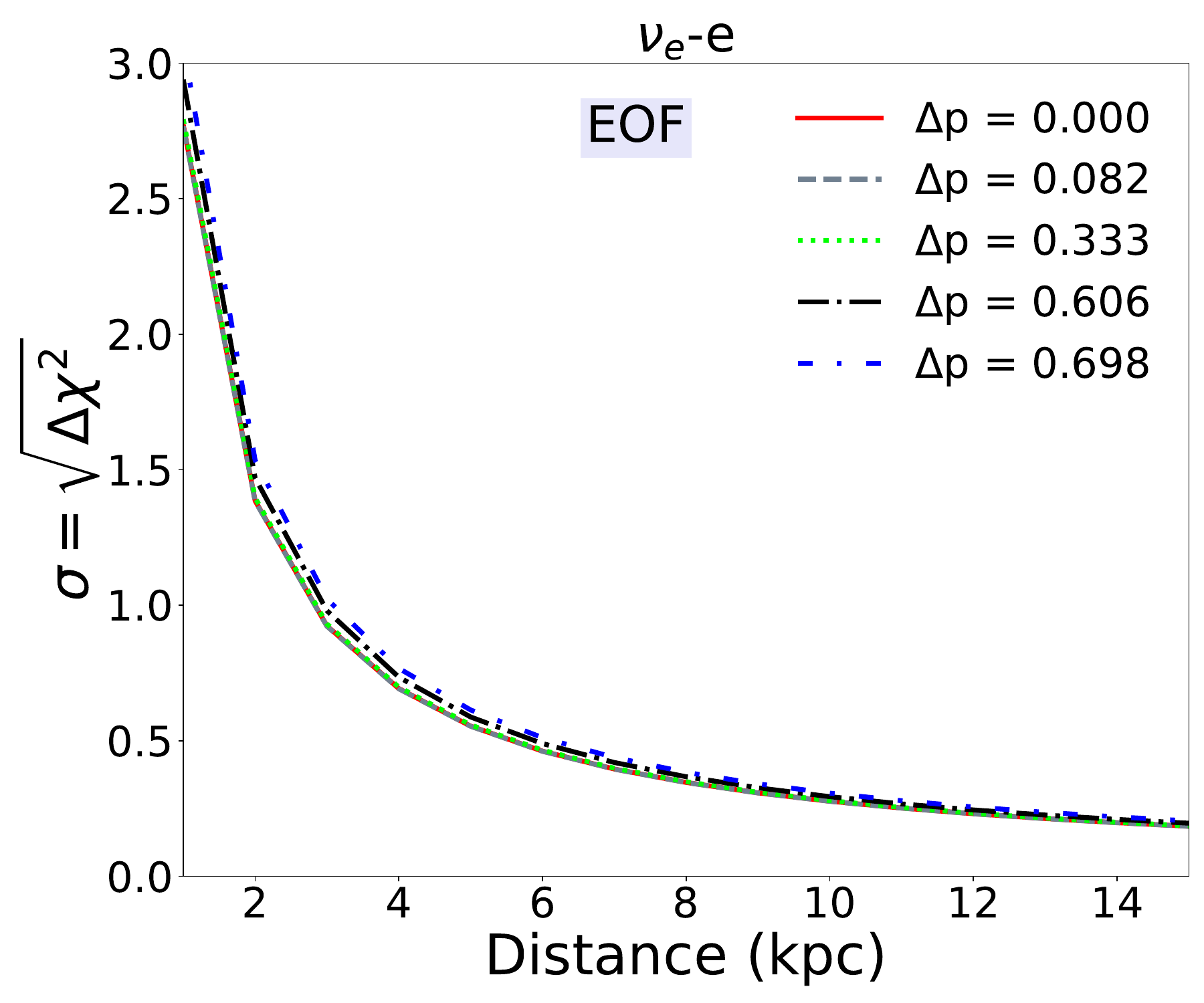}  \quad
    \includegraphics[width=0.48\linewidth]{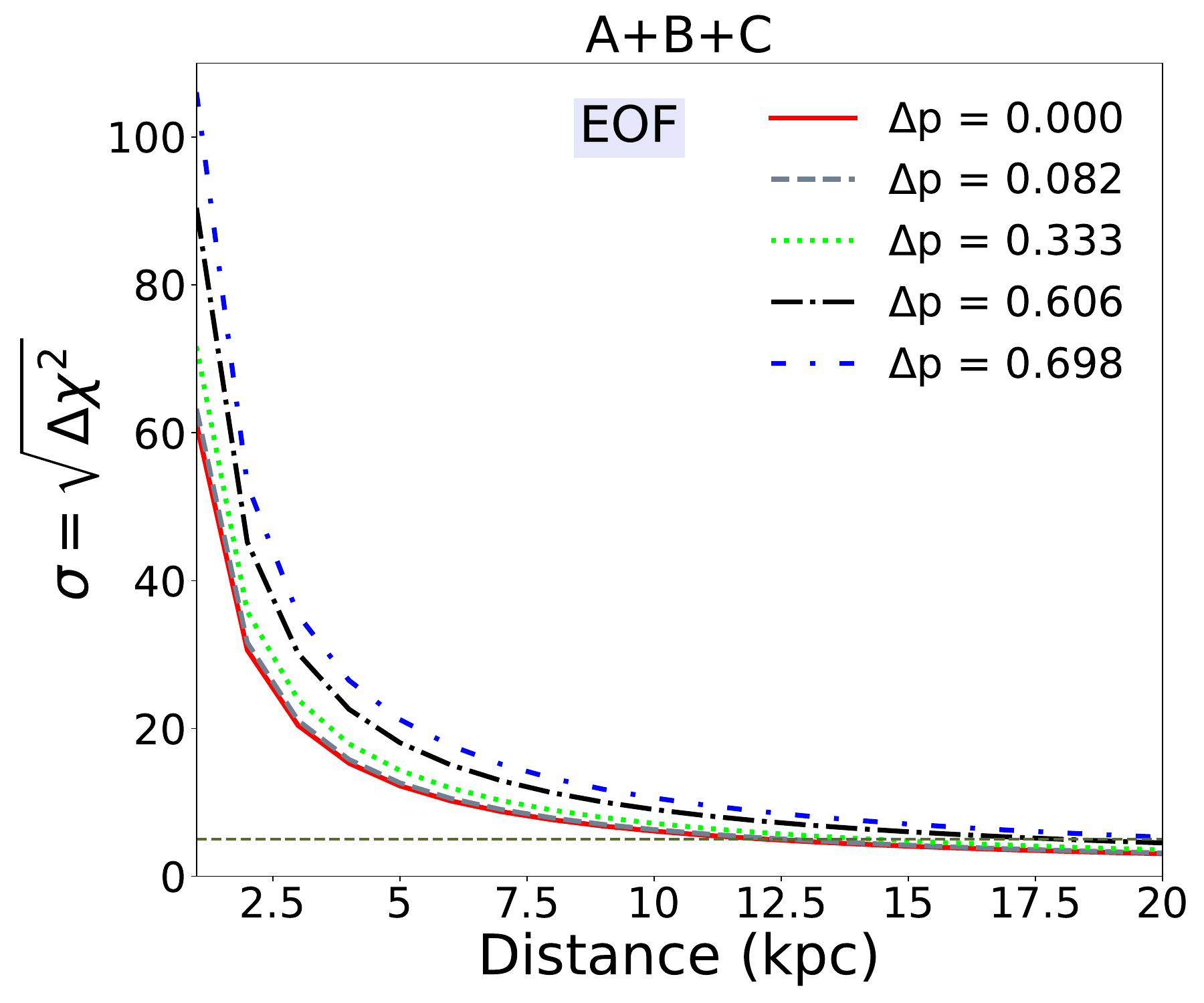}
\caption{Mass ordering sensitivity as a function of the supernova distance for the DUNE experiment, including the effects of neutrino entanglement quantified via the entanglement of formation. The upper left, upper right, lower left, and lower right panels correspond to Channel~A, Channel~B, Channel~C, and the combined (A+B+C) analysis, respectively. Results are shown for different benchmark values of the survival probability variation $\Delta p$.}
    \label{fig:MH-EOF}
\end{figure*}

The NH and IH produce distinct event rates across the three detection channels. The MH sensitivity is evaluated using Eq.~\ref{eq:chi2_stat}, based on the corresponding event spectra for NH and IH. Figure~\ref{fig:MH-EOF} shows the resulting MH sensitivity as a function of supernova distance for the three individual channels and for different values of the EOF, assuming zero systematic uncertainties. The upper left and upper right panels of Fig.~\ref{fig:MH-EOF} correspond to Channels A and B, respectively, while the lower left panel shows the sensitivity for Channel C. The lower right panel presents the synergistic MH sensitivity obtained by combining Channels A, B, and C. In addition to the standard case ($\Delta p = 0$), four representative nonzero values of $\Delta p$ (Table~\ref{tab:pee_entanglement}) are considered, corresponding to EOF $\geq 1$, for which significant deviations in the event rates arise due to their effect on EOF. The horizontal dark green dashed lines indicate the $5\sigma$ MH sensitivity. For all detection channels, the MH sensitivity decreases with increasing supernova distance, reflecting the reduction in neutrino flux.

For Channel A, the MH sensitivity reaches the $5\sigma$ level for a supernova located at approximately 12~kpc in the standard case ($\Delta p = 0$). When quantum entanglement among supernova neutrinos is included, corresponding to $\mathrm{EOF} \geq 1$, the distance at which $5\sigma$ sensitivity is achieved changes. For small entanglement ($\Delta p = 0.082$), the MH sensitivity closely follows the standard scenario. In contrast, for larger values of $\Delta p$ ($0.333$, $0.606$, and $0.698$), the MH sensitivity is significantly enhanced, allowing $5\sigma$ sensitivity to be reached at greater distances. Notably, for the maximum entanglement considered ($\Delta p = 0.698$), a $5\sigma$ MH sensitivity can be achieved for a supernova as far as 21~kpc at DUNE. This demonstrates that neutrino entanglement can substantially improve the MH sensitivity for Channel A. For Channel B, the MH sensitivity reaches the $5\sigma$ level for a supernova located at approximately 1~kpc in the standard case. The inclusion of neutrino entanglement slightly enhances the sensitivity, extending the distance at which $5\sigma$ can be achieved. In particular, for the maximum entanglement considered ($\Delta p = 0.698$), the $5\sigma$ MH sensitivity is reached at around 2~kpc. For intermediate values of $\Delta p$ ($0.333$ and $0.606$), a moderate improvement is observed, indicating that entanglement has a smaller but noticeable effect on the MH sensitivity for this channel. Owing to limited statistics, the MH sensitivity for Channel C remains below the $5\sigma$ level for all supernova distances considered. As in the other channels, the sensitivity decreases monotonically with increasing supernova distance. However, a combined analysis of all three channels significantly enhances the overall MH sensitivity, as shown in the lower right panel of Fig.~\ref{fig:MH-EOF}. For $\Delta p = 0.333$, $0.606$, and $0.698$, pronounced deviations from the standard case are observed, with $5\sigma$ sensitivity maintained up to supernova distances of approximately 15~kpc. Notably, for the maximum entanglement considered ($\Delta p = 0.698$), $5\sigma$ sensitivity can be achieved for a supernova as far as 21~kpc. The Table \ref{Tab:Sigma} shows the distance at which $5 \sigma$ C.L. can be achieved for MH sensitivity. 

\begin{figure*}[htbp]
    \centering
    \includegraphics[width=0.48\linewidth]{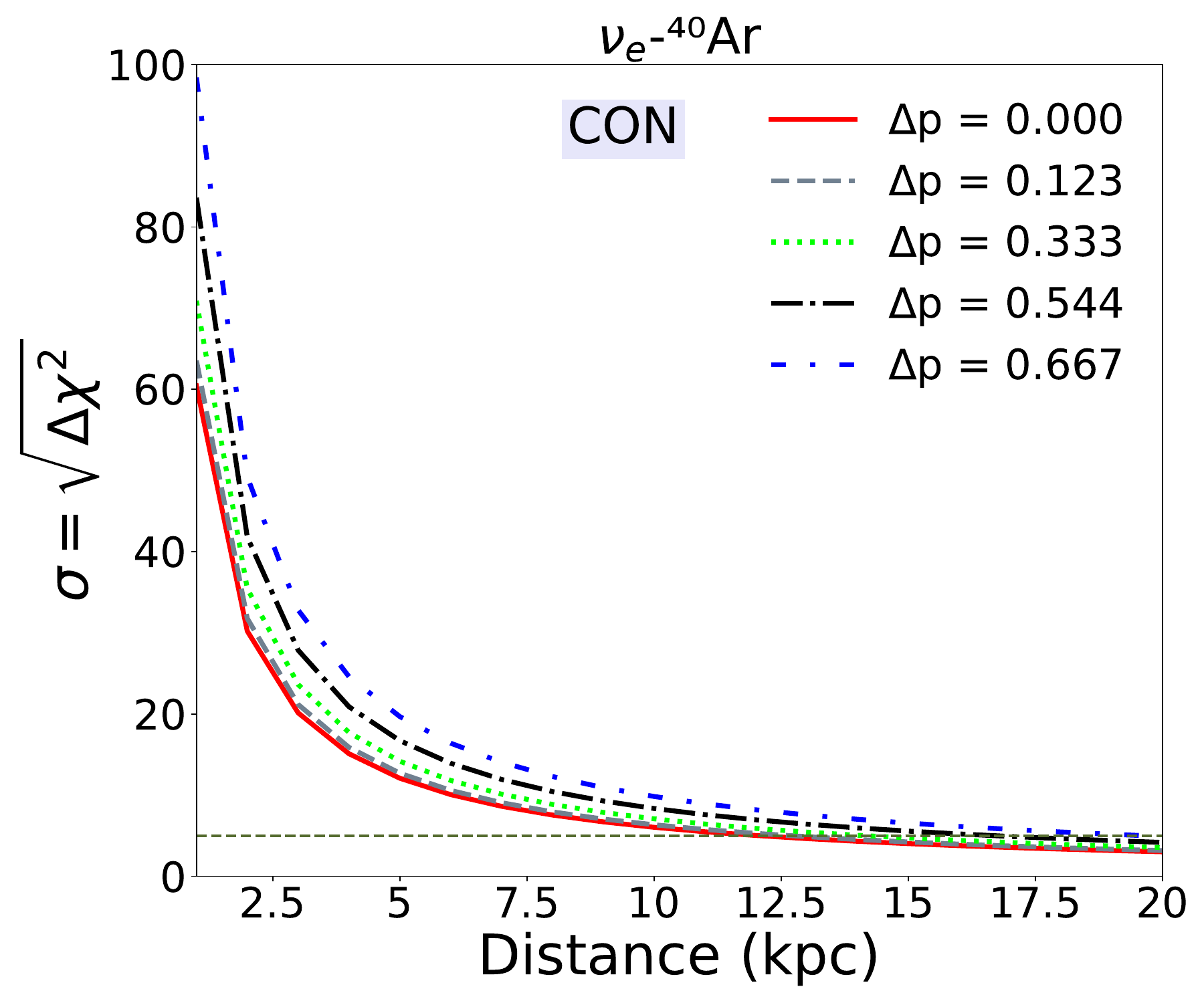}\quad
    \includegraphics[width=0.48\linewidth]{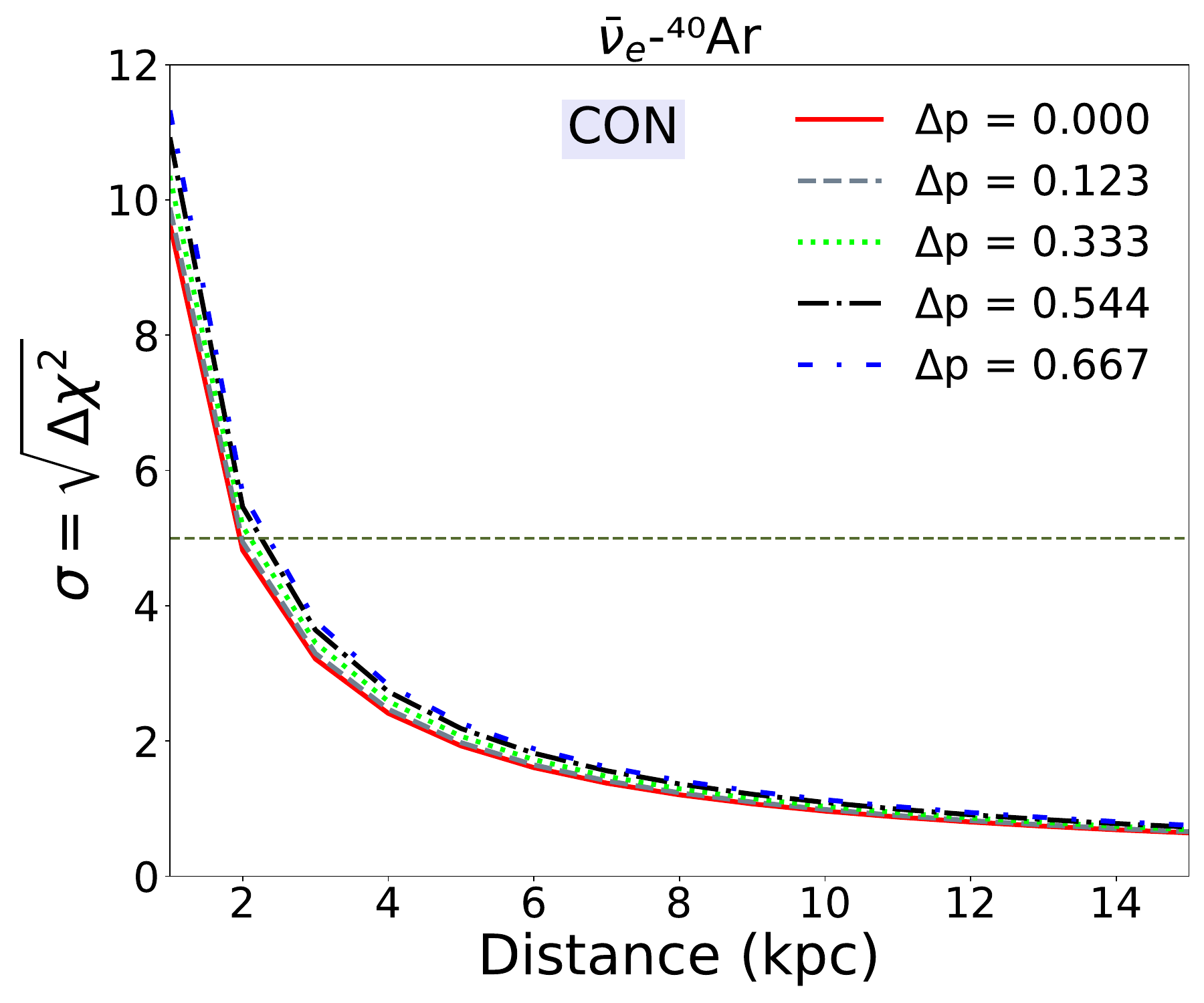}\\
    \includegraphics[width=0.48\linewidth]{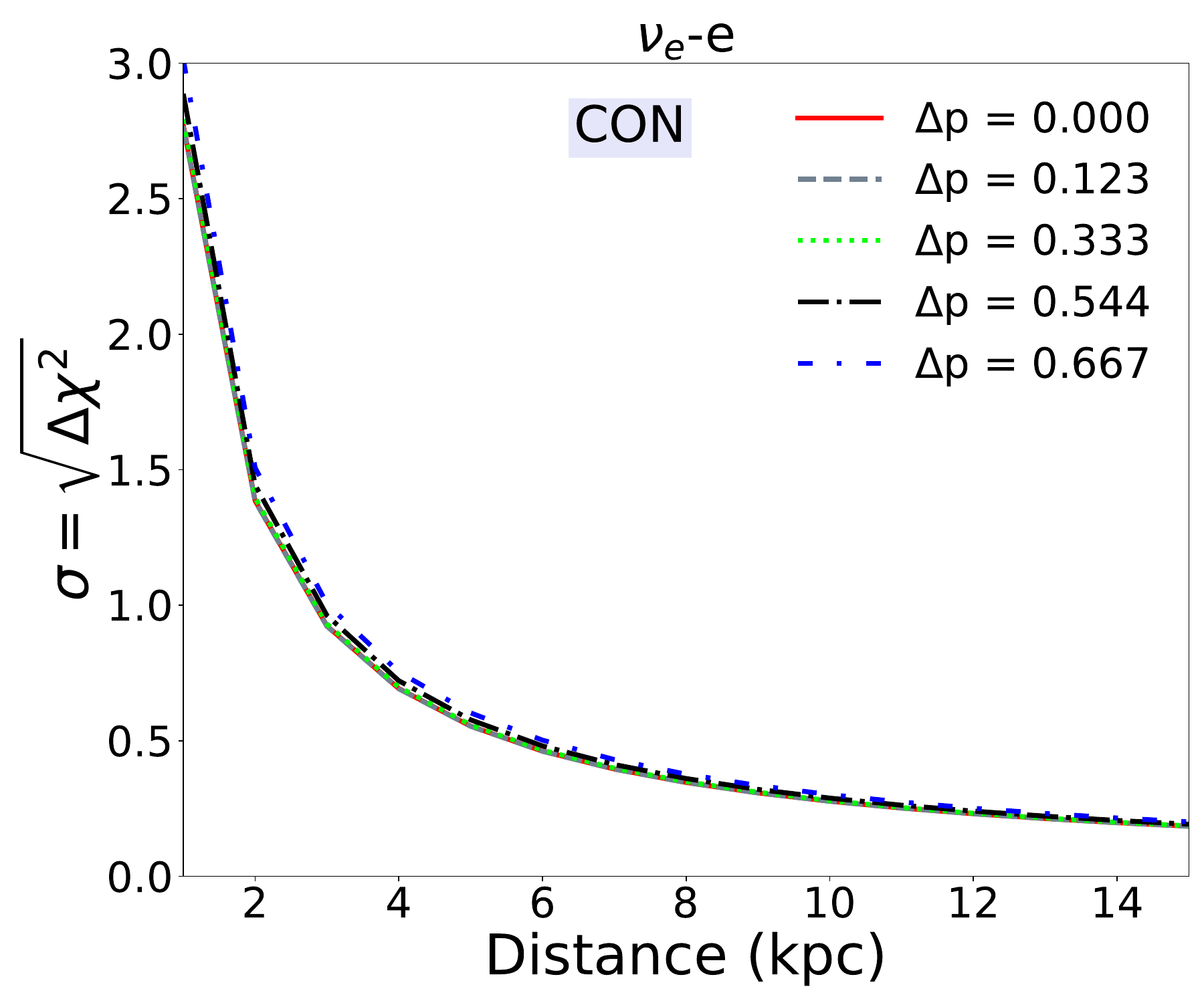}  \quad
\includegraphics[width=0.48\linewidth]{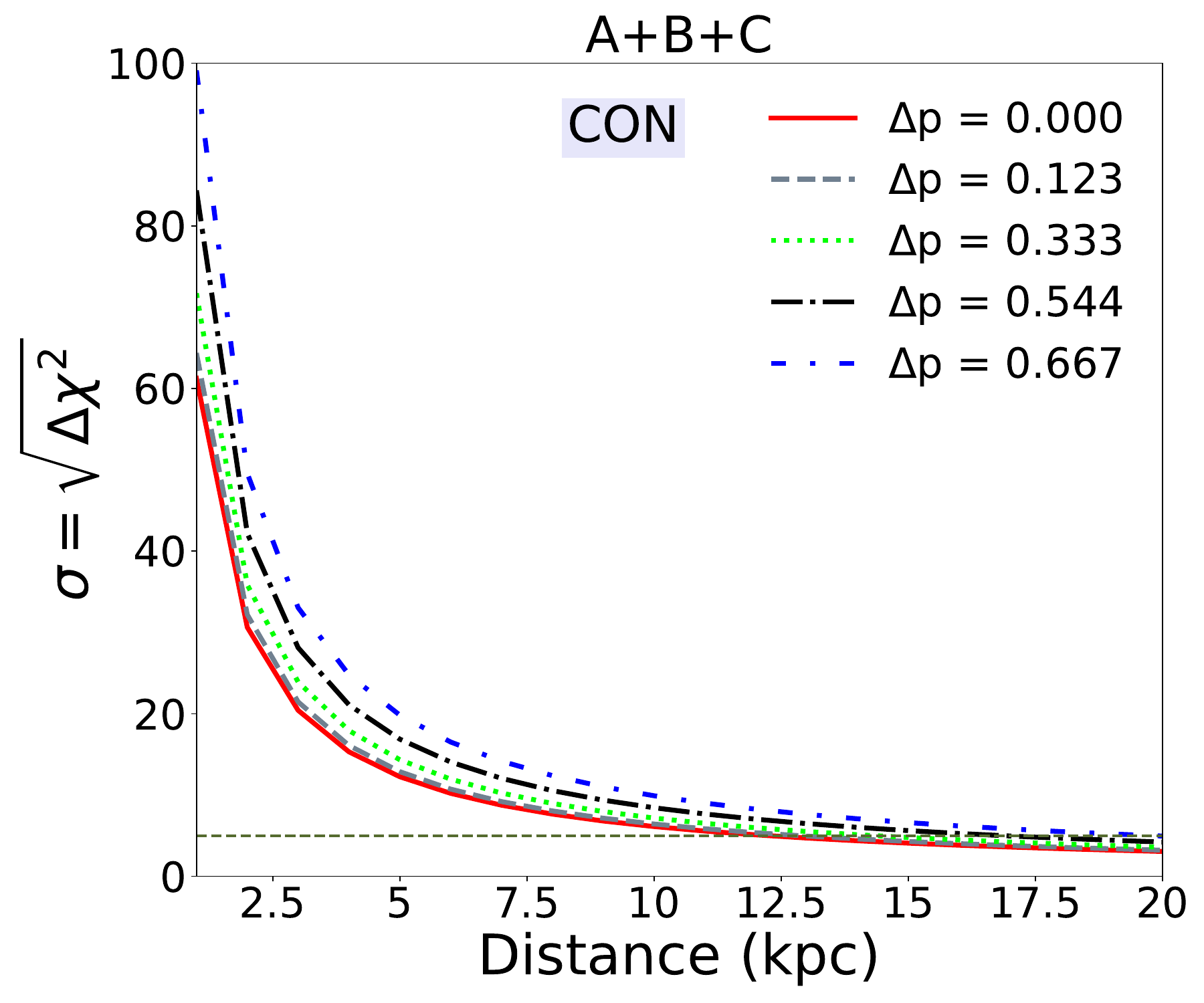}
\caption{Same as Fig.~\ref{fig:MH-EOF}, but showing results for different benchmark values of the survival-probability variation $\Delta p$ corresponding to concurrence.}
    \label{fig:MH-CON}
\end{figure*}

Figure~\ref{fig:MH-CON} shows the MH sensitivity as a function of supernova distance when neutrino entanglement is quantified using the concurrence. Similar to Fig.~\ref{fig:MH-EOF}, significant deviations from the standard case ($\Delta p = 0$) are observed for $\Delta p = 0.333$, $0.544$, and $0.667$. For Channel A, a pronounced enhancement in the sensitivity is seen, whereas for Channel B the effect is moderate, and for Channel C it remains minimal. Nevertheless, the combined analysis of all three channels improves the overall MH sensitivity both in the presence and absence of entanglement. For the maximum concurrence considered ($\Delta p = 0.667$), the $5\sigma$ MH sensitivity is achieved at approximately 19~kpc for Channel A, 2~kpc for Channel B, and 19~kpc for the combined A+B+C analysis.

\begin{figure*}[htbp]
    \centering
    \includegraphics[width=0.48\linewidth]{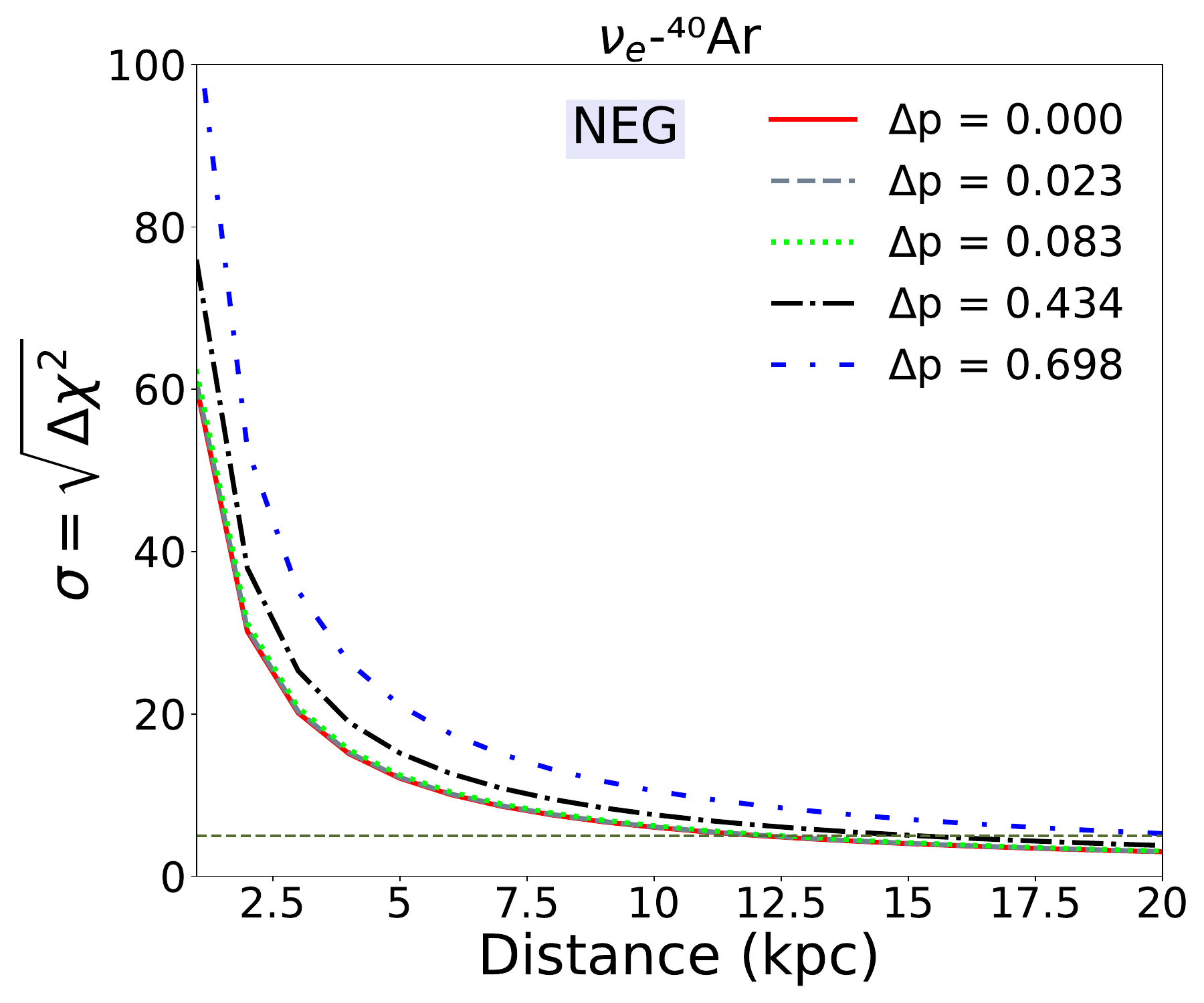}\quad
    \includegraphics[width=0.48\linewidth]{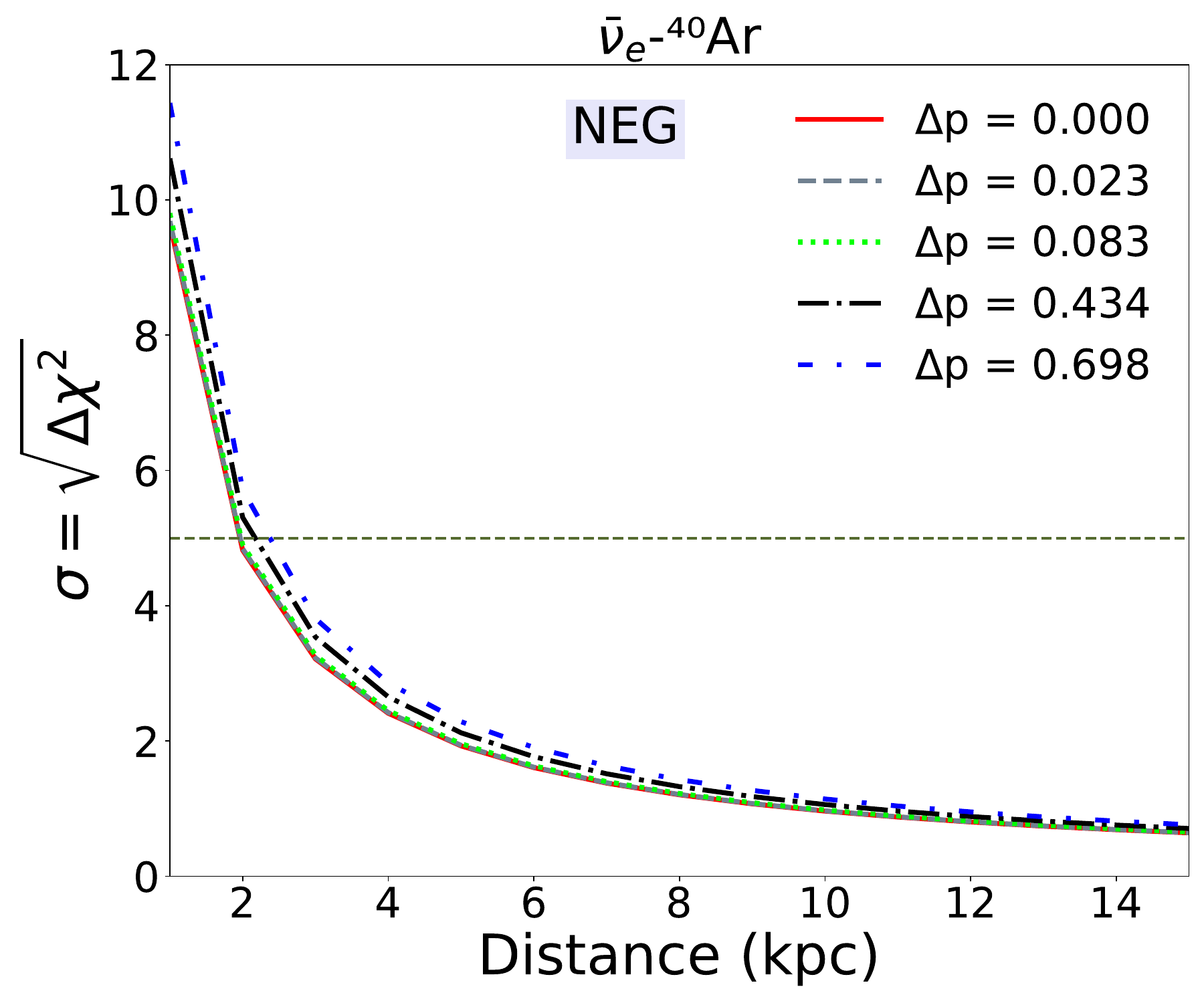}\\
    \includegraphics[width=0.48\linewidth]{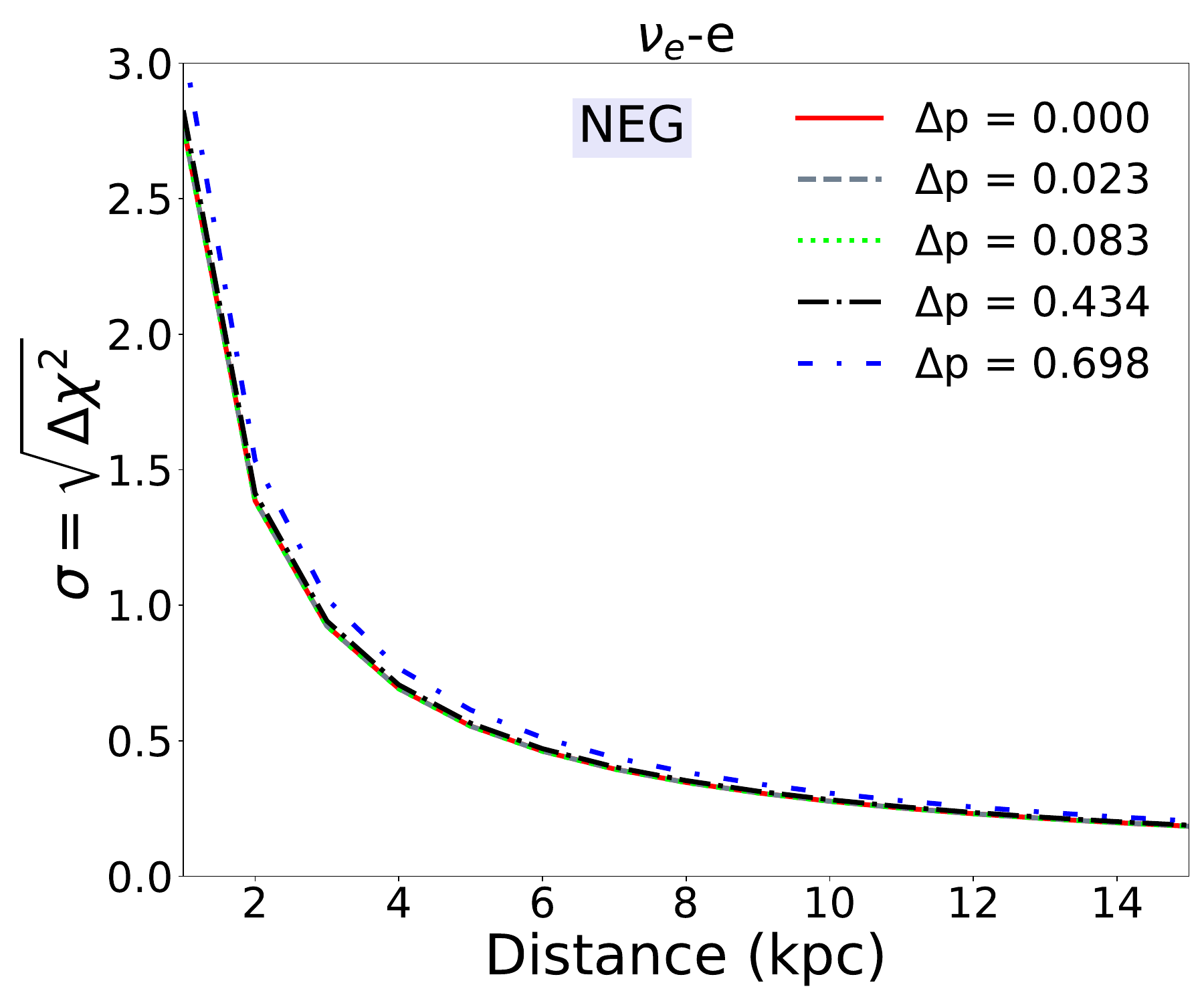}  \quad
\includegraphics[width=0.48\linewidth]{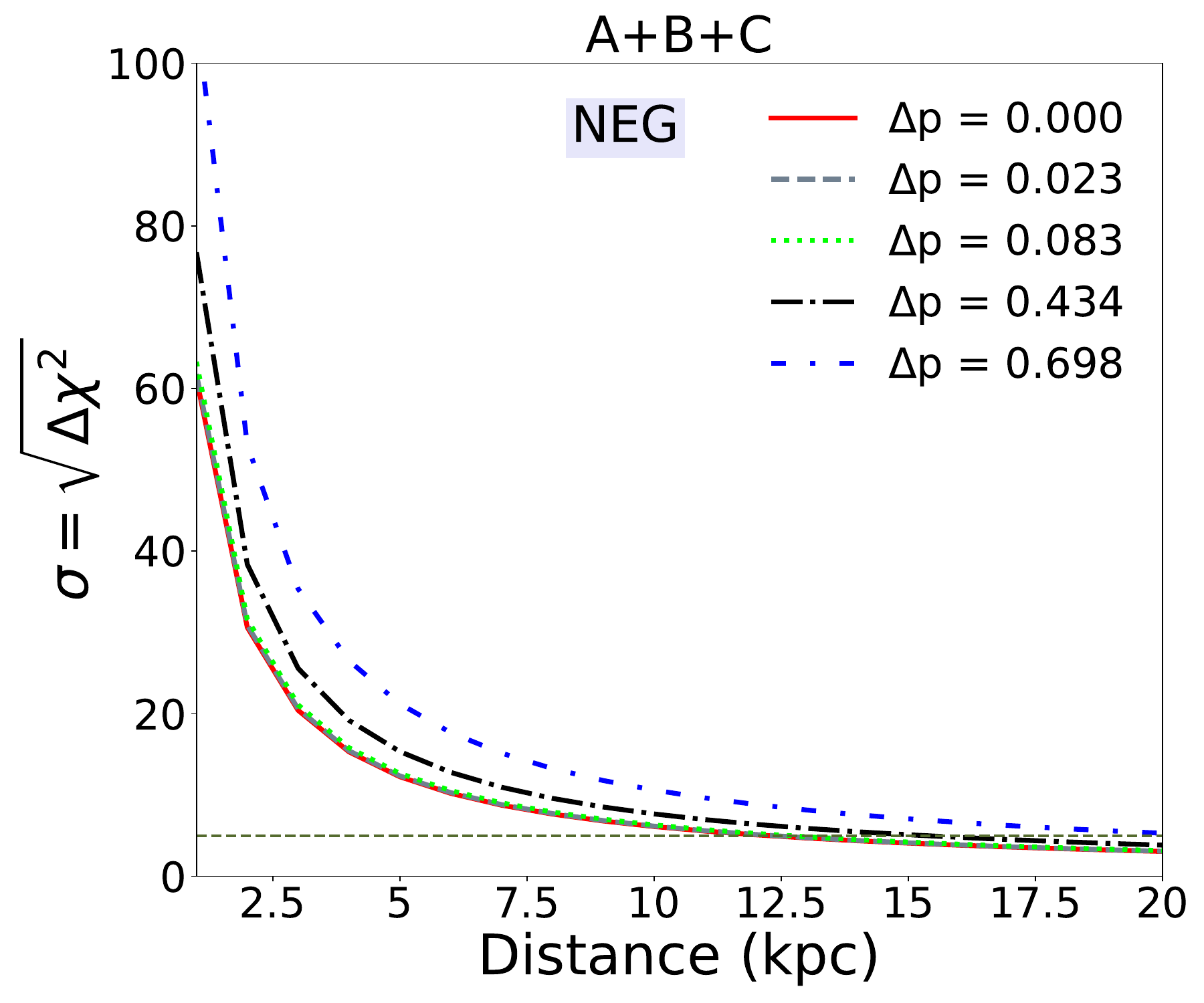} 
\caption{Same as Fig.~\ref{fig:MH-EOF}, but showing results for different benchmark values of the survival-probability variation $\Delta p$ corresponding to negativity.}
    \label{fig:MH-NEG}
\end{figure*}

Figure~\ref{fig:MH-NEG} shows the MH sensitivity as a function of supernova distance when neutrino entanglement is quantified using the negativity. Consistent with the trends observed for EOF and concurrence, the MH sensitivity exhibits clear deviations from the standard case ($\Delta p = 0$). In particular, for $\Delta p = 0.434$ and $0.698$, a noticeable enhancement in the MH sensitivity is observed compared to the unentangled scenario. 

These results suggest that the DUNE experiment has the potential not only to detect supernova neutrinos across Galactic distances but also to probe the effects of quantum entanglement among them through the resulting improvement in MH sensitivity.

\begin{table*}[htbp]
\centering
\footnotesize
\setlength{\tabcolsep}{4pt}
\renewcommand{\arraystretch}{1.0}

\begin{tabular}{|c|c|c|c|c|}
\hline
\textbf{Channels} & \textbf{Standard Case} & \textbf{Entanglement of Formation}  & \textbf{Concurrence} & \textbf{Negativity} \\
 & $\boldsymbol{\Delta p=0.000}$ & $\boldsymbol{\Delta p=0.698}$ & $\boldsymbol{\Delta p=0.667}$ & $\boldsymbol{\Delta p=0.698}$ \\
\hline
$\nu_e + {}^{40}\mathrm{Ar}$       & 12.09 & 21.09  & 19.70 & 21.09 \\
$\bar{\nu}_e + {}^{40}\mathrm{Ar}$ & 1.96 & 2.38  & 2.35 & 2.38 \\
A+B+C                              & 12.27 & 21.22  & 19.83 & 21.22 \\
\hline
\end{tabular}

\caption{Supernova distances (in kpc) at which a $5\sigma$ sensitivity to the neutrino mass ordering can be achieved at DUNE, for different entanglement measures and interaction channels.}
\label{Tab:Sigma}
\end{table*}

\begin{figure*}[htbp]
    \centering
    \includegraphics[width=0.48\linewidth]{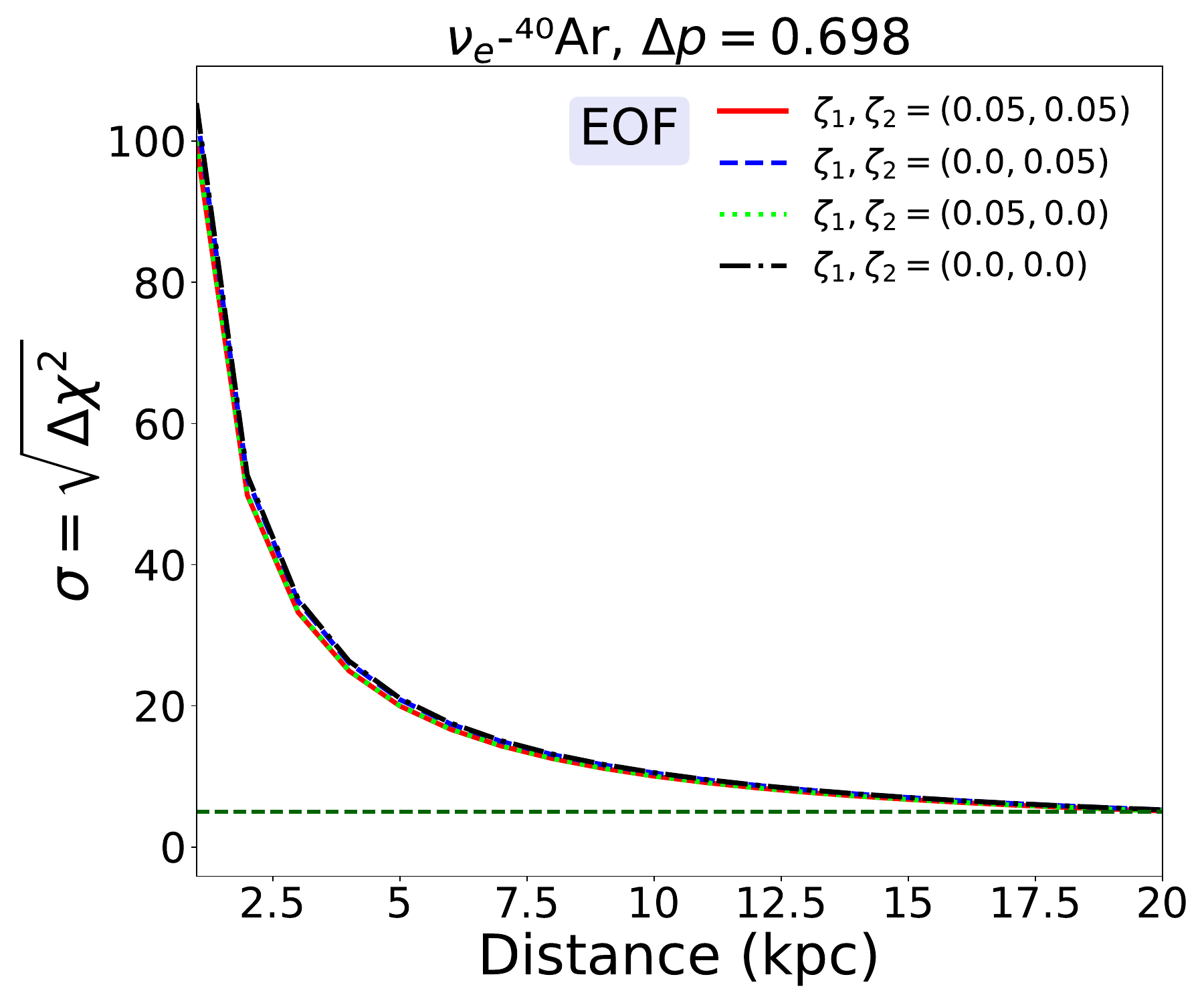}\quad
    \includegraphics[width=0.48\linewidth]{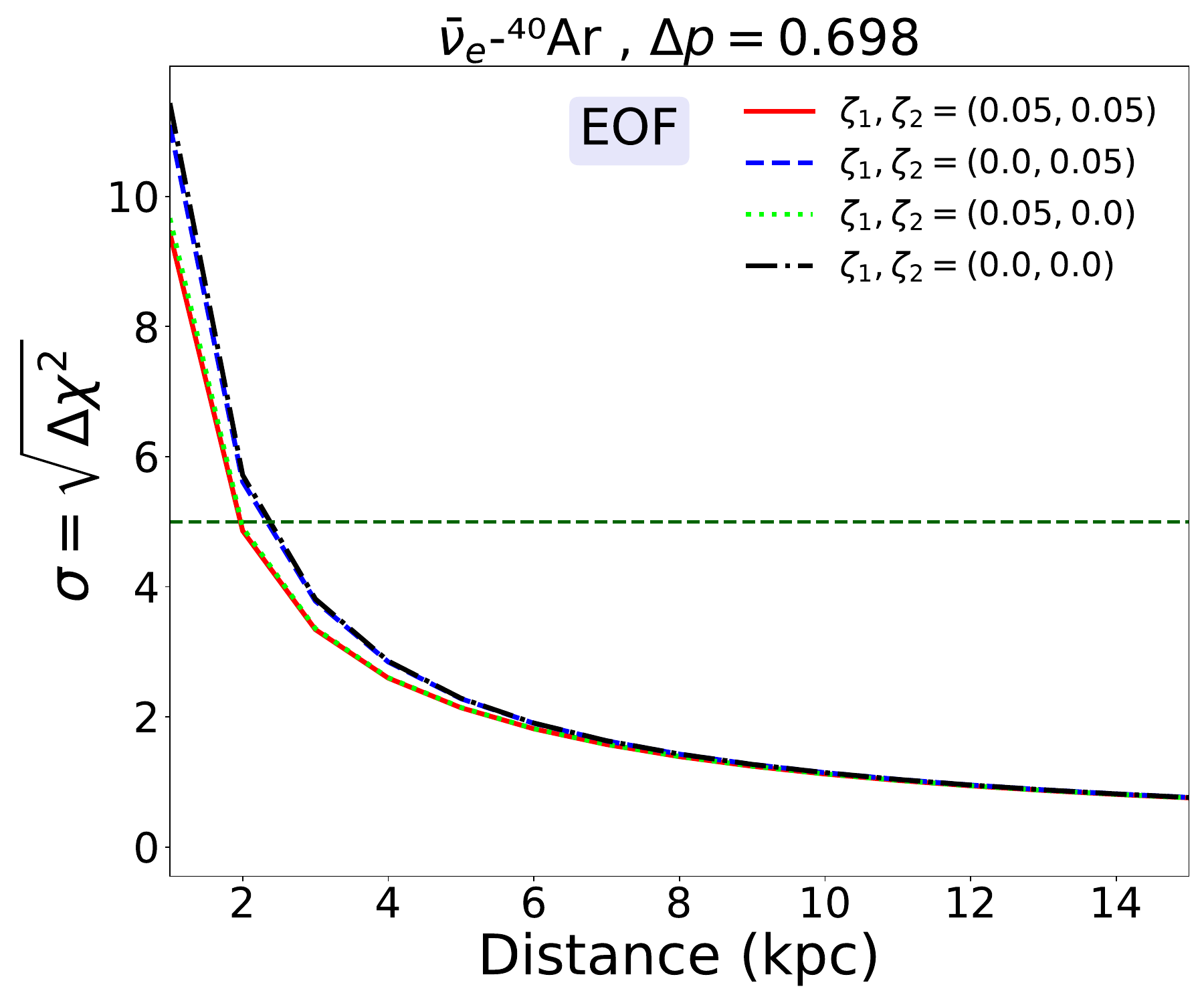}\quad
    \includegraphics[width=0.48\linewidth]{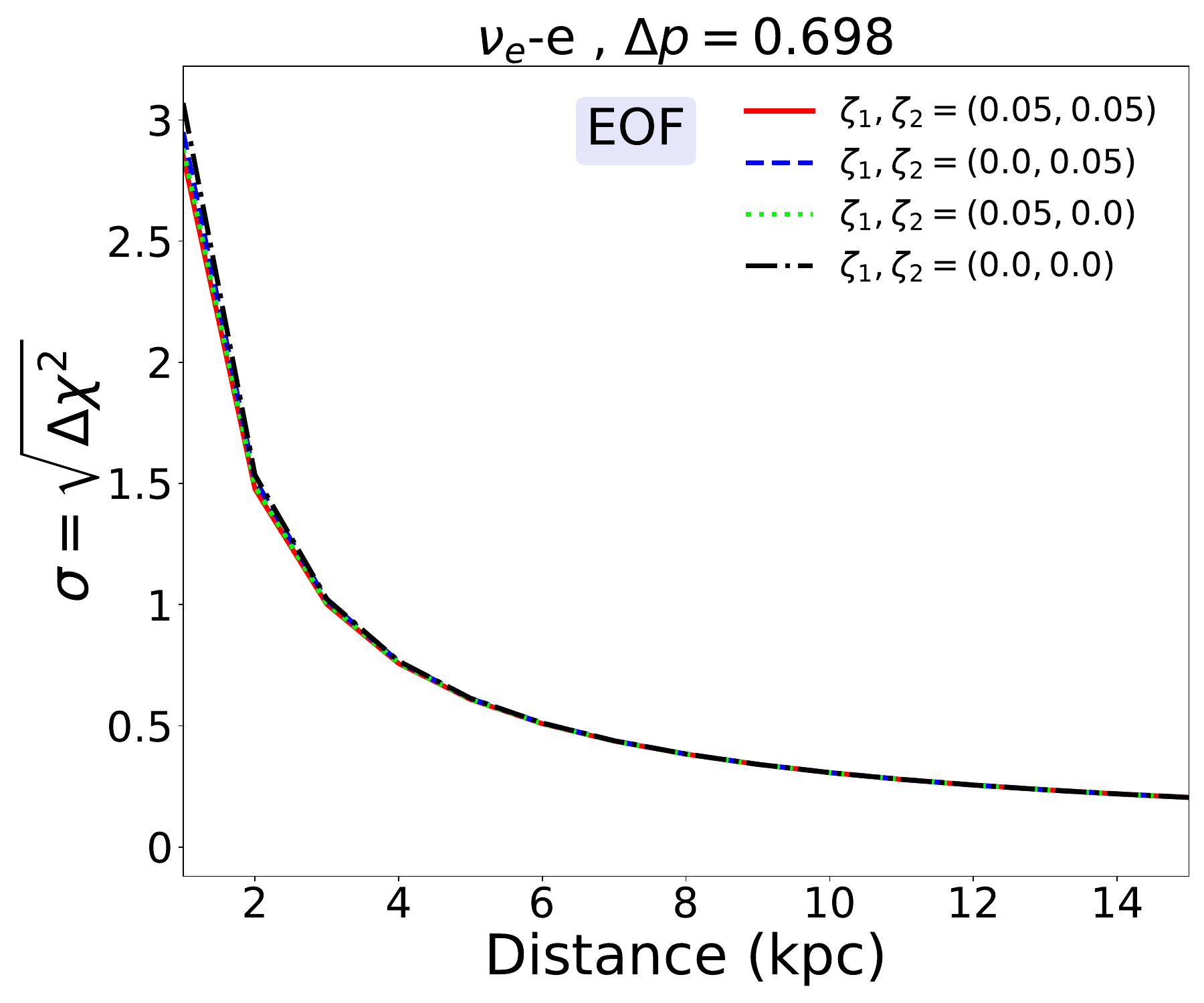}\quad
    \includegraphics[width=0.48\linewidth]{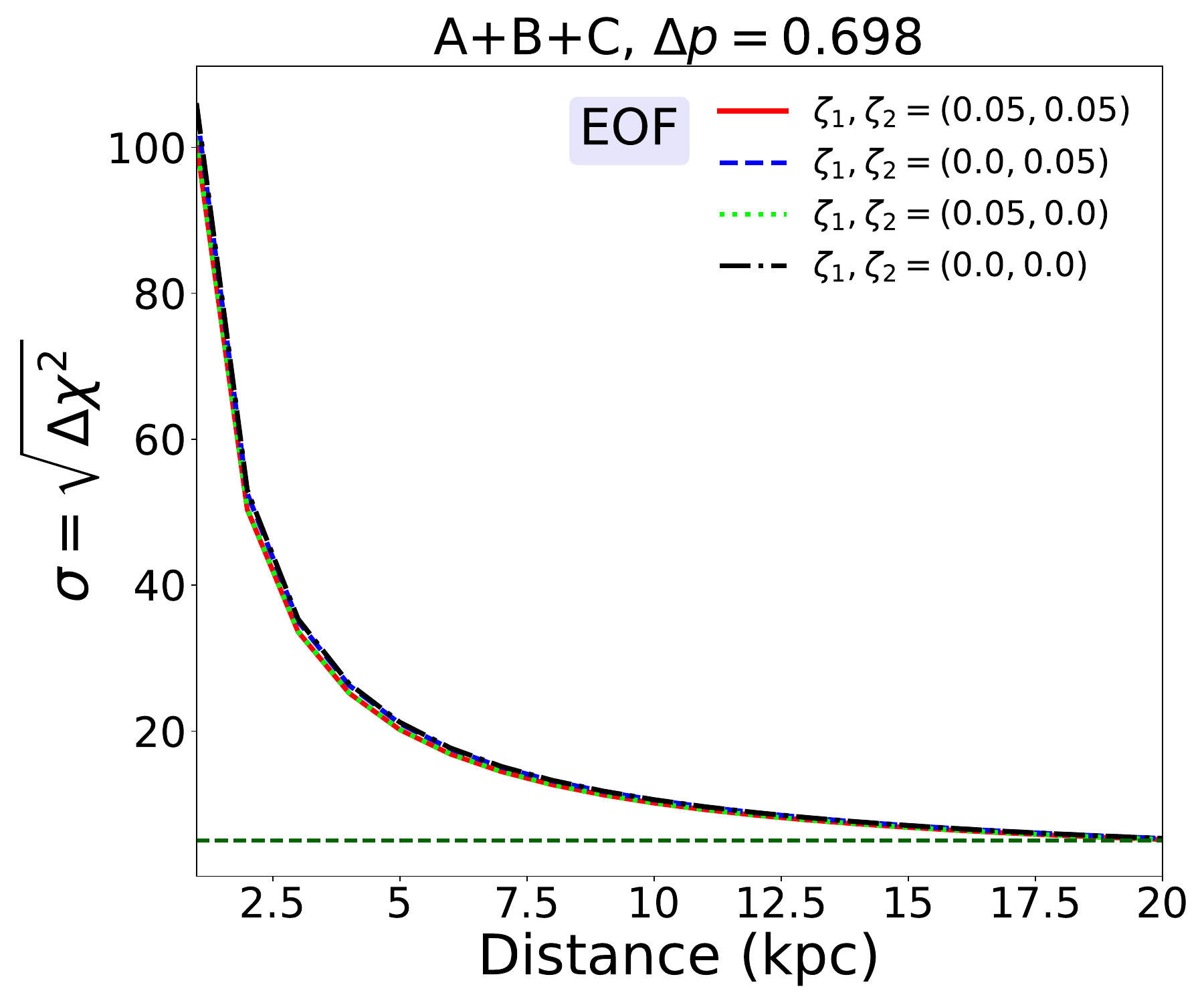}
\caption{Mass ordering sensitivity as a function of the supernova distance for a representative benchmark value of the survival probability variation $\Delta p$ associated with the entanglement of formation, including systematic uncertainties. The parameters $\xi_1$ and $\xi_2$ represent the normalization and energy calibration uncertainties, respectively, each taken to be 5\%. The upper left, upper right, lower left, and lower right panels correspond to Channel~A, Channel~B, Channel~C, and the combined (A+B+C) analysis, respectively.}
    \label{fig:MH-EOF-sys}
\end{figure*}

It is also important to study the impact of systematic uncertainties on the MH sensitivity. In this analysis, we consider a $5\%$ uncertainty in both the normalization and the energy calibration. Figure~\ref{fig:MH-EOF-sys} shows the MH sensitivity as a function of supernova distance for four cases: without systematic uncertainties, with only normalization uncertainty ($5\%$), with only energy calibration uncertainty ($5\%$), and with both uncertainties set to $5\%$. These cases are represented by the gray, cyan, blue, and red curves, respectively. We find that the effect of systematic uncertainties on the MH sensitivity is minimal for supernova distances up to approximately 20~kpc, with the overall behavior remaining similar to the scenario without uncertainties. A slight shift in sensitivity is observed for Channel B when both normalization and energy calibration uncertainties are included, or when only the $5\%$ normalization uncertainty is considered. In these cases, the $5\sigma$ MH sensitivity is reached at around 1.96~kpc for Channel B. Similar behavior is observed when neutrino entanglement is quantified using concurrence or negativity.

\section{Conclusion}
\label{sec:conclusion}

In this work, we have investigated the role of quantum entanglement among supernova neutrinos and its impact on observable signatures at the DUNE experiment. Neutrino entanglement has been quantified using three independent measures: the entanglement of formation, concurrence, and negativity. The electron neutrino survival probability $P_{ee}$ depends primarily on the mixing angles $\theta_{12}$ and $\theta_{13}$, which are precisely measured by solar and reactor experiments. In the presence of matter effects due to the MSW mechanism, this probability can deviate from the vacuum case. 

To parametrize such deviations, we introduced a phenomenological parameter $\Delta p$, which characterizes departures of $P_{ee}$ from the no-MSW scenario and serves as an effective probe of neutrino entanglement. We considered three detection channels at DUNE: channel A corresponding to $\nu_e$ charged current interactions on argon, channel B corresponding to $\bar{\nu}_e$ charged current interactions on argon, and channel C corresponding to elastic scattering on electrons. Benchmark values of $\Delta p$ associated with EOF, concurrence, and negativity were employed to study their effects on the fluence, event rates, and mass hierarchy sensitivity for the individual channels as well as for their synergistic analysis. All event rate and MH sensitivity calculations have been performed using the Garching flux model implemented in the \texttt{SNOwGLoBES} software package.

From the fluence analysis, we find that the $\nu_e$ fluence for the inverted hierarchy  is consistently higher than that for the normal hierarchy, enabling hierarchy discrimination in the absence of strong entanglement effects. However, for certain values of $\Delta p$, the NH and IH fluences can become comparable within the frameworks of EOF, concurrence, and negativity, thereby reducing the hierarchy discrimination capability. For the $\bar{\nu}_e$ fluence, the discrimination between NH and IH, as well as between vanishing and nonzero $\Delta p$, is found to be more challenging. We further studied the event rates as functions of neutrino energy for all three entanglement measures, for each detection channel and their combination. Channel A was found to dominate the event statistics, with higher event rates for the NH compared to the IH scenario. Since the event rate depends on both the supernova neutrino flux and the interaction cross section, we also examined the energy dependence of the relevant neutrino--argon cross sections for the three channels.

The MH sensitivity was analyzed as a function of the supernova distance for different values of $\Delta p$. We find that the MH sensitivity generally improves with increasing $\Delta p$ for all channels, with similar qualitative trends observed across EOF, concurrence, and negativity. We also examined the impact of systematic uncertainties on the MH sensitivity by considering normalization and energy calibration errors separately, as well as their combined effect. Our results indicate that the $5\sigma$ mass ordering sensitivity can be achieved for supernova distances of approximately 20~kpc for channel A and around 2~kpc for channel B.

In summary, our results indicate that the DUNE experiment has strong potential to detect neutrinos from a Galactic supernova and that quantum entanglement among neutrino states can leave observable imprints on the MH sensitivity. These effects, if present, could provide an additional window into the quantum nature of supernova neutrinos. More realistic supernova models and detailed detector effects can be included in future studies to better understand the impact of neutrino entanglement on observable signals. Combining data from DUNE with other neutrino detectors may further improve sensitivity to both the mass hierarchy and entanglement effects.

\acknowledgments
We gratefully acknowledge the use of the CUK HEP laboratory facility for carrying out the computational work. RM also acknowledges financial support from the Odisha State Higher Education Council (OSHEC) under the Mukhyamantri Research Innovation (MRI) Extramural Research Funding 2024.

\bibliography{biblio.bib}

\end{document}